\documentclass[preprint]{aastex}
\usepackage{multirow}
\usepackage{booktabs}
\usepackage{subfigure}
\usepackage{epic,eepic}
\usepackage{graphicx}
\usepackage{graphics}
\usepackage{hyperref}

\shorttitle{CHANDRA CATALOG}
\shortauthors{Wang et al.}

\begin{document}
\title{$CHANDRA$ ACIS SURVEY OF X-RAY POINT SOURCES: THE SOURCE CATALOG}

\author{Song Wang\altaffilmark{1}, Jifeng Liu\altaffilmark{1,2}, Yanli Qiu\altaffilmark{1,3},
Yu Bai\altaffilmark{1}, Huiqin Yang\altaffilmark{1,3},
Jincheng Guo\altaffilmark{1,3,4}, and Peng Zhang\altaffilmark{1,3,5}}

\altaffiltext{1}{Key Laboratory of Optical Astronomy, National Astronomical Observatories,
Chinese Academy of Sciences, Beijing 100012, China;\\
jfliu@bao.ac.cn, songw@bao.ac.cn}
\altaffiltext{2}{College of Astronomy and Space Sciences,
University of Chinese Academy of Sciences, Beijing 100049, China}
\altaffiltext{3}{University of Chinese Academy of Sciences, Beijing 100049, China}
\altaffiltext{4}{Harvard-Smithsonian Center for Astrophysics, 60 Garden Street, MS-10, Cambridge, MA, 02138, USA}
\altaffiltext{5}{Purple Mountain Observatory, Chinese Academy of Science, Nanjing 210008, China}

\begin{abstract}

The $Chandra$ archival data is a valuable resource for various studies on different topics of X-ray astronomy.
In this paper, we utilize this wealth and present a uniformly processed data set, which can be
used to address a wide range of scientific questions.
The data analysis procedures are applied to 10,029 ACIS observations,
which produces 363,530 source detections, belonging to 217,828 distinct X-ray sources.
This number is twice the size of the $Chandra$ Source Catalog (Version 1.1).
The catalogs in this paper provide abundant estimates of the detected X-ray source properties, including
source positions, counts, colors, fluxes, luminosities, variability statistics, etc.
Cross-correlation of these objects with galaxies shows 17,828 sources
are located within the $D_{25}$ isophotes of 1110 galaxies,
and 7504 sources are located between the $D_{25}$ and 2$D_{25}$ isophotes of 910 galaxies.
Contamination analysis with the log$N$--log$S$ relation indicates that 51.3\% of objects within 2$D_{25}$
isophotes are truly relevant to galaxies,
and the ``net'' source fraction increases to 58.9\%, 67.3\%, and 69.1\% for
sources with luminosities above $10^{37}$, $10^{38}$, and $10^{39}$ erg s$^{-1}$.
Among the possible scientific uses of this catalog,
we discuss the possibility to study intra-observation variability,
inter-observation variability, and supersoft sources (SSSs).
About 17,092 detected sources above 10 counts are classified as variable in individual observation with
the K-S criterion ($P_{\rm K-S} < 0.01$).
There are 99,647 sources observed more than once and 11,843 sources observed 10 times or more,
offering us a treasure of data to explore the long-term variability.
There are 1638 individual objects ($\sim$ 2350 detections) classified as SSSs.
As a quite interesting subclass,
detailed studies on X-ray spectra and optical spectroscopic follow-up are needed
to categorize these SSSs and pinpoint their properties.
In addition, this survey can enable a wide range of statistical studies,
such as X-ray activities in different types of stars,
X-ray luminosity functions in different types of galaxies,
and multi-wavelength identification and classification on different X-ray populations.

\end{abstract}

%\keywords{catalogs -- X-rays: general -- X-rays: binaries}
\keywords{catalogs -- X-rays: general}

\section{INTRODUCTION}
\label{intro.sec}

The study of X-ray astronomy has greatly advanced since the launch of the
$Chandra$ X-Ray Observatory. With the unparalleled subarcsecond spatial resolution
(e.g., 10 times superior to that of $ROSAT$ HRI), $Chandra$ can easily distinguish very closely
spaced point sources and differentiate them from ambient diffuse emission.
Simultaneously, with the lower background level and sensitivity limit, $Chandra$
provides a new and unique view of the X-ray sky 10--100 times deeper than previously \citep{Weisskopf2000}.
Observations using $Chandra$, in conjunction with other telescopes,
have deepened or revolutionized our understanding on many scientific
topics \citep[see][for a review]{Tananbaum2014},
from the nearest solar system objects \citep{Branduardi-Raymont2008} and exoplanets \citep{Cohen2011},
to the farthest supermassive black holes \citep{Young2005, Green2010}.
$Chandra$ observations have provided abundant information on many mysterious objects,
such as microquasars \citep{Pakull2010} and ultraluminous X-ray sources \citep[ULXs;][]{Feng2011}.
Recently, combining X-ray measures with $Chandra$, people put better constraints
on cosmological models and dark energy \citep{Vikhlinin2009}.

%For example, some recent $Chandra$ results have been reported on
%solar system objects \citep{Branduardi-Raymont2008},
%exoplanets and proplanetary disks \citep{Cohen2011},
%stellar Winds \citep{Schroter2011},
%neutron stars and matter at extreme densities \citep{Shternin2011},
%microquasars \citep{Pakull2010},
%ultraluminous x-ray sources \citep{Feng2011},
%planetary nebulae \citep{Kastner2012},
%supernova remnants as sites of particle acceleration \citep{Eriksen2011},
%young star clusters \citep{Townsley2011},
%growth of supermassive black holes (SMBHs) and host galaxies \citep{Green2010},
%accretion flow of SMBHs \citep{Young2005},
%feedback in groups and clusters of galaxies \citep{Blanton2011},
%and constraints on cosmological models and dark energy \citep{Vikhlinin2009}.

The $Chandra$ Data Archive is a wealth for various studies,
and a $Chandra$ source catalog, which include a uniform reduction of the mission data,
would be a valuable data set to address diverse scientific questions. Much work has been
done to provide catalogs of X-ray sources detected by $Chandra$.
The $Chandra$ Multiwavelength project \citep[ChaMP;][]{Kim2004} has presented an X-ray point source
catalog with $\sim$ 6800 X-ray sources detected in 149 archive observations \citep{Kim2007},
aiming to investigate the nature and evolution of quasars, galaxies, and clusters of galaxies.
The $Chandra$ Source Catalog \citep[CSC;][]{Evans2010} is a project
to uniformly reduce the $Chandra$ observations
for point source studies, including properties for 94,676 distinct X-ray sources
from Advanced CCD Imaging Spectrometer (ACIS) imaging observations during the first eight years of the $Chandra$ mission.
\citet{Liu2011} exploited the $Chandra$ archive to study X-ray point source populations
in 383 nearby galaxies within 40 Mpc with isophotal major axis above one arcmin,
leading to 17,599 independent sources from 626 public ACIS observations.
Two most sensitive surveys, $Chandra$ Deep Field-North \citep{Alexander2003}
and $Chandra$ Deep Field-South \citep{Xue2011}, are the deepest $Chandra$ surveys to date,
both of which are unique data set for studies on galactic nuclei and starburst galaxies.

\begin{figure*}[!htb]
%\figurenum{1}
\center
\includegraphics[width=0.8\textwidth]{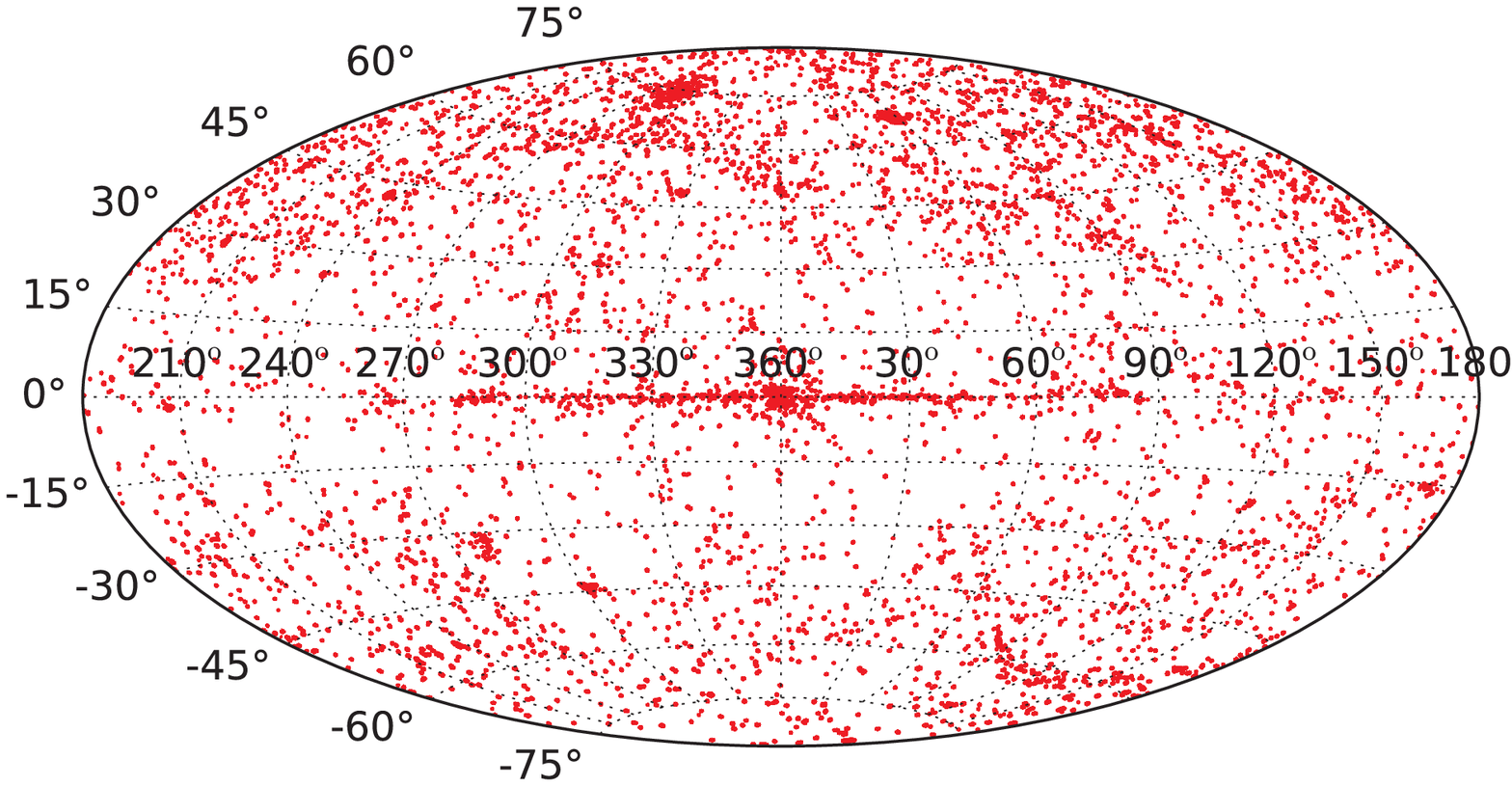}
\caption[]{The sky distribution of the 217,828 individual X-ray sources in this paper,
in Galactic coordinates.}
\label{source.fig}
\end{figure*}

In this paper, we use the $Chandra$ data archive ($\sim$ 14 years)
to provide a more complete catalog of X-ray point sources,
aiming to extract ample information of the X-ray sky and enable more constructive studies.
We present the procedures for data analysis and catalog construction following \citet{Liu2011},
at risk of repeating some structures of that paper in order to keep consistence in style with \citet{Liu2011}.
Section \ref{analysis.sec}, which comprises the bulk of the paper,
describes the uniform procedures applied to all ACIS observations,
including detecting and visually checking point sources,
estimating the source position uncertainty,
computing the source counts and colors, checking the source variability,
and extracting the spectrum and flux.
In Section \ref{multi.sec}, the cross-identification of sources in multiple
observations are described, and the upper limits are computed for sources observed but not detected.
In Section \ref{galaxy.sec}, the association of these sources with galaxies are shown,
with a contamination analysis using the log$N$--log$S$ relation.
The catalogs for 363,530 detected point sources and for 217,828 independent sources
are presented in Section \ref{catalog.sec},
and a comparison with previous catalogs is followed in Section \ref{com.sec}.
The distribution of the 217,828 independent sources on the sky is presented in \autoref{source.fig}.
In Section \ref{sci.sec}, we briefly discuss the scientific issues that can be tackled using our catalogs,
including intra-observation variability, inter-observation variability,
and the supersoft sources (SSSs).
Finally, we summarize our results in Section \ref{summary.sec}.

\section{ANALYSIS OF ACIS OBSERVATIONS}
\label{analysis.sec}

To present a uniform data set, the same procedures are applied for all ACIS observations,
including detecting and visually checking point sources,
estimating the source position uncertainty,
computing the source counts and colors, checking the source variability,
and extracting the spectrum and flux \citep{Liu2011}.
The ACIS observations were downloaded from the $Chandra$ Data Archive on December 4, 2014, which
yields 10,047 ACIS observations.
Eighteen observations with PI as ``Calibration'' or Exposure as zero were excluded (e.g., 963, 1093, 1265, and 1309).
Finally, there are 10,029 ACIS observations containing 4146 ACIS-I observations and 5883 ACIS-S observations in our sample.
For each observation, the on-axis chips are used, including all four I chips if the aimpoint is on an I chip,
and both S2 and S3 chips if the aimpoint is on S3 chip.
The exposure times for the selected observations cover a range from 50 s to 190 ks,
with a total of 221,851 ks.
All these observations were analyzed using CIAO 4.6.

\subsection{Detection and Visual Examination}
\label{detec.sec}

A wavelet detection algorithm, called {\tt wavdetect}, is used for point source detection,
which is available in the CIAO software package and
is largely used for $Chandra$ observations \citep{Freeman2002}.
This tool is more reliable in separating closely spaced point sources,
identifying extended sources, and recognizing diffuse emission
than the traditional celldetect algorithm.
Firstly, the aspect histogram, instrument map, exposure map, and PSF map are
created using {\tt asphist}, {\tt mkinstmap}, {\tt mkexpmap}, and {\tt mkpsfmap}, respectively.
The {\tt wavdetect} is then run on each on-axis chip with scales of $1^{\prime\prime}$,
$2^{\prime\prime}$, $4^{\prime\prime}$, and $8^{\prime\prime}$ in the 0.3--8 keV
band, with one normalized background created simultaneously.
The significance threshold is set to $10^{-6}$, equivalent to one
possibly spurious pixel in one CCD.
For the remaining parameters, we used the default values given in CIAO.

\begin{figure*}[!htb]
%\figurenum{2}
\center
\subfigure{\includegraphics[width=0.3\textwidth]{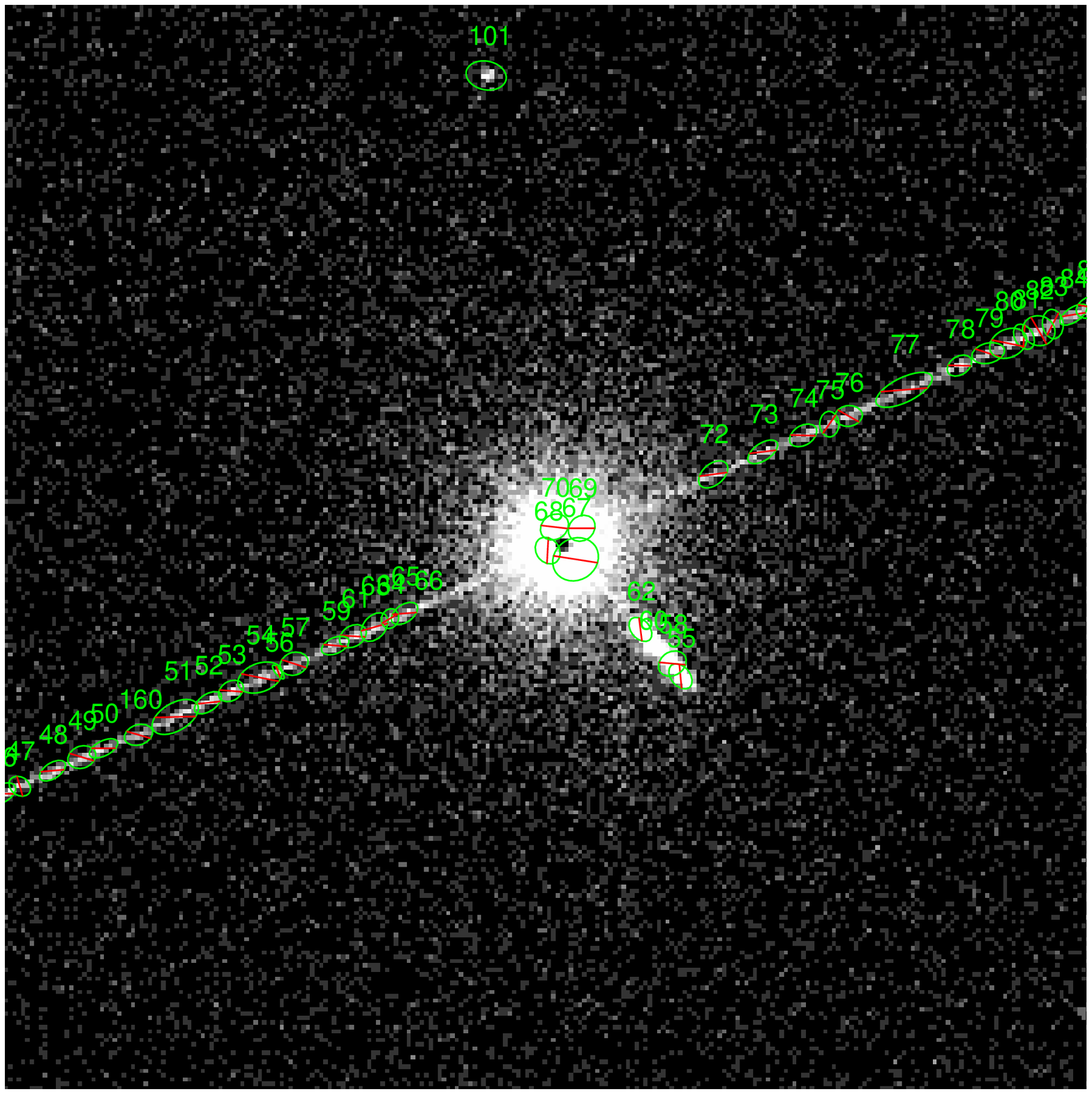}}
\put (-80,-15) {$(1)$}
\hspace{2 mm}
\subfigure{\includegraphics[width=0.3\textwidth]{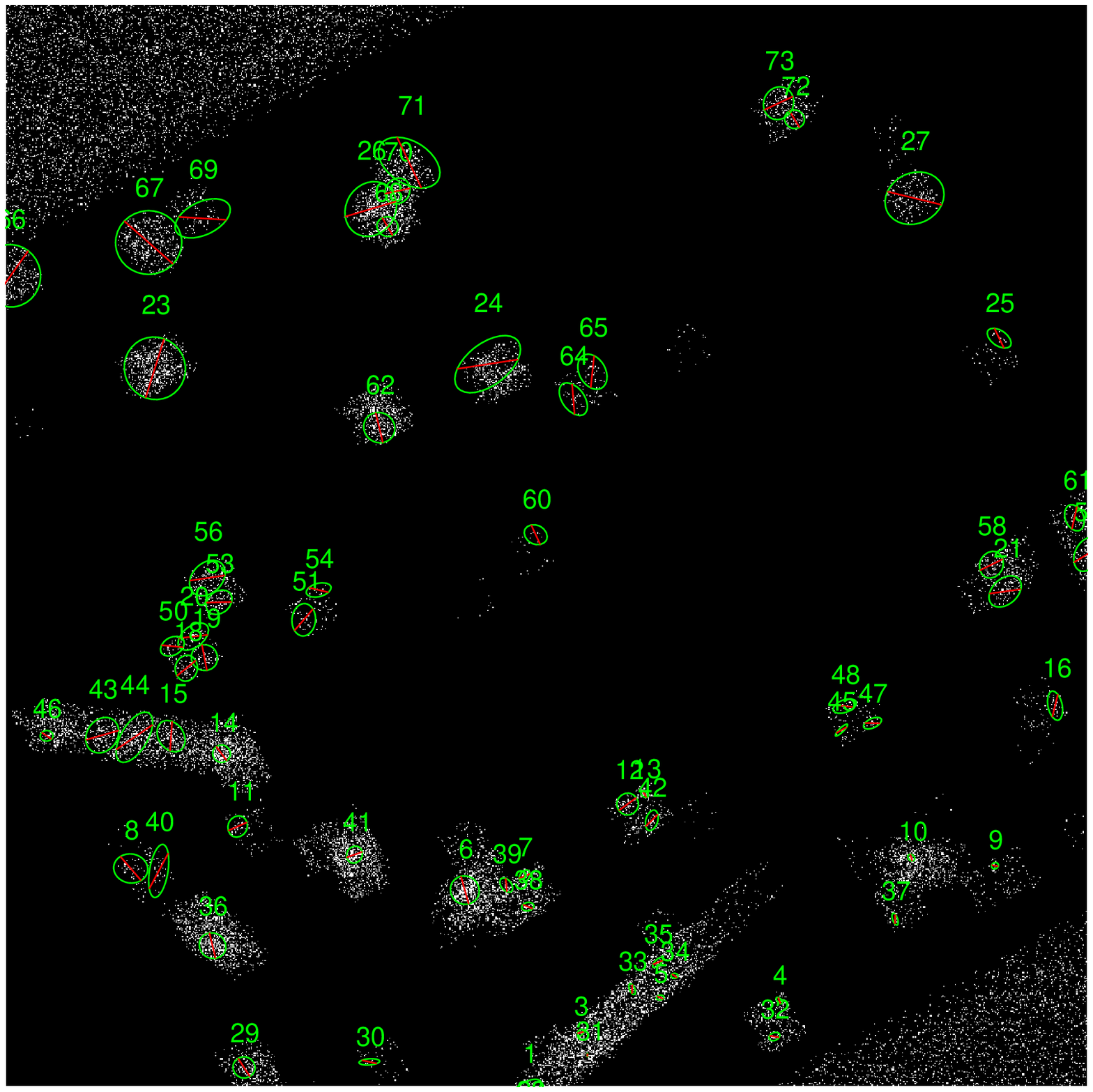}}
\put (-80,-15) {$(2)$}
\hspace{2 mm}
\subfigure{\includegraphics[width=0.3\textwidth]{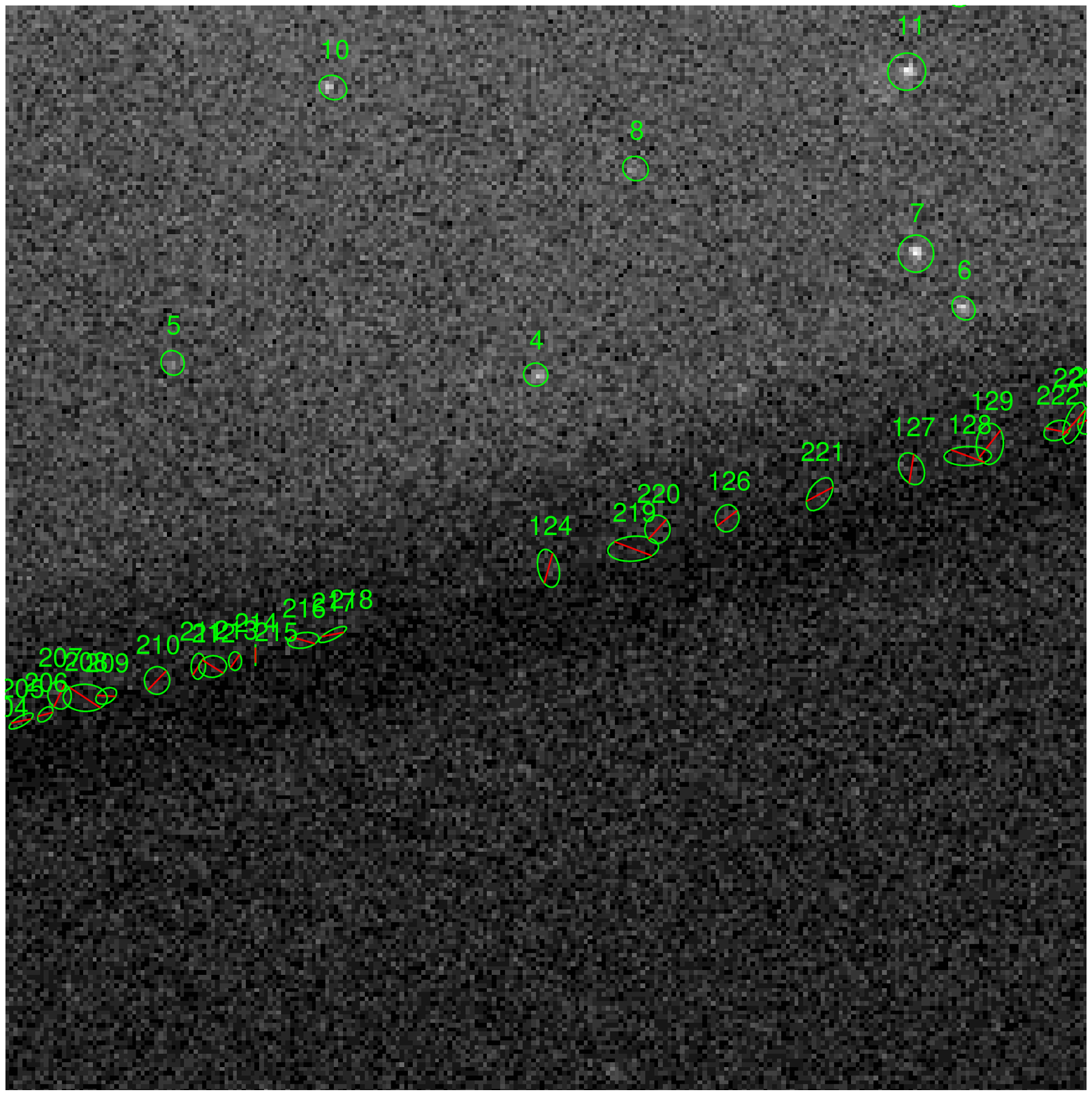}}
\put (-80,-15) {$(3)$} \\
\subfigure{\includegraphics[width=0.3\textwidth]{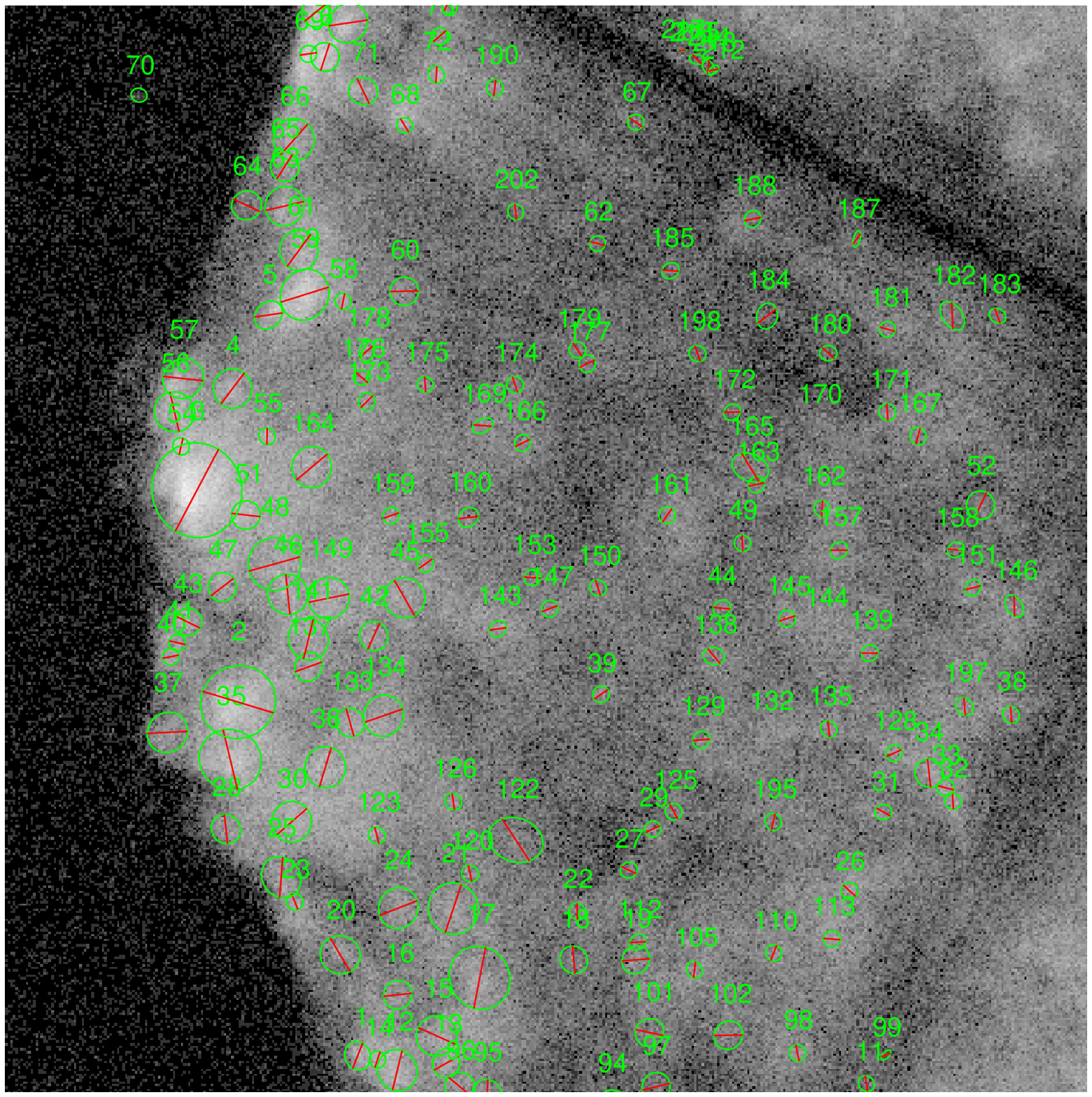}}
\put (-80,-15) {$(4)$}
\hspace{2 mm}
\subfigure{\includegraphics[width=0.3\textwidth]{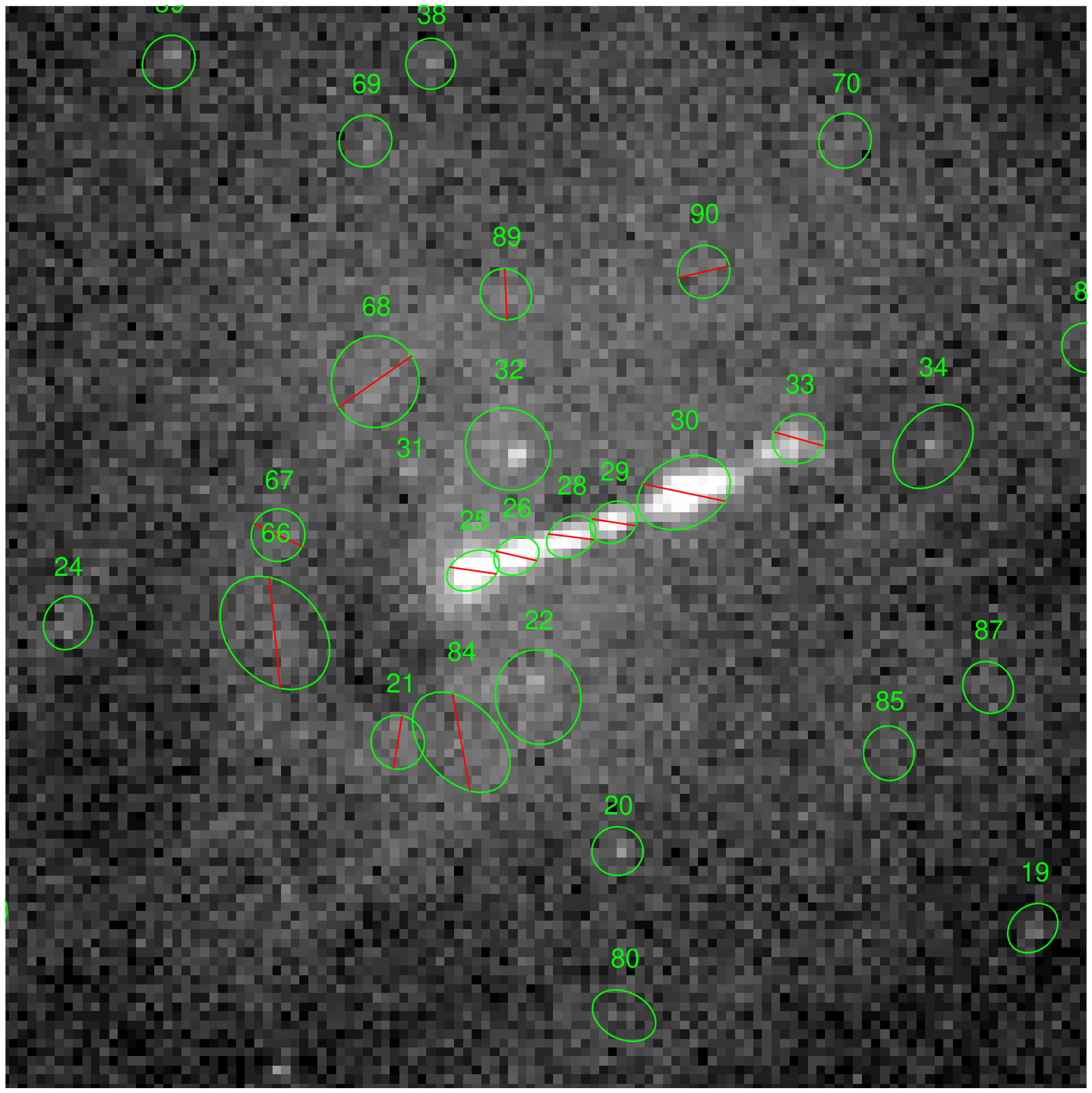}}
\put (-80,-15) {$(5)$}
\hspace{2 mm}
\subfigure{\includegraphics[width=0.3\textwidth]{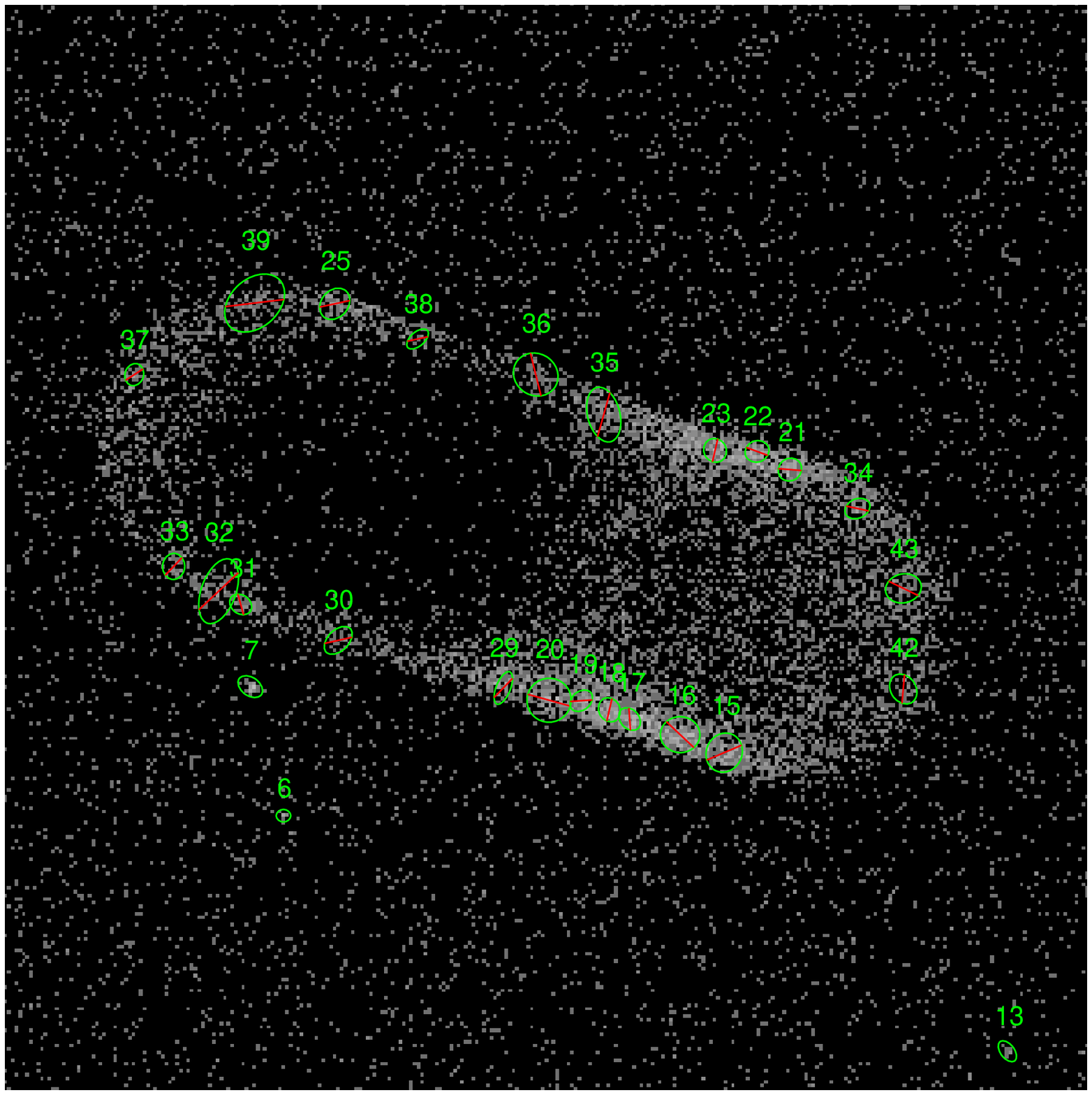}}
\put (-80,-15) {$(6)$}
\caption[]{Spurious point sources removed with visual examinations:
(1) false detections in the ACIS readout streak;
(2) bad detections due to the spacecraft dither motion in Lissajous patterns;
(3) a stream of sources on the CCD edges;
(4) sources as diffuse emission in SNR or starburst regions;
(5) sources as bright knots from the X-ray jets;
(6) sources in the Solar system.}
\label{spurious.fig}
\end{figure*}

Although {\tt wavdetect} performs well in identifying both point sources and extended sources,
some instrumental or observational issues may result in spurious sources (\autoref{spurious.fig}),\\
(1) False detections in the ACIS readout streak.\\
(2) Bad detections due to the spacecraft dither motion in Lissajous patterns, which
aims to minimize the effect of bad detector pixels,
and to avoid possible burn-in degradation of the camera by bright X-ray sources \citep{Evans2010}.\\
(3) A stream of sources on the CCD edges.\\
(4) Sources as diffuse emission in supernova remnant (SNR) or starburst regions (e.g., Crab Nebula).\\
(5) Sources as the bright knots from the X-ray jets (e.g., M87).\\
(6) Sources in the Solar system (e.g., Jupiter).\\
All above spurious sources are removed from the source lists
after visually examining the detected sources overlaid on the X-ray images.
An automatic procedure to help visual examination is developed based on XPA and DS9.
When the source regions are automaically loaded to each image, we identify the false detections,
and input the numbers of the false detections to the terminal or select them on the image,
then these detections would be automatically removed from the source list.
The visual examination is a tremendous and job.
Four reviewers are trained by senior researchers about which cases are false detections.
First, the false detections are introduced and explained by the senior researchers,
and an exercise using a sample of observations are performed for the four reviewers.
The exercise results are shown and discussed by the reviewers,
and the same criteria for excluding spurious sources are established.
For example, all detections but the central point source (in some cases) and surrounding isolated sources in the diffuse emission region
are removed from the source list, such as the green ellipses with red lines in Figure \ref{spurious.fig}(4). 
To be conservative, each observation was visually checked twice by different reviewers to avoid missing
false detections. It takes us about one month to complete the visual examination, and
we believe most of the sources in these six cases are removed from the source lists.

On a rare occasion, {\tt wavdetect} may incorrectly split a point source affected by serious pile-up
into several sources.
In such cases, the split sources are merged as single sources.
Finally, 363,530 point sources are detected and verified from the 10,029 observations.

\subsection{Position Uncertainties}
\label{poserr.sec}

$Chandra$ provides high on-axis positional accuracy less than $1^{\prime\prime}$.
For each source detection, {\tt wavdetect} presents the positional uncertainty
based on a statistical moments analysis, but without
instrumental effects (e.g., pixelization, aspect-included blur)
considered \citep{Evans2010}.
Therefore, the position errors estimated by {\tt wavdetect} would be underestimated
for sources at large off-axis angles (OAAs).

\begin{figure*}[!htb]
%\figurenum{3}
\center
\includegraphics[width=0.7\textwidth]{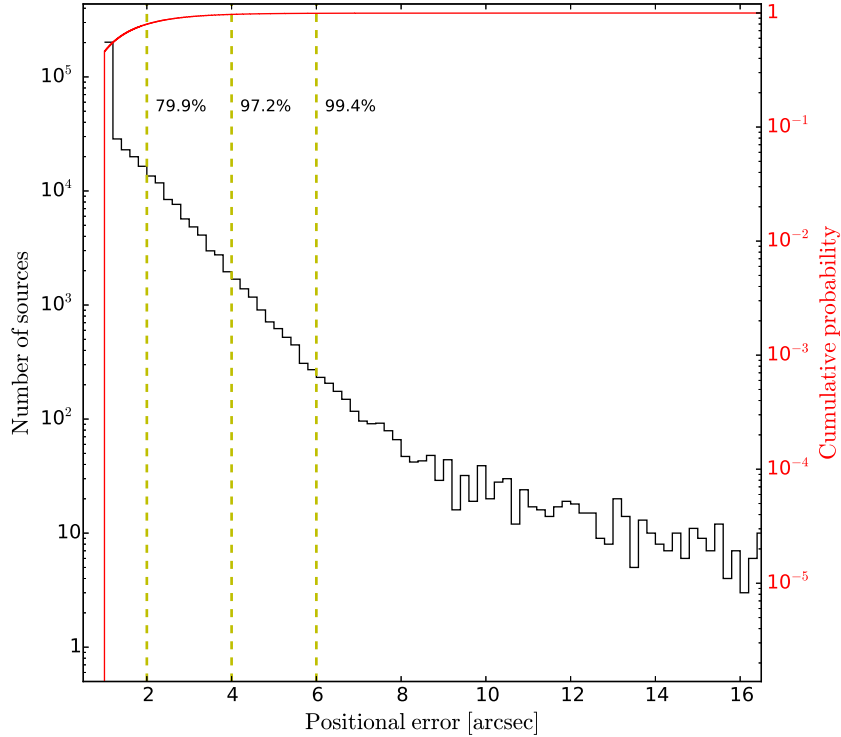}
\caption[]{Histogram of the positional errors for the 363,530 X-ray point sources, which
are calculated following \citet{Kim2007} with a minimum value of 1$''$.
The positional errors are smaller than 2$''$/4$''$/6$''$ for 79.9\%/97.2\%/99.4\% of these sources.}
\label{poserr.fig}
\end{figure*}

Simulations by \citet{Kim2004} showed the positional error is usually less than
$1^{\prime\prime}$ for a bright source, regardless of its OAA;
while for a weak source, it can increase to $4^{\prime\prime}$
at a large OAA (${\rm OAA} > 8^{\prime}$).
\citet{Kim2007} provided empirical relationships
for positional error as a function of both net counts and OAA,
with the formulae as,
\begin{equation}
%\footnotesize
{\rm
log (P.E.) = \left\{
\begin{array}{lcl}
0.1145 \times {\rm OAA} - 0.4958 \times {\rm log(C)} + 0.1932, & & 0.0000 < {\rm log(C)} \leq 2.1393, \\
0.0968 \times {\rm OAA} - 0.2064 \times {\rm log(C)} - 0.4260, & & 2.1393 < {\rm log(C)} \leq 3.3000.
\end{array}\right.}
\end{equation}
Here the formulae are adopted to calculate positional errors for the 363,530 detections.
Although there are 2658 sources with counts more than 10$^{3.3}$,
beyond the valid range of the formulae,
their positional errors are still estimated using the same formula as counts in the
range of 10$^{2.1393}$ to 10$^{3.3}$.
In addition, a minimum value of 1$^{\prime\prime}$ is applied for these sources.
The positional errors for the detected sources are shown in \autoref{poserr.fig},
with 79.9\% smaller than $2^{\prime\prime}$ and 99.4\% smaller than $6^{\prime\prime}$.

\subsection{Counts}
\label{count.sec}

The photon counts of sources are computed in two ways, {\tt wavdetect} and aperture photometry.
{\tt wavdetect} computes the net source counts as the total counts in the image pixels included in the source cell
minus the total estimated background counts in the image pixels included in the source cell.
The aperture photometry is performed by summing the
photons within the $3\sigma$ elliptical source region, which is constructed by {\tt wavdetect} by
fitting a 2D elliptical Gaussian to the distribution of (observed) counts in the source cell,
with background counts subtracted.
The background region is taken as an elliptical annulus around the source region
with the inner/outer radius as 2/4 times the radius of the source ellipse.
Therefore, the net source count from aperture photometry is computed as $C_N = C_s - C_b\times A_s/A_b$,
and the error is computed as $C_E = 1 + \sqrt{0.75 + C_s + C_b \times (A_s/A_b)^2}$ \citep{Gehrels1986},
where $C_s$ is the raw source count, $C_b$ is the background count,
$A_s$ is the source region area, and $A_b$ is the background region area.
Although \citet{Ebeling2003} proposed improved numerical approximations to the Poissonian confidence limits
for small numbers $n$ of observed events, the \citet{Gehrels1986} expression were still used
for its structurally simple formulae and sufficient accuracy in most cases.
In addition, symmetrical errors are applied for all the detections.

\begin{figure*}[!htb]
%\figurenum{4}
\center
\includegraphics[width=0.7\textwidth]{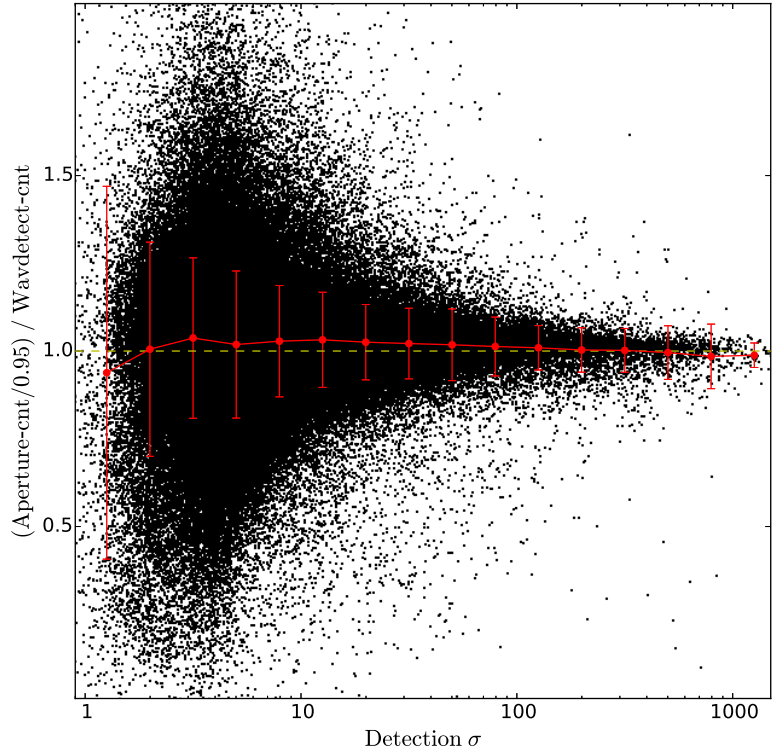}
\caption[]{Comparison of the photon counts from {\tt wavdetect} and from aperture photometry.
The red circles and the error bars indicate the average ratio and dispersion for detection $\sigma$ intervals,
and the yellow dotted line indicates a ratio of 1.}
\label{wav_aper.fig}
\end{figure*}

The comparison of the source counts obtained from the two methods are shown in \autoref{wav_aper.fig}.
For sources with detection significance $\sigma>20$, the computed counts from
aperture photometry agree well with 95\% of the {\tt wavdetect} counts,
with 1$\sigma$ dispersion of 3\%--10\%.
However, for sources at lower detection significance ($\sigma<3$), the computed counts from aperture photometry could
significantly deviate from the {\tt wavdetect} counts, for example, the average counts may deviate
more than 5\% for those detections with $\sigma<2$.

\begin{figure*}[!htb]
%\figurenum{5}
\center
\includegraphics[width=0.95\textwidth]{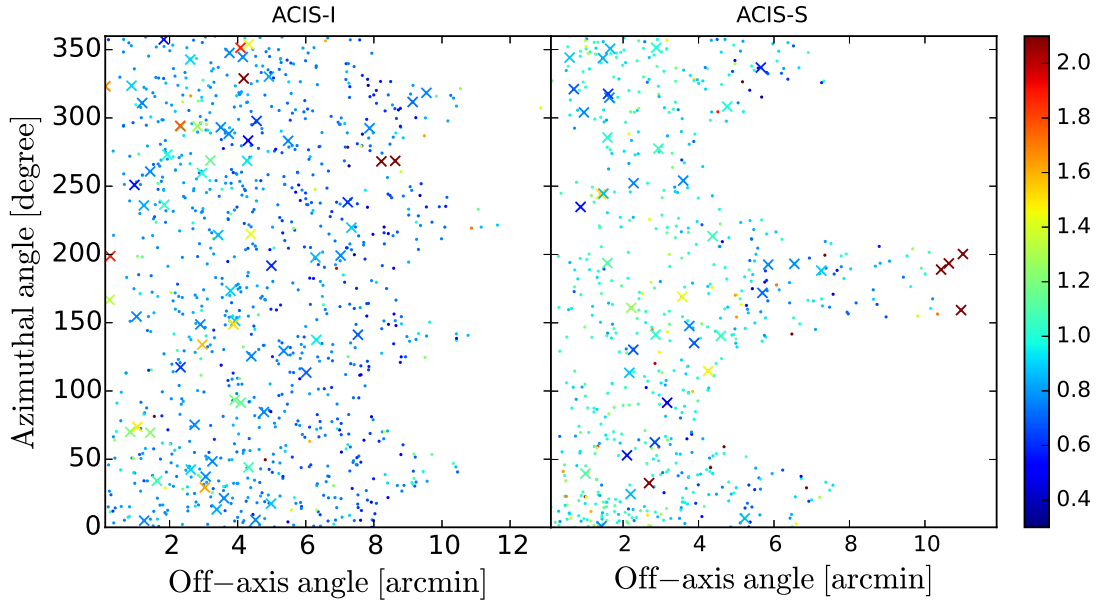}
\caption[]{The ratio of the events number from the circle region ({\tt psfsize\_srcs})
to that from the $3\sigma$ elliptical region ({\tt wavdetect}) as a function of OAA and azimuthal angle,
for ACIS-I and ACIS-S respectively.
The dots represent all the detections, while the crosses represent faint detections with $\sigma < 3$.}
\label{ec95IS.fig}
\end{figure*}

\begin{figure*}[!htb]
%\figurenum{3}
\center
\includegraphics[width=0.7\textwidth]{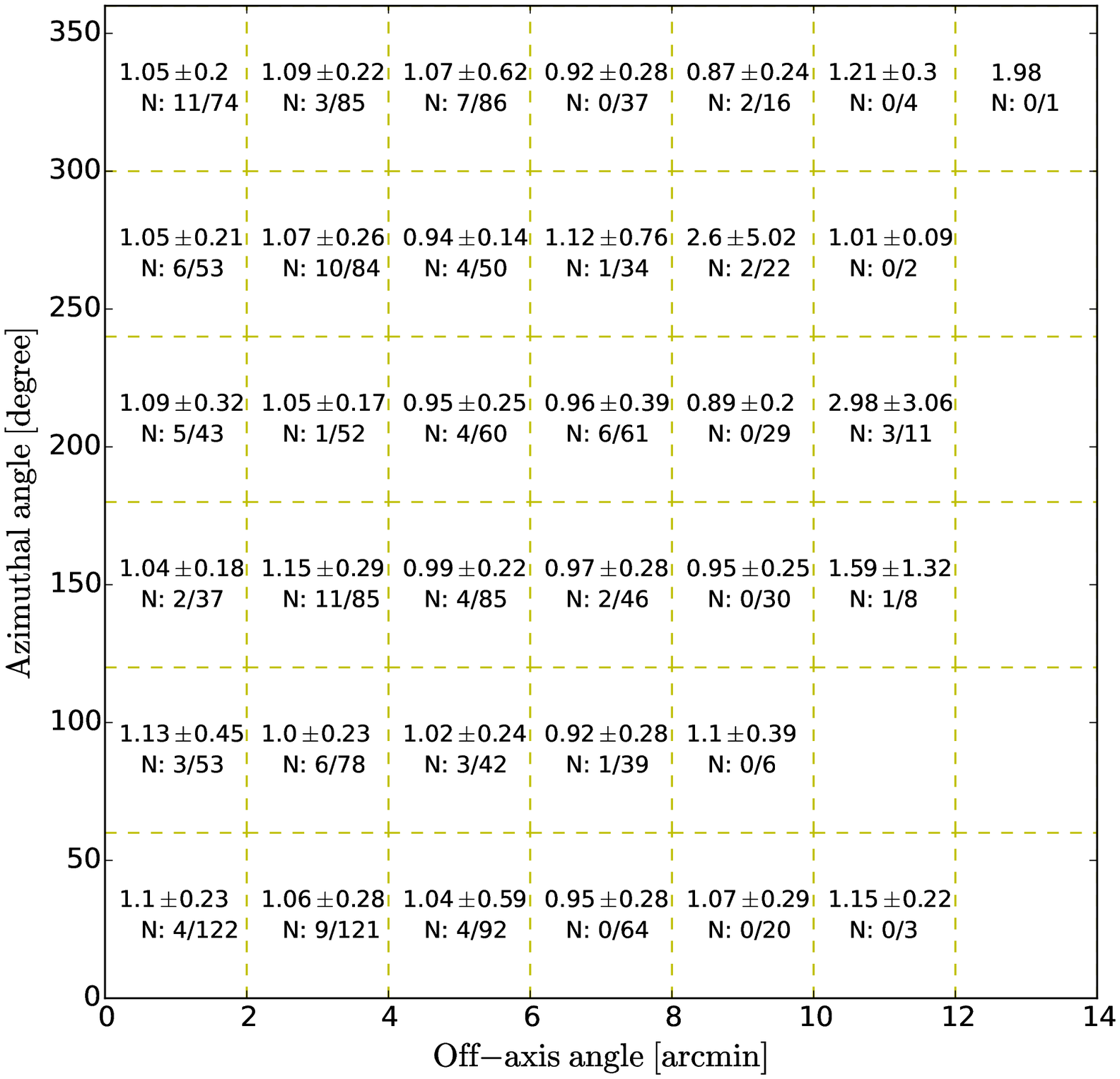}
\caption[]{The average ratio of the events number from the circle region ({\tt psfsize\_srcs})
to that from the $3\sigma$ elliptical region ({\tt wavdetect}) for different groups of OAAs and azimuthal angles.
The upper number in each square shows the averaged ratio for each group, while the lower number shows the number of faint detections
($sigma < 3$) and all detections.}
\label{ec95.fig}
\end{figure*}

The discrepancy of the source counts in the low-$\sigma$ side may be due to two reasons \citep{Yang2004}.
First, for {\tt wavdetect},
the construction of source cells is carried out by convolving the source image with wavelet functions,
and some sources may display multiple peaks in the convolved image due to
statistical fluctuations in the source photon distribution.
This may result in an incorrect estimation of source counts
especially if the source is quite off-axis and the point-spread function (PSF) shape can not be expressed by a Gaussian.
For aperture photometry,
generally, the $3\sigma$ elliptical source region contains 95\% of the total counts for an assumed 2D
Gaussian distribution. However, this assumption is not robust since the distribution of the observed counts
in the source cell may be non-gaussian, particularly for low-count detections.
Here we make some evaluation of the uncertainty of the 95\% counts fraction,
by comparing the events with those extracted from a circle enclosing 95\% fraction of PSF using {\tt psfsize\_srcs},
which has also been used as a photometry tool \citep{Feng2015}.
Twenty randomly selected observations (10 ACIS-I and 10 ACIS-S configurations) are used in the evaluation,
which include 1047 and 688 detections, respectively.
The ratio of the events number from the circle region ({\tt psfsize\_srcs})
to that from the $3\sigma$ elliptical region ({\tt wavdetect}) are plotted as a function of OAA and azimuthal angle (\autoref{ec95IS.fig}).
The detections with largest discrepancy (reddest) are mostly faint detections ($\sigma<3$),
which means the aperture photometry with the $3\sigma$ elliptical region ({\tt wavdetect})
are underestimated for faint detections, which can be clearly seen in \autoref{wav_aper.fig}.
To have a quantitative view, we compute the average ratio for different groups of OAAs and
azimuthal angles with all the 1735 detections (\autoref{ec95.fig}).
The discrepancy seems large for detections with OAA more than 10$'$, while no clear trend can be seen
for the ratio with azimuthal angle.
%These ratios should be folded into the error estimate of the aperture photometry with different OAAs and azimuthal angles.
An average ratio as $1.07\pm0.74$ is derived for all these detections, indicating the aperture photometry
with the {\tt wavdetect} $3\sigma$ region are statistically underestimated, compared to the photometry
with the {\tt psfsize\_srcs} circle region.
It should be noted that the $Chandra$ PSF becomes highly elliptical with increasing OAAs.
The {\tt psfsize\_srcs} script determines a circular region that encloses specified fraction of the PSF,
using the CALDB radially enclosed energy fraction file.
However, this calibration file was created assuming a flat detector,
an accurate assumption for ACIS-S3 but not for other tilted ACIS CCDs.

Second, different background determination would result in different net counts. For example,
if the background region is too close to the source,
the PSF wing from {\tt wavdetect} may be taken as background by mistake,
which could cause an oversubtraction of the background and thus an underestimation of source counts.
The {\tt wavdetect} counts are used to compute the fluxes and luminosities in the
0.3--8 keV band for easy comparisons with previous studies.

\begin{figure*}[!htb]
%\figurenum{7}
\center
\includegraphics[width=0.7\textwidth]{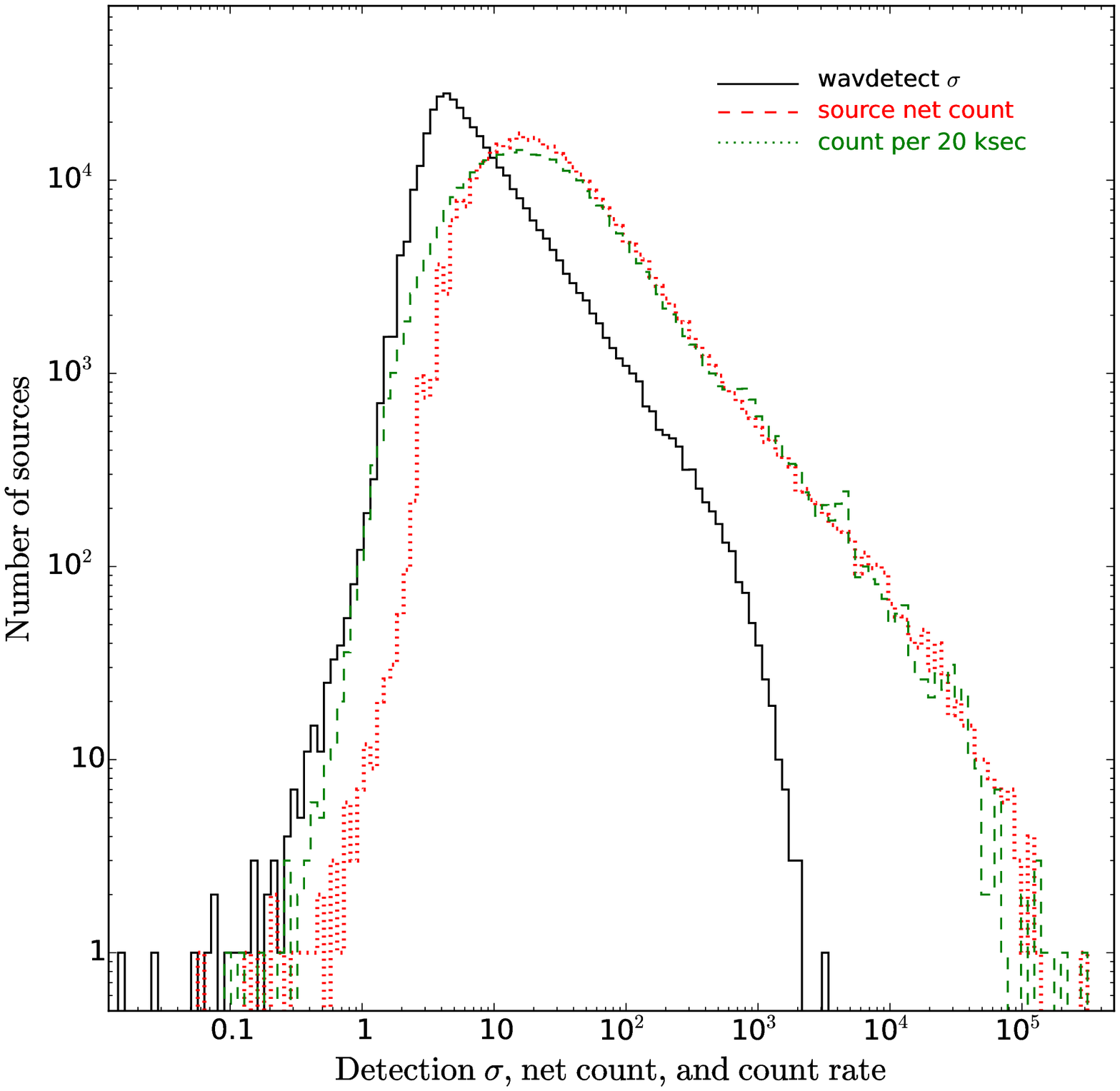}
\caption[]{Histograms of the {\tt wavdetect} $\sigma$, source net count, and count rate for sources detected in this survey.
About 2.0\%/10.5\%/40.5\% are detected with $\sigma$ below 2/3/5.
The source distributions display peaks around 4.1 $\sigma$, 14.8 counts after background subtraction,
and $7.4\times10^{-3}$ counts s$^{-1}$ after vignetting correction.
If the outliers in the two sides of the distributions are ignored, the source count ranges from $\sim$ 1 to $10^5$,
and the count rate ranges from $10^{-5}$ to 5 counts s$^{-1}$.}
\label{counts.fig}
\end{figure*}

Among the 363,530 detections, there are 7421 sources (2.0\%) with detection significance $\sigma < 2$,
38,111 sources (10.5\%) with detection significance $\sigma < 3$,
and 147,383 sources (40.5\%) with $\sigma < 5$.
As shown in \autoref{counts.fig},
the source distributions display peaks around 4.1 $\sigma$, 14.8 counts after background subtraction,
and $7.4\times10^{-3}$ counts s$^{-1}$ after vignetting correction.
If the outliers in the two sides of the distributions are excluded,
in this composite survey, the source count ranges from $\sim$ 1 to $10^5$,
and the count rate ranges from $10^{-5}$ to 5 counts s$^{-1}$.

\subsection{Colors}
\label{color.sec}

\begin{figure*}[!htb]
%\figurenum{6}
\center
\includegraphics[width=0.7\textwidth]{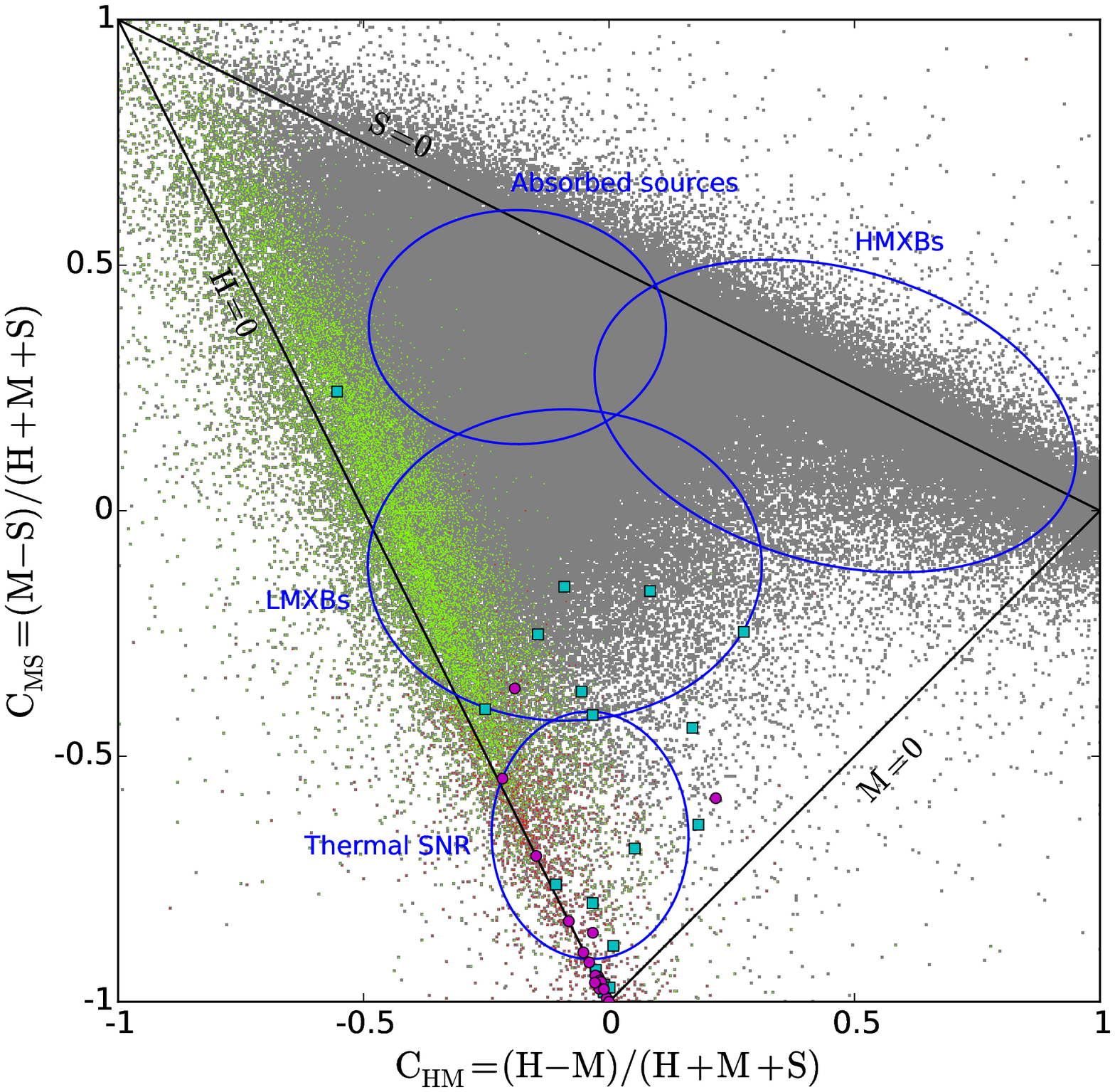}
\caption[]{The X-ray color-color diagram for sources above 10 counts. In general, the
sources fall into the triangle whose sides represent zero counts in the soft
(S:0.3--1 keV), medium (M:1--2 keV), and hard (H:2--8 keV) bands.
These sources located out of the triangle are those with net counts below zero in
soft/medium/hard bands caused by background oversubtraction.
%The four regions named with thermal SNR, LMXB, HMXB, and absorbed sources are taken from \citet{Prestwich2003}.
The red dots clustered around $C_{MS} = -1$ and $C_{HM} = 0$ represent SSSs, while the green dots
along ${\rm H=0}$ represent QSSs.
M81-ULS1 and M101 X-1 are plotted with magenta circles and cyan squares, respectively.}
\label{color.fig}
\end{figure*}

The combination of the X-ray colors, $C_{MS} = (M-S)/(H+M+S)$ and $C_{HM} =
(H-M)/(H+M+S)$, may simply recognize LMXBs, HMXBs, AGNs, supernovae,
and stars, respectively, since they occupy different locations in the
color-color diagram \citep{Prestwich2003}.
Here we compute the photon counts with aperture photometry for three bands following
\citet{Prestwich2003}, i.e., the soft band (S:0.3--1.0 keV), the medium band
(M:1.0--2.0 keV), and the hard band (H:2.0--8.0 keV).
\autoref{color.fig} shows all detected sources above 10 counts from this survey in the
color-color diagram. About 1\% , 27\%, 19\%, and 42\% fall into the regions of
SNR, HMXB, LMXB, and absorbed sources, respectively.
\citet{Liu2011} reported that approximately 11\% falls into the
HMXB region and 42\% falls into the LMXB region, which are inconsistent with
those in this paper.
This may be caused by the different sample, since \citet{Liu2011}
only included X-ray sources in nearby galaxies.
Some sources located out of the triangle are those with net counts below zero in
different bands caused by background oversubtraction,
which is mostly due to the statistical fluctuation of the background,
especially for low-count detections.
It should be noted that many sources show prominent variability
in different observations, and thus fall into different regions.
M81-ULS1 \citep{Bai2015} and M101 X-1 \citep{Liu2013},
two famous ultraluminous supersoft sources (ULSs), are plotted as examples.
In addition, true-color images with S/M/H bands are generated as
red/green/blue channels for visualizing the source colors.

Many galaxies house a significant population of very soft X-ray sources,
including SSSs and quasi-soft sources (QSSs),
while the nature of these sources are still unknown.
\citet{Di Stefano2003a} employed strict hardness ratio (HR) criteria to identify SSSs
using three energy bins to define HRs: soft (0.1--1.1 keV), medium (1.1--2 keV), and hard (2--7 keV).
In this paper, the photon counts are calculated using the same energy bands,
and the hierarchical classification scheme \citep{Di Stefano2003b} is used
to classify whether a source is SSS, QSS, hard, or dim
(i.e., less than 10 counts).
Finally, about 0.6\%, 6.6\%, 71.3\%, and 21.5\% of all detected sources are classified
as SSS, QSS, hard, and dim, respectively.
As shown in \autoref{color.fig}, SSSs are clustered around $C_{MS} = -1$ and $C_{HM} = 0$,
while QSSs distribute along $H=0$.

\subsection{Variability Analysis}
\label{shortvar.sec}

The $Chandra$ observations have been proved powerful in testing
short timescale variability and the pulsation signal \citep{Esposito2013a, Esposito2013b} of the sources,
and the discovery of coherent X-ray pulsations, in particular, is a key element to understand the nature of a source.

For each source, the event list is extracted from the $3\sigma$ elliptical source
region, and a binned light curve is constructed to visualize its variability.
For sources with different exposures and photon counts,
the light curves are binned with different bin widths;
for one detection, the time bin is uniform.
The standard nonparametric Kolmogorov-Smirnov (K-S) test is used
to quantitatively test the source variability within an observation.
A source can be viewed as variable if the null hypothesis probability $P_{\rm K-S}$
is much smaller than one, otherwise it may be viewed as constant.
To be conservative, here a source is defined as variable if $P_{\rm K-S}<0.01$ \citep{Liu2011},
indicating the source varies during an observation with a significance in excess of 99\%.
This leads to $\sim$ 17,092 detected sources above 10 counts
classified as variable in an observation.
However, the standard K-S test may be not appropriate
for Poisson distributed variables, and the results can be misleading in the extreme Poisson regime.
Here we regard the K-S test results as conservatively valid for sources counts above 100,
since the Poisson(100) distribution may be considered approximately Gaussian.

On the other hand, fourier power spectra are computed for 24,247 light curves with more than 200 photons,
which could be powerful in discovering coherent or quasi-coherent signals.
We should remind that some variability signatures or possible period signals
may be caused by spacecraft dither. For example, if a source dithers across a bad pixel or column,
or dithers off the edge of a CCD or between two adjacent CCDs,
%(especially for ACIS-S if a source dithers between a front-illuminated and back-illuminated CCD),
a strong signal at (a harmonic of) the dither frequency (e.g., 707 s, 1000 s)
can be seen in the power spectrum.

\subsection{Spectrum and Flux}
\label{spec.sec}

\begin{figure*}[!htb]
%\figurenum{7}
\center
\includegraphics[width=0.7\textwidth]{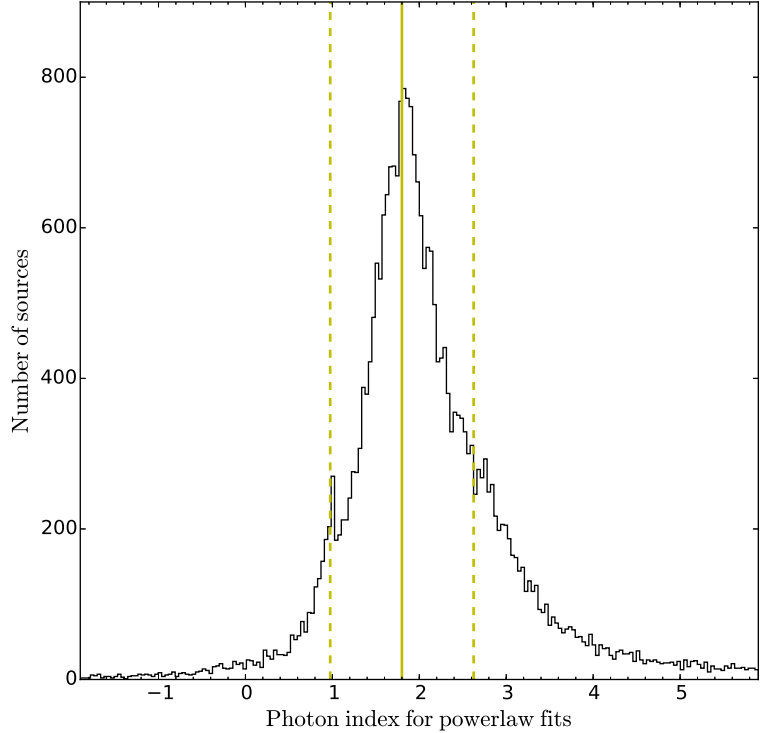}
\caption[]{Histogram of the photon indices for reasonable power-law fits (reduced $\chi^2 < 1.5$)
to 44,740 spectra from this survey. The vertical solid line marks the peak ($\sim$ 1.8) of the distribution,
while the dashed lines display the 68.3\% of all photon indices enclosed between 0.98 and 2.63.}
\label{phoIndex.fig}
\end{figure*}

The spectrum is extracted from the $3\sigma$ elliptical source region with {\tt specextract}
for each bright source with counts $> 100$ and $D_{\rm edge} > 4R_{\rm semi-major}$,
where $D_{\rm edge}$ is the distance from the source center to the chip edge,
and $R_{\rm semi-major}$ is the semi-major axis of the source ellipse.
The corresponding background spectrum is extracted from its local background region.
Finally, the spectra of 44,740 sources are extracted in this work.
Each spectrum is grouped using {\tt specextract}, with the grouptype set as ``NUM\_CTS''.
Then an absorbed power-law model is fitted to the grouped spectrum using {\tt xspec},
with the fit statistic option being ``chi-squared''.
The Galactic $n_H$ is set as the minimum absorption column density in the fitting process.
The majority (85\%) of the spectra can be fitted by the absorbed power-law model
with the photon index between 0 and 4, and
about 63.6\% (28,440) of the spectra are fitted with reduced $\chi^2 < 1.5$.
As shown in \autoref{phoIndex.fig}, the photon index
distribution shows a peak at $\sim$ 1.8, with 68.3\% enclosed between 0.98 and 2.63.

\begin{figure*}[!htb]
%\figurenum{8}
\center
\includegraphics[width=0.7\textwidth]{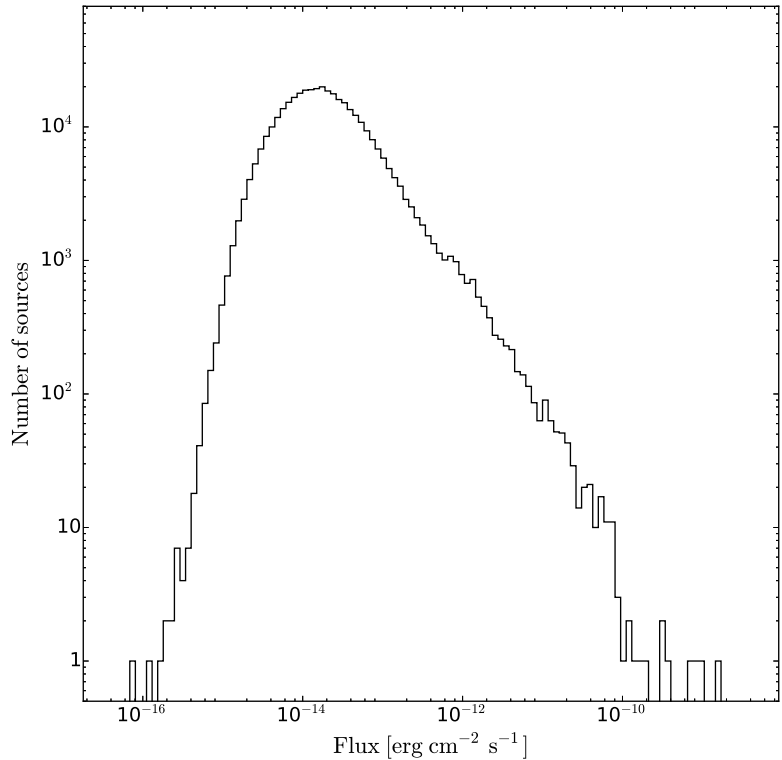}
\caption[]{Histograms of the flux for $\sim$ 363,500 detections.
This survey covers a range of six orders of magnitudes in the source flux
(from $10^{-16}$ to $10^{-10}$ erg cm$^{-2}$ s$^{-1}$),
with a peak at $1.4\times10^{-14}$ erg cm$^{-2}$ s$^{-1}$.}
\label{flux.fig}
\end{figure*}

The source flux (0.3--8 keV) is calculated for each source from its count rate,
which is computed from the source count and the exposure time
and corrected by a vignetting factor.
The vignetting factor is derived from the exposure map as the
ratio between the local and the maximum map value.
However, no vignetting information could be obtained for seven observations
(Obsid 750, 1105, 1194, 1195, 10644, 10655, and 13095) because they were
performed using ACIS subarrays and the aimpoint is not on the used chip,
meaning the maximum map value can not be obtained.
We compute the energy conversion factor (ECF) between the count rate and flux with {\tt xspec}
using the response matrix file (RMF) of the chip center, assuming Galactic
absorption and a power-law spectrum with a photon index $\Gamma=1.7$, which is
generally adopted for $Chandra$ sources.
Finally, the fluxes (${\rm F_{ECF}}$) are computed for about 363,500 detected sources.
The flux distribution (\autoref{flux.fig}) peaks around $1.4\times10^{-14}$ erg cm$^{-2}$ s$^{-1}$,
with a range from $10^{-16}$ to $10^{-10}$ erg cm$^{-2}$ s$^{-1}$.

Although the vignetting effect has been corrected,
the ECF may vary with the observation date and pixel position due to the
temporal and spatial variations of the CCD quantum efficiency \citep{Kim2007} and the charge transfer inefficiency.
%The spatial variation of ECF caused by quantum efficiency is less than 5\% \citep{Kim2007}.
%Using the RMF for the chip center may introduce an error in the flux estimates,
%compared to computing the RMF at the source location, since the charge transfer inefficiency (CTI),
%which has been corrected in the standard $Chandra$ data processing
%using {\tt acis\_process\_events} although,
%can produce a spatially varying change in the response function.
To investigate the spatial variation of the ECF and evaluate the flux estimate errors
by using the RMF of the chip center,
here we compare ECFs derived from the RMFs at the chip center (${\rm C1}$)
and at the locations of the sources (${\rm C2}$), using several randomly selected observations.
The relative offset for each source is determined as ${\rm (C1-C2)/C2}$ and the average
relative offset is computed (Table \ref{confactor.tab}).
This method, by using the chip center response,
may lead to an approximate uncertainty in the flux estimates up to 15\%.

\begin{figure*}[!htb]
%\figurenum{10}
\center
\subfigure{\includegraphics[width=0.5\textwidth]{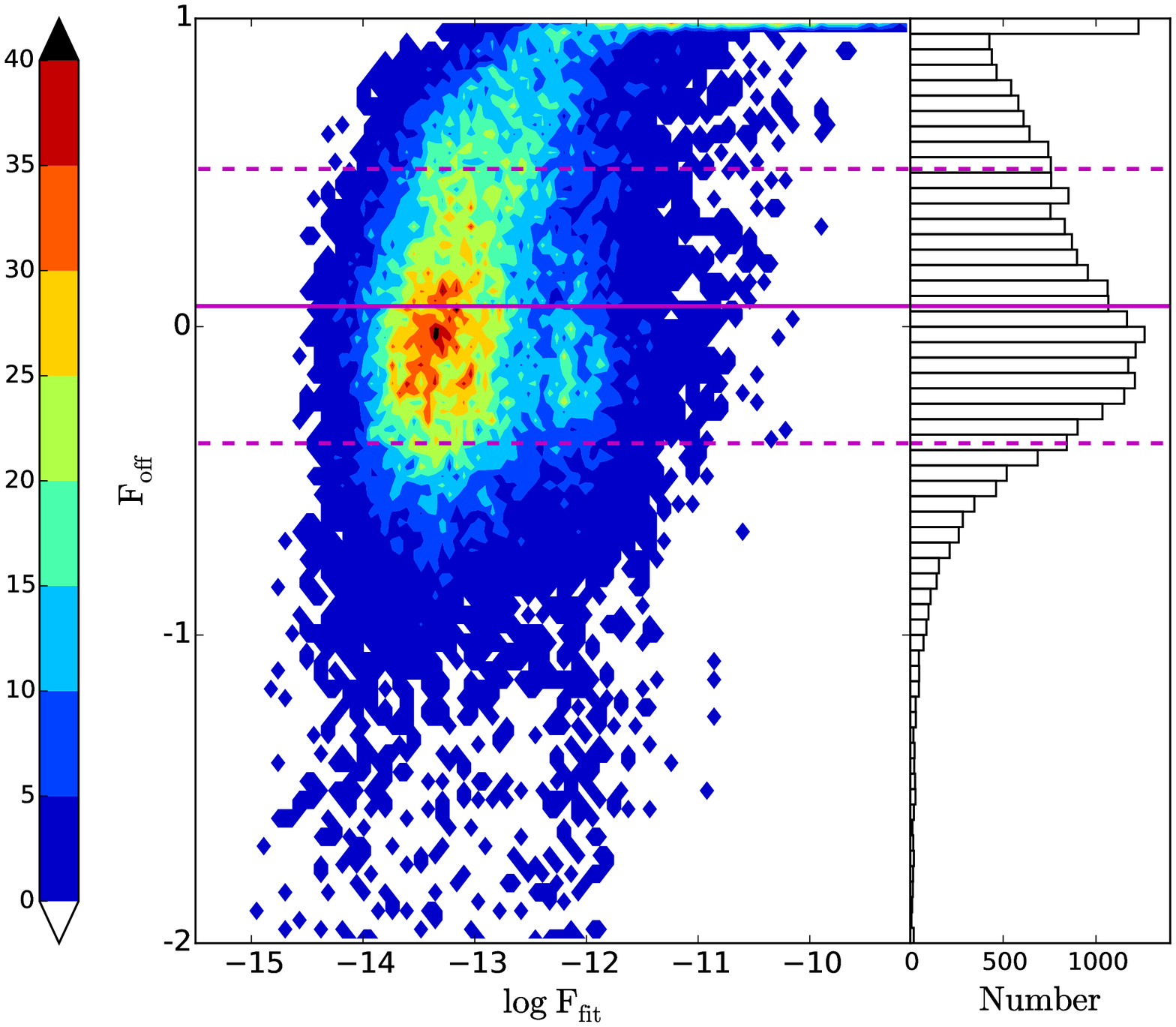}}
\put (-110,-10) {$(a)$}
%\hspace{2 mm}
\subfigure{\includegraphics[width=0.5\textwidth]{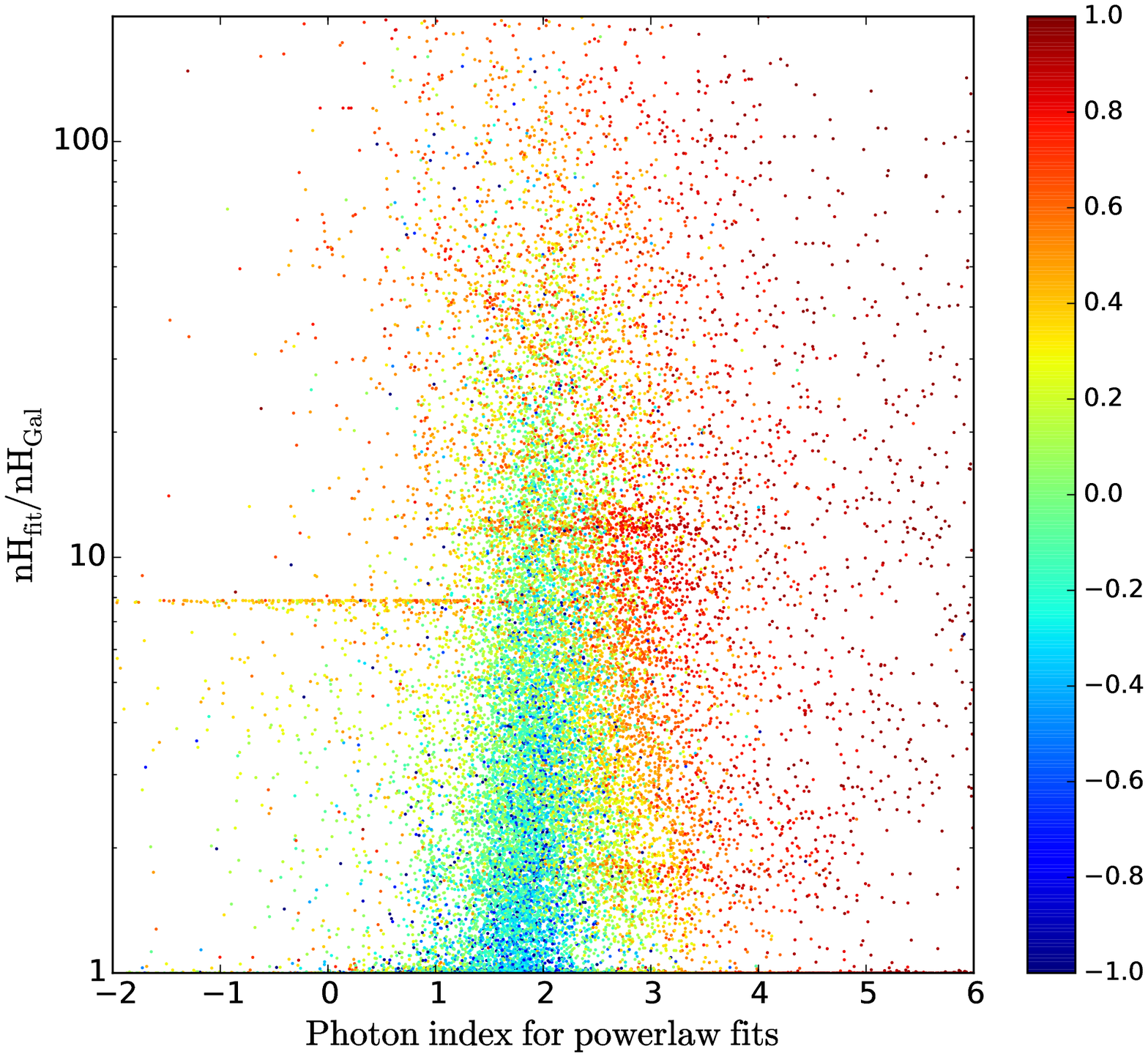}}
\put (-140,-10) {$(b)$}
\caption[]{
(a) Distribution of the relative flux offset ${\rm F_{off}}$ as a function of ${\rm F_{fit}}$.
The histogram of ${\rm F_{off}}$ shows an average at $\sim$ 0.07, with 68.3\% enclosed between -0.378 and 0.512.
(b) The distribution of  $\Gamma$ and ${\rm nH_{fit}/nH_{Gal}}$ for detections with reasonable spectrum fitting,
with the colorbar indicating the relative flux offset.}
\label{fluxcom.fig}
\end{figure*}

Considering that ${\rm F_{ECF}}$ may be seriously miscalculated
if the photon indice $\Gamma$ and absorption are overestimated/underestimated \citep{Liu2011},
we want to provide a description of the offset in the flux estimates
by comparing ${\rm F_{ECF}}$ with the fluxes ${\rm F_{fit}}$ derived from power-law fits,
for the 28,440 detections with reasonable spectrum fitting (reduced $\chi^2 < 1.5$).
We determine the relative flux offset as ${\rm F_{off} = (F_{fit}-F_{ECF})/F_{fit}}$.
\autoref{fluxcom.fig}(a) displays the distribution of the relative flux offset,
which shows an average at $\sim$ 0.07, with 68.3\% enclosed between -0.378 and 0.512.
That means for 68.3\% detections, the offset range is -37.8\% (overestimated) to 51.2\% (underestimated).
\autoref{fluxcom.fig}(b) shows the distributions of $\Gamma$ and ${\rm nH_{fit}/nH_{Gal}}$
for the detections with reasonable spectrum fitting,
with the colorbar indicating the relative flux offset.
It is clear that when $\Gamma > 2.5$, ${\rm F_{fit}}$ is nearly almost larger than ${\rm F_{ECF}}$,
and the offset increases with nH when ${\rm nH_{fit}/nH_{Gal}} > 10$.
Although the $\Gamma$ seems have more significant effect on the flux than nH,
it is hard to quantitatively sperate the uncertainties introduced by $\Gamma$, nH, and the ECF.
We should remind that people should be careful when using these fluxes in their studies
\citep[also reported in][]{Liu2011}.

\section{SOURCES IN MULTIPLE OBSERVATIONS}
\label{multi.sec}

\subsection{Matching Source Detections from Multiple Observations}
\label{match.sec}

Each source record in the catalog should be constructed by combining source detections in
multiple observations, which
requires matching the source detections from all the observations
that include the same region of the sky \citep{Evans2010}.
A first step here is dividing the 10,029 ACIS observations into 4683 groups
based on the proximity of the pointings.
Therefore, the observations in the same group with overlapping fields of view
are appropriate to be studied together.
There are 1145 groups with two or more $Chandra$ ACIS observations, including seven groups
(group 1, 2, 5, 7, 11, 13, and 28) with more than 100 observations.
To determine which detections in different observations are associated with the same
sources, we cross-correlate the $3\sigma$ source ellipse regions
from multiple observations in the same group, and recognize the correlated
source detections as the same sources \citep{Liu2011}.

\begin{figure*}[!htb]
%\figurenum{9}
\center
\includegraphics[width=0.95\textwidth]{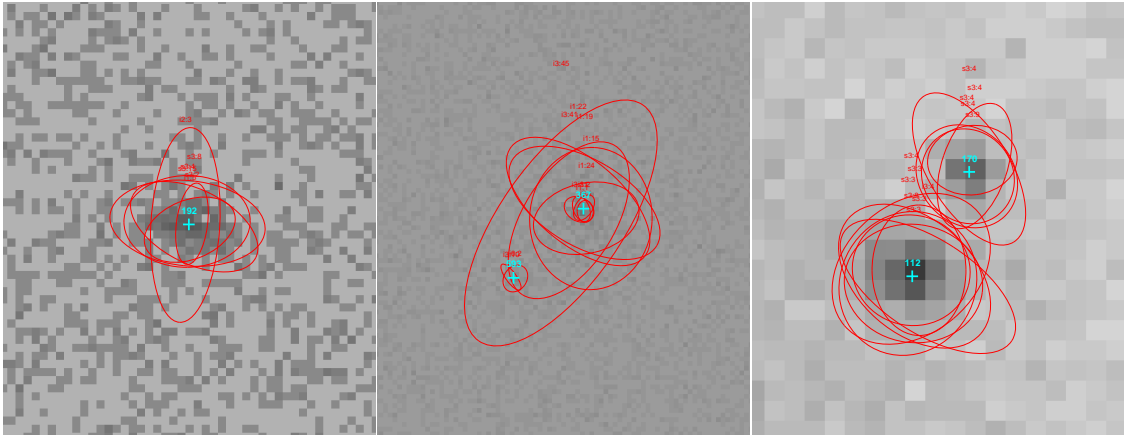}
\caption[]{Cross-identification of sources detected in different observations.
The source ellipses presented by {\tt wavdetect} in each observation
are overlaid on the merged X-ray image with their labels, while the single sources on the sky
are marked as crosses.
Left: The common source matching case where the source detections from the individual observations
all uniquely match a single source on the sky.
Middle: The off-axis source region computed from one detection overlaps
multiple source regions from other observations.
Right: The ellipse regions from two nearby sources overlap each other,
which shape a confused ``pair of pairs'' \citep{Evans2010}.}
\label{match.fig}
\end{figure*}

The actual matching procedure is quite complex \citep{Evans2010, Liu2011}.
The most common situation is that the source detections with overlapping source ellipses
all uniquely match a single source on the sky
due to the excellent $Chandra$ spatial resolution, thus they are identified as the same source,
as shown in \autoref{match.fig} (left).
However, due to the strong dependence of the PSF size with OAA,
in some observations, two close sources at large OAAs cannot be
resolved and may be detected as a single source, as shown in \autoref{match.fig} (middle).
In this case, the counts (and flux) of the detected source will be split into two parts, and the respective
fractions are determined from its separations from the centers of the two sources
based on other observations.
This method assumes that these resolved sources have constant count ratios and fluxes in different observations, and the local PSF is circularly symmetric.
In a few cases, source ellipses from two nearby sources may overlap, as
illustrated by the two sources on the right panel of \autoref{match.fig}.
The majority of these confusions can be removed if we shrink the source ellipses to about the PSF
sizes, while in rare cases of extreme confusion, human judgment is required to complete a match.
%This again presents the crucial role of visual inspection in this work.
For each source, when the individual detections are determined, we compute the final position
by averaging the positions of individual detections with
the detection significance as weights.
This process only combines positions from resolved detections and does not included unresolved detections,
which have been split in previous steps.
The minimum positional error from individual detections is taken as the
positional error.

\subsection{Upper Limits}
\label{uplimit.sec}

In some cases, a source is detected by {\tt wavdetect} in one observation
but not detected in another observation.
We compute the upper limit as the background-subtracted counts within
the circular regions that encloses 95\% of the energy at 1.5 keV,
with the extraction radii computed using the HRMA PSF models.
The background values are
estimated from a neighboring annulus without detected sources.
The source count $C_N$ and error $C_E$ are computed as described in Section \ref{count.sec}.
A minimum of one photon is set for $C_N$ if it is less than unity.
The source significance is defined as $\sigma = C_N / C_E$, which is different
from the detection significance as reported by {\tt wavdetect}.
%Therefore, these computed photon counts range from a few up to hundreds depending on the
%off-axis angles and the environments of the sources.

\section{ASSOCIATION WITH GALAXIES AND CONTAMINATION ANALYSIS}
\label{galaxy.sec}

Studies of X-ray sources in multiple galaxies can provide important information on their formation and
evolution in different environments.
To check whether a source belongs to a galaxy,
the separation $\alpha$ between the galaxy center and the source is computed and compared to the elliptical
radius $R_{25}$ of the $D_{25}$ isophote along the great arc connecting the galaxy center and
the source.
Generally a source is regarded as associated with a galaxy if it is located within the
$D_{25}$ isophote ($\alpha < R_{25}$),
but here a source with $R_{25} < \alpha < 2R_{25}$ is also considered belonging to the galaxy
to avoid missing any galactic sources \citep[see][for details]{Liu2005}.
All the galaxy information (e.g., coordinate, $R_{25}$, distance) are from
the Third Reference Catalog of Galaxies \citep[RC3;][]{Vaucouleurs1991}.
Cross-correlation of these sources with galaxy isophotes yields
17,828 sources within the $D_{25}$ isophotes of 1110 galaxies,
and 7504 sources between the $D_{25}$ and 2$D_{25}$ isophotes of 910 galaxies.
The X-ray luminosity is determined for a source with its flux and the distance of the host galaxy.
If the distance of the host galaxy is unknown, or one X-ray source is not associated with a galaxy,
the distance for the source is assumed to be 1 Mpc.

When studying X-ray point sources associated with galaxies, one needs to
exclude the foreground stars and background QSO/AGNs projected into the host
galaxies by chance.
However, the identification of foreground/background objects is quite complex
and is beyond the scope of this paper.
Here we estimate the contamination rate for the point sources associated with galaxies,
using the log$N$--log$S$ relation that predicts the number of X-ray
sources per deg$^2$ $N$ as a function of flux $S$.
\citet{Hasinger1993, Hasinger1998} derived a log$N$--log$S$ relation based on $ROSAT$
observations of the Lockman Hole region, with the differential form as $dN/dS = N_1 S^{-\beta_1}$ for
$S>S_b$ and $dN/dS = N_2 S^{-\beta_2}$ for $S<S_b$,
with $S$ in unit of $10^{-14}$ erg cm$^{-2}$ s$^{-1}$ in the 0.5-2 keV band,
$S_b=2.66$, $N_1 = 238.1$, $\beta_1 = 2.72$, $N_2 = 111.0$, and $\beta_2 = 1.94$.
\citet{Mushotzky2000} derived a log$N$--log$S$ relation based on $Chandra$
ACIS observation, where the number of sources over the flux range
(2.3-70)$\times10^{-16}$ erg cm$^{-2}$ s$^{-1}$ are given by
$N(>S) = 185(S/(7\times10^{-15}))^{-0.7\pm0.2}$, with $S$ as the 0.5-2 keV flux.
In this study, we adopt the log$N$--log$S$ relation derived from $ROSAT$ observations
\citep{Hasinger1993, Hasinger1998} complemented at low fluxes by the log$N$--log$S$ relation
derived from $Chandra$ ACIS observations \citep{Mushotzky2000}.

\begin{figure*}[!htb]
%\figurenum{10}
\center
\subfigure{\includegraphics[width=0.5\textwidth]{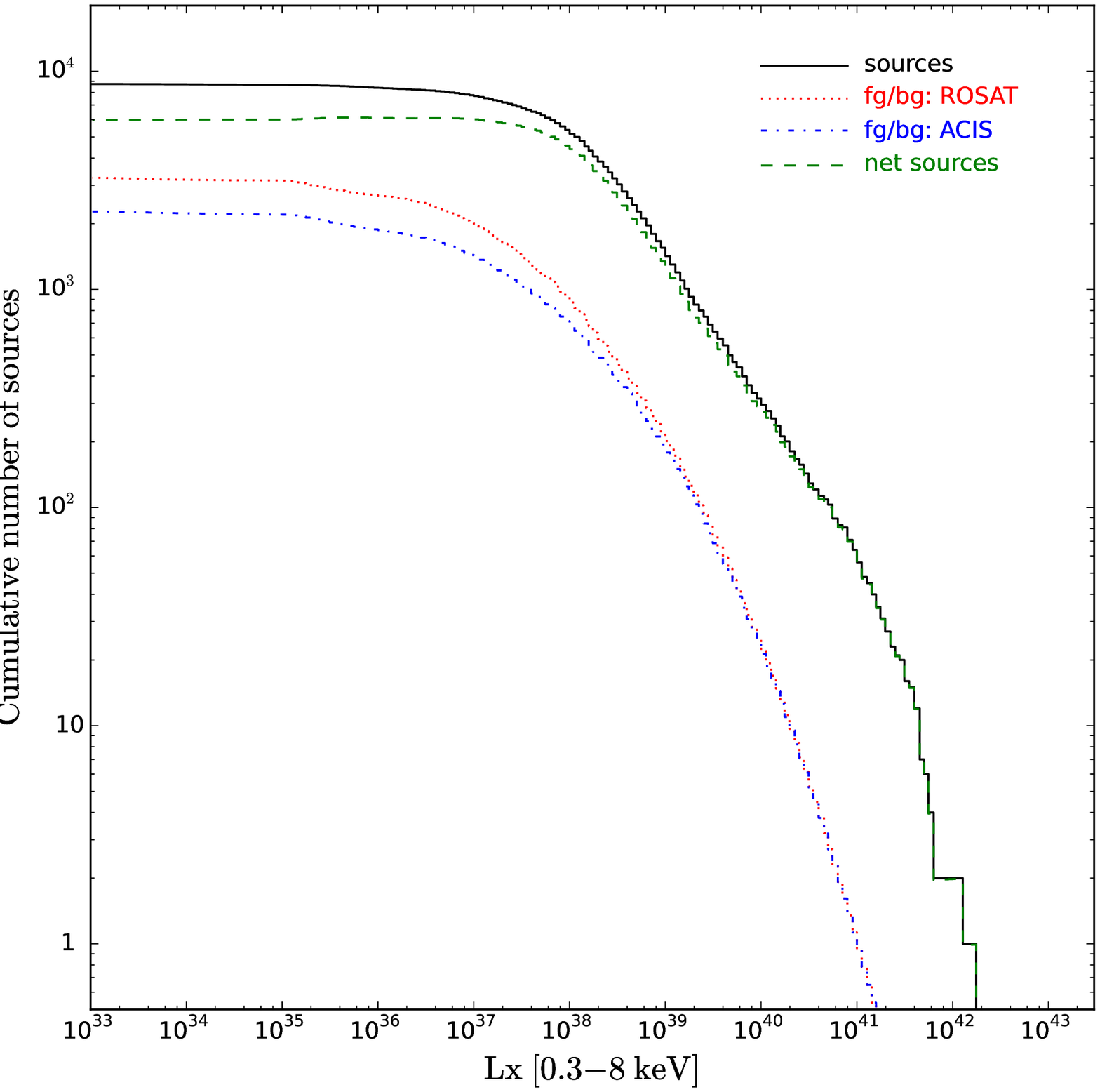}}
\put (-110,-10) {$(a)$}
%\hspace{2 mm}
\subfigure{\includegraphics[width=0.5\textwidth]{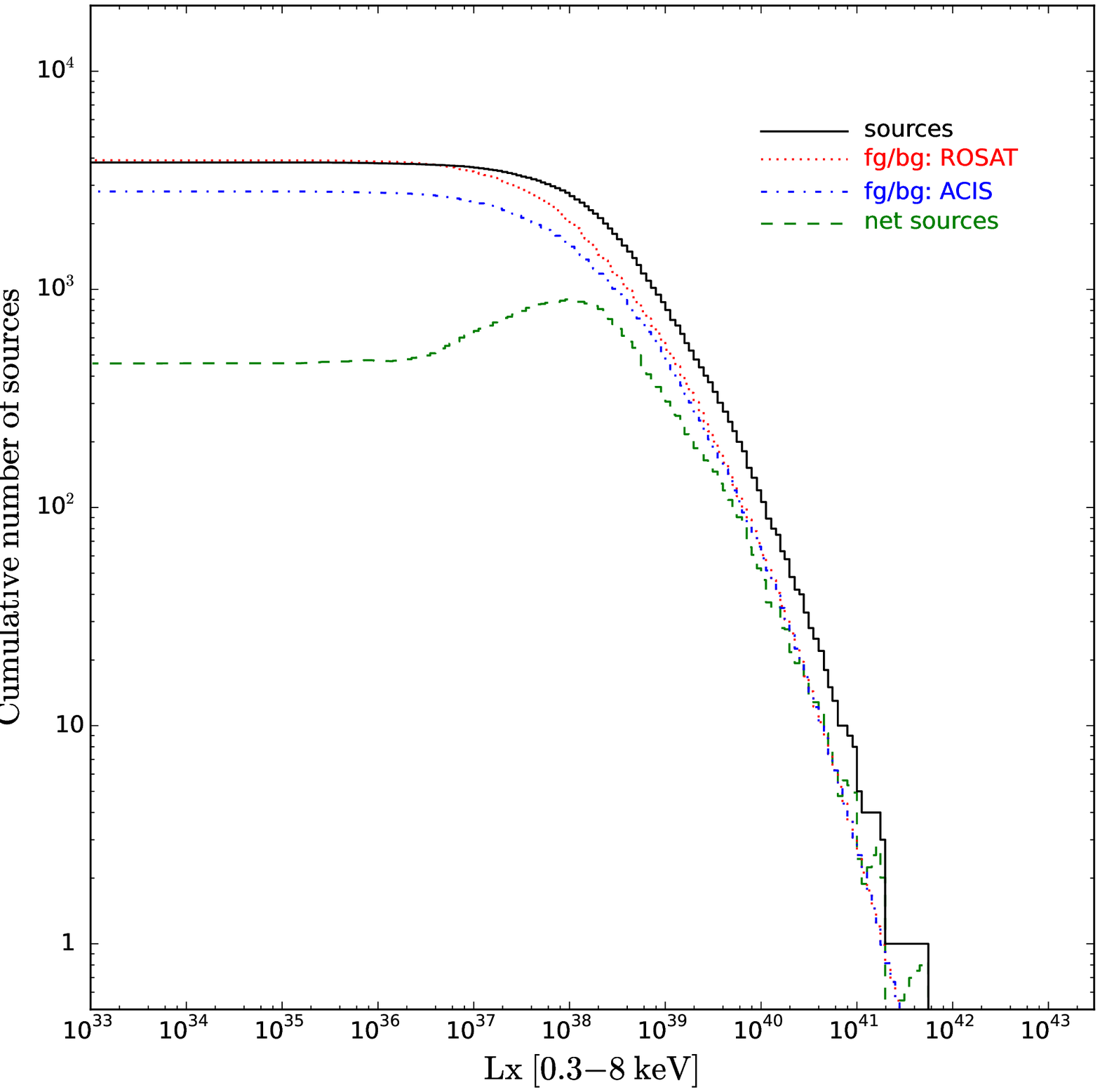}}
\put (-110,-10) {$(b)$}
\caption[]{
(a) The cumulative numbers of sources within the $D_{25}$ isophotes of the sample galaxies.
These lines plot the detected sources (solid line),
the foreground/background contaminating sources predicted by the $ROSAT$ log$N$--log$S$ relation
(dotted line) and the ACIS log$N$--log$S$ relation (dash-dotted line),
and the ``net'' sources (dashed line), which are computed as subtracting the contaminating sources
(average of the two log$N$--log$S$ predictions) from the detected sources.
(b) The cumulative numbers of sources located out of $D_{25}$ but in 2$D_{25}$ isophotes. }
\label{pnet.fig}
\end{figure*}

A sample of $\sim$ 500 galaxies with isophotal major axis between 1 and 50 arcmin
are selected to estimate the contamination rate.
For each galaxy, we calculate the surveyed area curve $A(>S)$, the (cumulative)
number of sources with detection significance $\sigma>3$, the number of predicted
contaminating sources, and the number of ``net'' sources truly
relevant to the host galaxy \citep[see][for details]{Liu2011}.
Briefly, the surveyed area curve $A(>S)$ is computed by summing up the area of the pixels for
which the detection thresholds correspond to flux less than $S$,
then the number of contaminating sources in a flux interval can be calculated
with the differential form of the log$N$--log$S$ relation.
Estimates for individual galaxies are summed up to estimate the contamination
rate for the extragalactic point sources in the whole catalog.
%The surveyed area curves for such a total survey are plotted in \autoref{parea}.
%The $D_{25}$ isophotes of these sample galaxies cover about 2.8 deg$^2$ of the
%sky, while the 2$D_{25}$ isophotes cover about 7.1 deg$^2$ of the sky.
The cumulative numbers of detected sources, predicted contaminating sources,
and the ``net'' sources are plotted in \autoref{pnet.fig}.
For all detected sources within the $D_{25}$ isophotes, about 31.6\% are
contaminating objects, and 68.4\% are ``net''
sources truly relevant to the studied galaxies.
At higher luminosities, the net source fraction increases to 78.2\%, 85.0\%, 87.9\% for
detected sources above $10^{37}$, $10^{38}$, $10^{39}$ erg s$^{-1}$.
For all detected sources between $D_{25}$ and 2$D_{25}$ isophotes, the total ``net''
source fraction is about 12.0\%, while the ``net'' source fraction increases to 17.8\%, 33.2\%, 38.1\%
for sources above $10^{37}$, $10^{38}$, and $10^{39}$ erg s$^{-1}$.
In view of sources within 2$D_{25}$ isophotes, the total ``net'' source fraction is
about 51.3\% for all 11,824 sources, and for sources with luminosities
above $10^{37}$, $10^{38}$, and $10^{39}$ erg s$^{-1}$,
the fraction increases to 58.9\%, 67.3\%, and 69.1\%.

%\begin{figure*}[!htb]
%\figurenum{11}
%\center
%\includegraphics[width=0.7\textwidth]{plotarea.eps}
%\caption[]{Surveyed area curves for the $D_{25}$ isophotes (solid line), regions
%outside $D_{25}$ but inside 2$D_{25}$ isophotes (dotted line), and 2$D_{25}$
%isophotes of 439 galaxies observed in this ACIS survey.}
%\label{parea}
%\end{figure*}

%In addition, the ULXs, which are extranuclear sources with an observed luminosity
%$L_{X}$ (0.3--8keV) in excess of $2\times10^{39}$ erg s$^{-1}$, also presented in Table \ref{independent.tab}.
%Furthermore, these sources are classified as '1ULX' if they are within $D_{25}$ isophotes,
%and `2ULX' if they are between $D_{25}$ and 2$D_{25}$ isophotes of the host galaxies,
%Some extreme ULXs with maximum $L_{X}$(0.3--8keV) above $10^{40}$ erg s$^{-1}$ are designated as 'EULX'.
%This leads to 1992 1ULXs (including 897 1EULXs) and 1200 2ULXs (including 418 2EULXs) within 981 host galaxies.
%Among thsee ULXs inside the 2$D_{25}$ isophotes,
%227 ones are projected within two galaxies and nine ones are projected within three galaxies.

\section{A CATALOG OF $Chandra$ X-RAY SOURCES}
\label{catalog.sec}

\subsection{Catalog Construction}
\label{catcon.sec}

Here we summarize the steps for data analysis and catalog construction. $(i)$ For each observation, we first calculate chip dimensions, acquire nominal pointing, exposure times, observation date, and readout subarrays, construct binned images for on-axis chips, and create the aspect histogram, instrument map, exposure map, and PSF map. The {\tt wavdetect} is applied to detect sources on each on-axis chip, and numbered source regions are determined for visual examination. After visual check, these ``good'' sources in different chips are assembled for each observation. The distances for sources away from four chip edges are computed. The event lists are extracted for all sources using the 3$\sigma$ regions from {\tt wavdetect}, and the light curves are created for them. The standard nonparametric K-S test is applied to check the source variability, and the fast fourier transform (FFT) algorithm is applied to detect possible period signal for sources with counts more than 200. The counts in different bands are extracted, and simultaneously the X-ray colors are computed. Then the SSS/QSS/hard types based on S/M/H counts and errors are determined. The vignetting factors for sources are calculated using the exposure maps. The X-ray spectra for sources (counts $> 100$, $D_{\rm edge} > 4R_{\rm semi-major}$) are extracted with {\tt specextract}, and simple power-law models are fitted to the spectra using {\tt xspec}. For each chip, the response at the chip center is calculated, and a simple $\Gamma = 1.7$ power-law model is fitted, assuming the Galactic absorption. The unabsorbed fluxes are then computed for sources with the ECF determined from the response at the chip center and the source counts (from {\tt wavdetect}) with vignetting correction. The position errors are computed for each source. The DSS images are linked to $Chandra$ observations, and the X-ray source positions are automatically marked on DSS images for visual examination of possible optical counterparts. The true color images for source are created for each observation. $(ii)$ To combine source detections in multiple observations, we first divide the ACIS observations into 4683 groups based on the proximity of the pointings. The source detections within one group are then cross correlated using their 3$\sigma$ ellipse regions. For detections identified as the same source, the information on these detections are assembled, and the final position for each independent source is computed by averaging the positions of individual detections with the detection significance as weights. This final position is included in the source name, as shown in following tables. That whether a source belongs to a galaxy is investigated using the galaxy $D_{25}$ isophote (from RC3), and sources associated with a galaxy are ranked using the maximum $\sigma$ value. The galactic source nuclear separation is then estimated. The X-ray luminosity is determined for a source detection with its flux and the distance of the host galaxy. If the distance of the host galaxy is unknown, or one X-ray source is not associated with a galaxy, the distance for the source is assumed to be 1 Mpc. Finally, the averaged positional error, maximum detection significance, maximum counts, maximum luminosity, averaged flux, and flux ratio $F_{max}/F_{min}$ are all computed or extracted for each independent source.

The analysis of 10,029 ACIS observations produces 217,828 independent
sources from 363,530 detections. With above procedures, we obtain uniform data products
and abundant information for these sources \citep{Liu2011}.
The individual observations for these X-ray sources are listed in Table \ref{all.tab};
the contents are:\\
Column 1: group number.\\
Column 2: source names composed from their positions following CXO naming conventions.\\
Column 3: source name in individual observations in form of iiii.ss.nn, where ``iiii'' is the Obsid, ``ss'' is the CCD chip, and ``nn'' is the source number from {\tt wavdetect}. If the source is observed but not detected in an observation, ``nn'' is marked as ``u''. Enclosed in parenthesis is the split fraction, which is less than unity if a source is a split part of a merged source in an observation as described in Section \ref{match.sec}, and the true counts and flux in this observation can be derived from the listed counts and flux multiplied by this fraction.\\
Column 4: source exposure time after deadtime correction.\\
Column 5: Modified Julian Date for the beginning point of the observation.\\
Column 6: distance from the source center to the nearest chip edge.\\
Column 7: OAA in arcsecond.\\
Column 8: vignetting factor calculated from the exposure map being the ratio between the local and the maximum map value.\\
Column 9: statistical positional error along Right Ascension and Declination.\\
Column 10: source positional uncertainties calculated from the scheme of \citet{Kim2007}.\\
Column 11: detection significance from {\tt wavdetect}. If the source is not detected in an observation, it is computed as (net count)/(count error) (Section \ref{uplimit.sec}) and prefixed with a negative sign.\\
Column 12: background-subtracted photon counts with its uncertainty in parenthesis from {\tt wavdetect}, or calculated from a circular region equal to the PSF enclosing 95\% of the energy at 1.5 keV for the upper limit.\\
Column 13: background counts within the source region. \\
Column 14: X-ray color $C_{MS} = (M-S)/(H+M+S)$ with uncertainty in parenthesis. S/M/H represents background-subtracted counts in soft (0.3-1 keV), medium (1-2 keV), and hard (2-8 keV) bands. \\
Column 15: X-ray color $C_{HM} = (H-M)/(H+M+S)$ with uncertainty in parenthesis.\\
Column 16: source flux calculated from the vignetting corrected count rate in 0.3--8 keV, assuming Galactic $n_H$ absorption and a power law of $\Gamma = 1.7$.\\
Column 17: K-S probability that a source is constant during an observation. Here a source is regarded as variable if $P_{\rm K-S}<0.01$.\\
Column 18: phase (SSS, QSS, hard, and dim) classification for the source during an observation. \\
Column 19: period significance and the most possible period (in second) from FFT algorithm.\\
Column 20: number of possible afterglow events for sources with total events less than 100.\\

The counts in different bands for the 363,530 detections are listed in Table \ref{band.tab}:\\
Column 1: group number.\\
Column 2: source names composed from their positions following CXO naming conventions.\\
Column 3: source name in individual observations.\\
Column 4--18: source net counts with its uncertainty in parenthesis in different bands,
which is computed from aperture photometry with the {\tt wavdetect} $3\sigma$ region.\\

The independent X-ray sources and their average properties in multiple observations
are listed in Table \ref{independent.tab}, ordered by group number and position.
The main entries in Table \ref{independent.tab} are, for each source:\\
Column 1: group number.\\
Column 2: source names composed from their positions following CXO naming conventions. There are 926 sources marked with ``*'', which are projected within two or more galaxies with overlapping domains.\\
Column 3: source positional errors calculated from the scheme of \citet{Kim2007}.\\
Column 4: galactic source names. In each galaxy, the X-ray sources are numbered sequentially based on their maximum detection significance.\\
Column 5: galactic source nuclear separation in arcminute.\\
Column 6: galactic source nuclear separation in unit of the elliptical radius.\\
Column 7: distance in Mpc for the X-ray source.\\
Column 8: numbers that the source has been observed/detected.\\
Column 9: maximum luminosity in erg s$^{-1}$ among the detections.\\
Column 10: average flux in erg cm$^{-2}$ s$^{-1}$ calculated from detections.\\
Column 11: $F_{max}/F_{min}$ ratio as an extreme variability indicator, with $F_{\rm max}$ as the maximum 0.3--8 keV flux from all detections, and $F_{\rm min}$ as the minimum flux from all detections or upper limits. The ratio is set to one without an error if $F_{\rm max}$ equals to $F_{\rm min}$.\\
Column 12: maximum detection significance.\\
Column 13: maximum photon counts among the detections.\\
Column 14: statistics for SSS/QSS/hard/dim phases. For example, s1 means SSS in one observation, q2 means QSS in two observations, h3 means hard in three observations, and d4 means dim (below 10 counts) in four observations.\\
Column 15: statistics for variability during individual observations using the K-S criterion. For example, v:2 means the source is regarded as variable (i.e., $P_{\rm K-S}<0.01$, counts $>$ 10) in two observations.\\
Column 16: source identification. ULX: ultraluminous X-ray sources with observed $L_{X}$ (0.3--8keV) in excess of $2\times10^{39}$ erg s$^{-1}$ (EULX: maximum $L_{X}$ $\geq~10^{40}$ erg s$^{-1}$; 1ULX/1EULX: inside $D_{25}$ isophotes; 2ULX/2EULX: between $D_{25}$ and 2$D_{25}$ isophotes), without nuclear sources excluded; ULS: ultraluminous supersoft sources with at least one SSS phase in which $L_{X}$ (0.3--8 keV) exceeds $2\times10^{38}$ erg s$^{-1}$ \citep{Liu2011}.\\

\subsection{Afterglows}
\label{ag.sec}

The afterglow events, which arise from the energy deposited into the CCD by a cosmic ray,
can appear in a single pixel for several consecutive CCD frame readouts.
In this paper, we do not remove these events with CIAO tool {\tt acis\_find\_afterglow},
but instead, we evaluate the afterglow rate and flag these events for faint detections in the catalogs.
First, The afterglow rate is assessed using the background observations following \citet{Xue2011}.
Given an approximate nominal field background rate of 0.3 counts per second per chip from the $Chandra$ POG
\footnote{http://cxc.harvard.edu/proposer/POG/html/chap6.html\#tth\_sEc6.16.2},
which corresponds to a count rate of $\sim 5.72\times10^{-6}$ counts per 20 s per pixel,
a probability of $\sim 3.27\times10^{-11}$ can be estimated for
three or more counts occurring on a single pixel with in 20 s (hereafter ``repeated events'').
This probability can be negligible, and such events occurred in backgrounds are flagged as afterglow events.
We selected ten observations (five ACIS-I and five ACIS-S configurations) to derive an averaged afterglow event rate
with the backgrounds, by excluding all the sources identified in these observations.
This leads to an afterglow event rate of $(4.01\pm1.15)\times10^{-10}$ per second per pixel.
Then, about 6282 sources with ``repeated events'' are picked out from the 363,530
detections (excluding the seriously pile-uped sources).
After that, we evaluate the expected afterglow events for the 6282 sources,
using the afterglow event rate estimated from backgrounds and the area of each source region.
The evaluated expected afterglow events cover a range of $6\times10^{-7}$ to 0.3, which
are extremely lower than the ``repeated events'' for these 6282 objects.
This means most of the ``repeated events'' are actually photons from the objects, particularly for bright objects.
Although it seems that the afterglow events have little influence on our detections,
the statistical method may be lost for faint ones, since some faint detections are actually false detections
caused by afterglows. For example, some objects with all the events identified as ``repeated events'' may be false detections.
Therefore, we flag the number of ``repeated events'' for objects with total events less than 100.
User should be careful with these objects.

\subsection{False Source Rate}
\label{fsr.sec}

The false source rates in the catalog are estimated using simulations of empty fields, with different exposure, chip location, and detector configuration \citep{Primini2011}.
These simulations are constructed containing background only (``blank-sky''), with appropriate background data sets for the active chips derived using {\tt acis\_bkgrnd\_lookup}.
We should remind that for each observation in this paper, only the on-axis chips are used, including all four I chips if the aimpoint is on an I chip,
and both S2 and S3 chips if the aimpoint is on S3 chip.
The expected number of background events for each chip are firstly estimated from the chip nominal field background rate
%\footnote{http://cxc.harvard.edu/proposer/POG/html/chap6.html#tth\_sEc6.16.2}
and different exposure times ($\sim$ 10, 30, 60, and 120 ks),
and the final numbers of events and their positions are simulated by random sampling, with the
ratio of the expected number to the number of events in the corresponding blank-sky data set
\footnote{http://space.mit.edu/cxc/marx/examples/background/background.html}.
There are no nominal field background rate in 0.3--8 keV from the $Chandra$ POG, and we use the estimates
in the 0.3--10 keV, which may lead to an overestimated false source rate.
The sky coordinates for each chip are re-computed and re-projected to the observed position on the sky with {\tt reproject\_events},
and all these chips are re-assembled into a single event list using {\tt dmmerge}.

The same Obsids in \citet{Primini2011} are chosen for simple comparisons.
All simulated event lists are processed using the detection methods in this paper, and
the false source rates derived from these simulations are shown in Table \ref{fsr.tab}.
The false source rate is very high in each exposure for detections with low significance ($\sigma < 3$),
while for detections with 3 $\leq \sigma <$ 5, it becomes appreciable when exposures are longer than $\sim$ 50 ks.
The is no false detections with  $\sigma \geq$ 5.

\begin{figure*}[!htb]
%\figurenum{9}
\center
\includegraphics[width=0.95\textwidth]{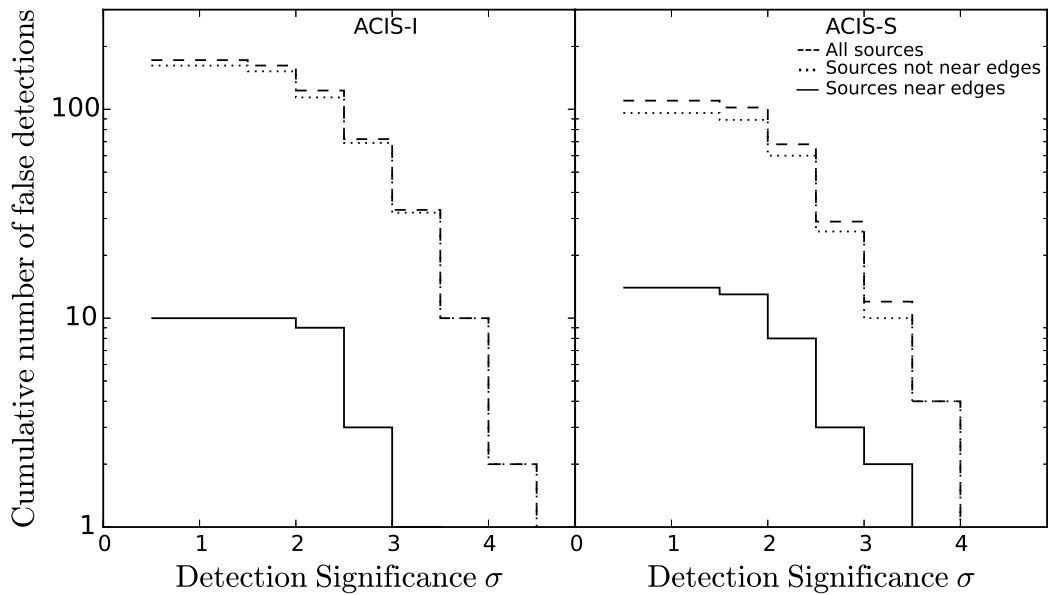}
\caption[]{The cumulative number of false detections as a function of detection significance $\sigma$ for the ACIS-I and ACIS-S
configurations, respectively.}
\label{fsr.fig}
\end{figure*}

As discussed in Section \ref{detec.sec}, sometimes there is a stream of false detections along the chip edges, and these
detections are removed in the virtual examination process.
We plot the cumulative number of false detections as a function of detection significance
for sources near and not near chips edges, respectively (\autoref{fsr.fig}).
Here we define a source near chip edge if it is within about thirty-two pixels (16'') of any edge of a CCD,
in which case the source may be dithered off the CCD during part of an observation.
The false detections near edges are mostly faint detections ($\sigma < 3$).

\section{COMPARISON WITH PREVIOUS CATALOGS}
\label{com.sec}
\begin{figure*}[!htb]
%\figurenum{11}
\center
\includegraphics[width=0.7\textwidth]{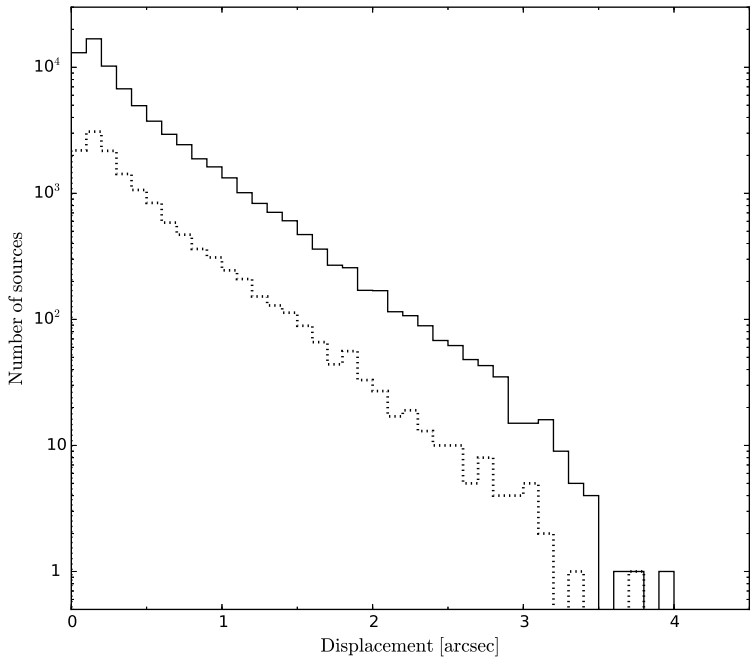}
\caption[]{Comparison of the source positions between our results and previous catalogs.
The solid line shows the comparison results for matches between this paper and the CSC,
while the dotted line shows the comparison results between this paper and \citet{Liu2011}.
}
\label{displace.fig}
\end{figure*}

To examine the quality and reliability of our catalog, we compared some properties of the X-ray sources
obtained here with previous studies.
The CSC \citep{Evans2010} is a project aiming to provide an all-inclusive, uniformly processed data set
from $Chandra$ archive observations.
The latest version of CSC (V1.1) presents 106,586 X-ray sources from ACIS and High Resolution Camera imaging observations,
about half of the number in this paper.
Cross-correlation between our sources and the CSC sources yields $\sim$ 71,200 matches, and
%71,236
the results from our survey and the CSC project are compared to illustrate the differences.
\autoref{displace.fig} shows the offsets between our source positions and CSC source positions (solid line),
with 90\% less than $0.98^{\prime\prime}$, and 99.9\% less than $3^{\prime\prime}$.
The results from this paper are also compared to those from \citet{Liu2011}, which leads to about 13,700 matches.
%13,772
The position offsets between these sources are very small,
with 90\% less than $0.96^{\prime\prime}$, and 99.9\% less than $3^{\prime\prime}$.
Therefore, the source positions derived in this paper are consistent with those from previous catalogs.

Individual observation can be used to scrutinize the detection differences, e.g.,
source numbers detected in the same observation.
Five randomly selected observations (Obsid 88, 4499, 4931, 6792, and 7227)
performed using ACIS-I chips are used to make the comparisons.
The analysis in this paper yields 308 (100, 16, 43, 104, and 45 for individual observation)
point sources from the five observations, and
there are 178 sources with detection significance $\sigma \geq$ 5, 109 sources with 3 $\leq \sigma <$ 5,
and 21 sources with $\sigma <$ 3.
CSC presents 233 (67, 14, 40, 93, and 19) sources,
containing 208 sources with detection significance $\sigma \geq$ 5 and  25 sources with 3 $\leq \sigma <$ 5.
Cross-correlation between these sources from this paper and CSC leads to 206 matches.
For bright sources, CSC detects 167 among 178 sources with detection significance $\sigma \geq$ 5 in this paper, and
our survey detects 192 among 208 sources with detection significance $\sigma \geq$ 5 in CSC.
For faint sources, CSC only detects 38 among 109 our sources with 3 $\leq \sigma <$ 5,
and 1 among 21 our sources with $\sigma <$ 3;
our survey detects 12 among 25 CSC sources with 3 $\leq \sigma <$ 5.
From the matched sources, we can find that 25 sources with $\sigma <$ 5 in our paper are classified as $\sigma \geq$ 5 in CSC.
The most striking thing is we find more objects than CSC in the 3 $\leq \sigma <$ 5 realm.
The false detections in these five observations are less than one using the false source rate shown in Table \ref{fsr.tab},
meaning that most of these sources (3 $\leq \sigma <$ 5) detected only in our paper are true sources.
In summary, our survey detects almost all CSC bright ($\sigma \geq$ 5) sources,
and detects more faint sources than CSC does \citep[also reported in][]{Liu2011}.
In addition, for the 206 matches, the aperture-corrected net count rates from this paper are on average
slightly higher (2\%$\pm$13\%) than those from CSC, which may be due to the wider energy bands adopted
in this paper (0.3--8 keV) than CSC (0.5--7 keV), or different source region and background region apertures.

%The solid and dotted lines show the comparison results for all matches and matches with
%detection $\sigma < 3$, respectively.
%It can be seen that most of the matches with displacements larger than 3$''$ are
%from the sources with detection significance $\sigma < 3$.
%Actually, many matches with offset more than $3^{\prime\prime}$ are mismatches, for example,
%one source detected in this work is not detected by CSC, then this source will match other
%nearby sources.
%Therefore, the source positions derived in this paper are consistent with those from previous catalogs.
%\autoref{displace} also shows the offsets between our source positions and the source positions from \citet{Liu2011},
%with 90\% less than $1.2^{\prime\prime}$, and 99\% less than $3^{\prime\prime}$.
%For individual observation, \citet{Liu2011} have compared the results from their survey and the CSC project,
%and found that their survey detects more faint sources than CSC does.

\section{SCIENTIFIC USE OF THE CATALOG}
\label{sci.sec}

A full exploitation of the data presented in this work is far beyond the scope of
the present paper. Here we would like to draw the reader's attention to some of the scientific
topics that can be addressed using the catalogs.

\subsection{Intra-observation Variability}
\label{intra.sec}

\begin{figure*}[!htb]
%\setlength{\unitlength}{1mm}
%\figurenum{12}
\center
\subfigure{\includegraphics[width=0.45\textwidth]{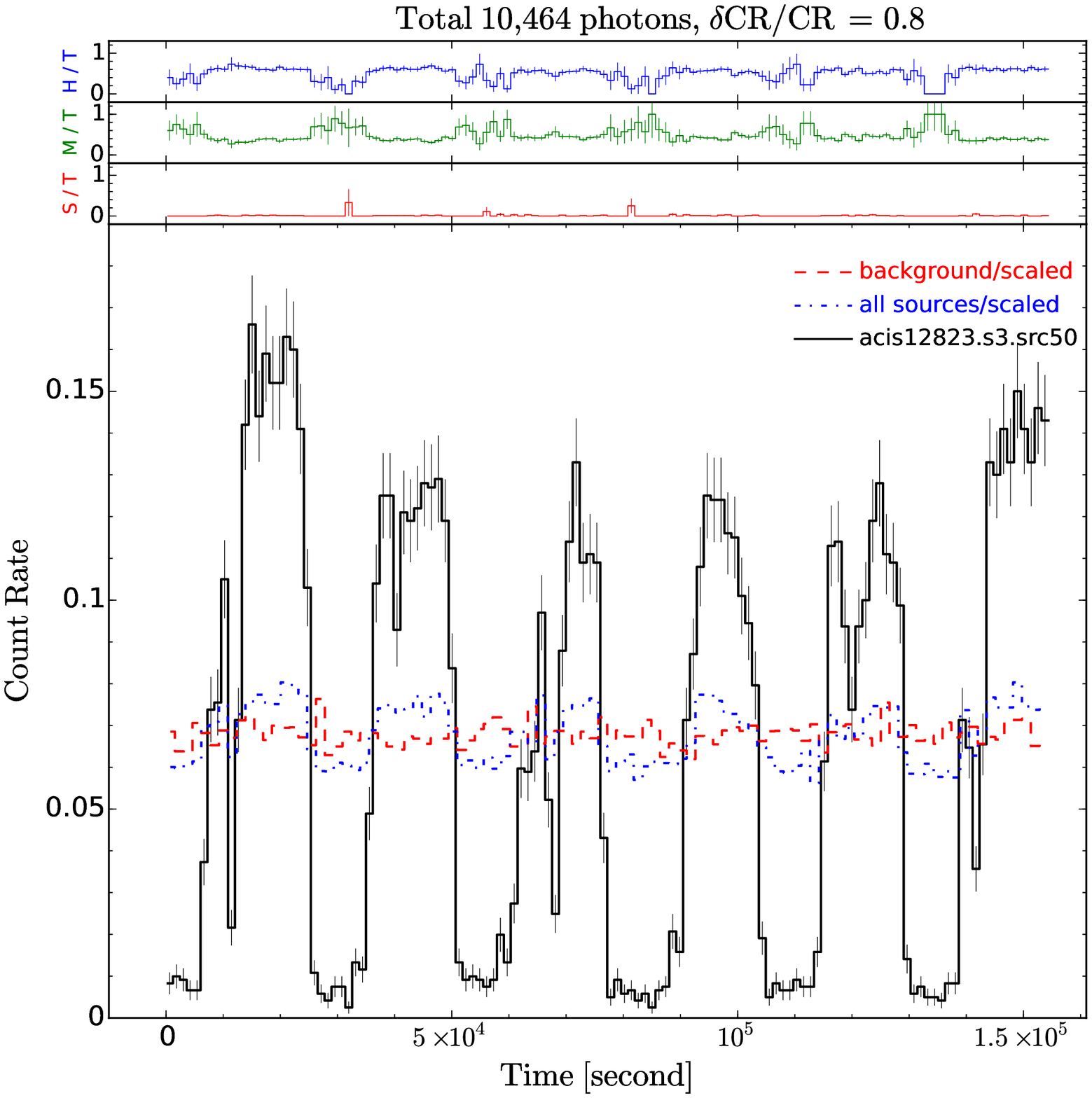}}
\put (-110,-10) {$(1)$}
\hspace{2 mm}
\subfigure{\includegraphics[width=0.45\textwidth]{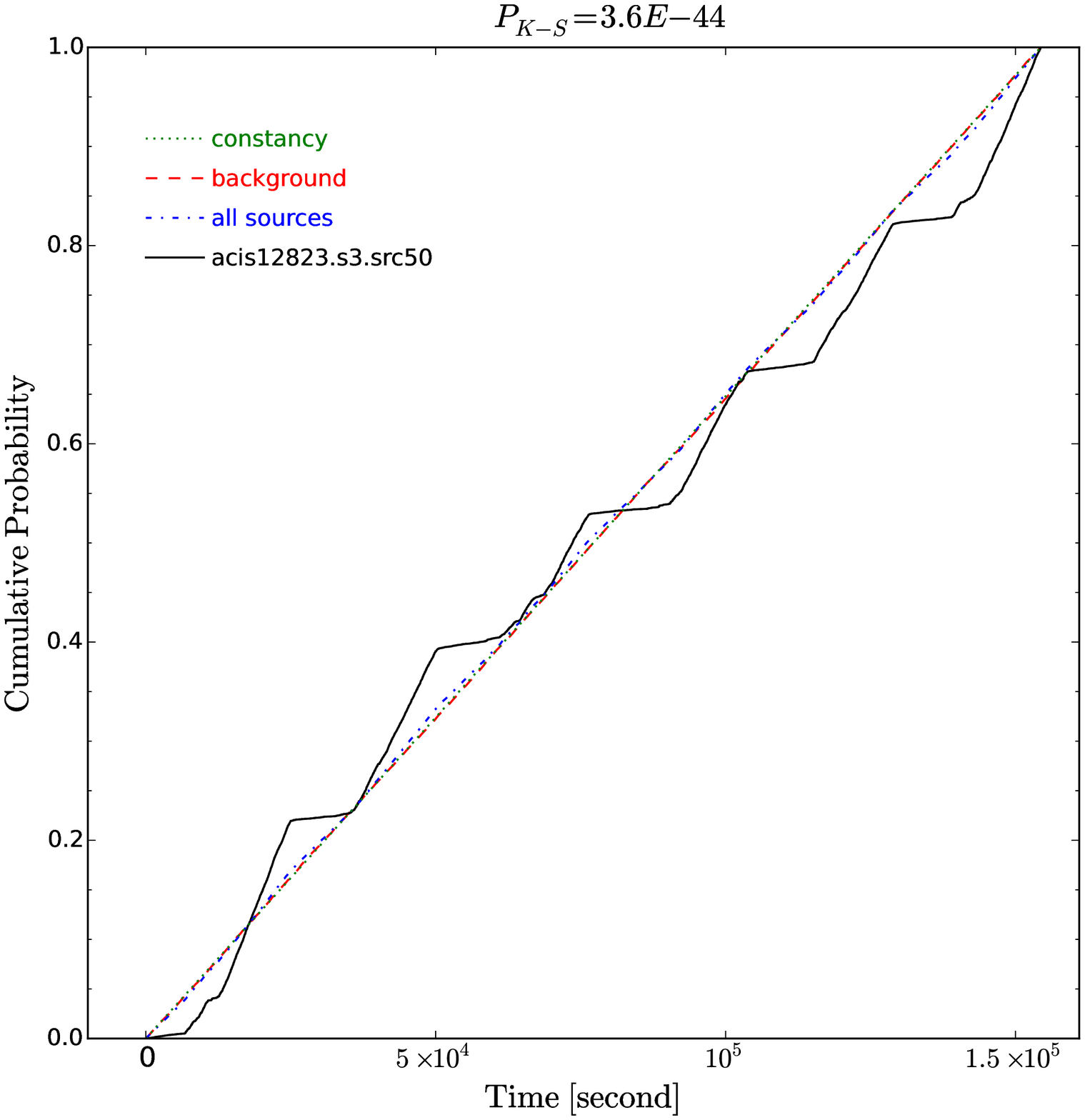}}
\put (-110,-10) {$(2)$} \\
\vspace{-3 mm}
\subfigure{\includegraphics[width=0.45\textwidth]{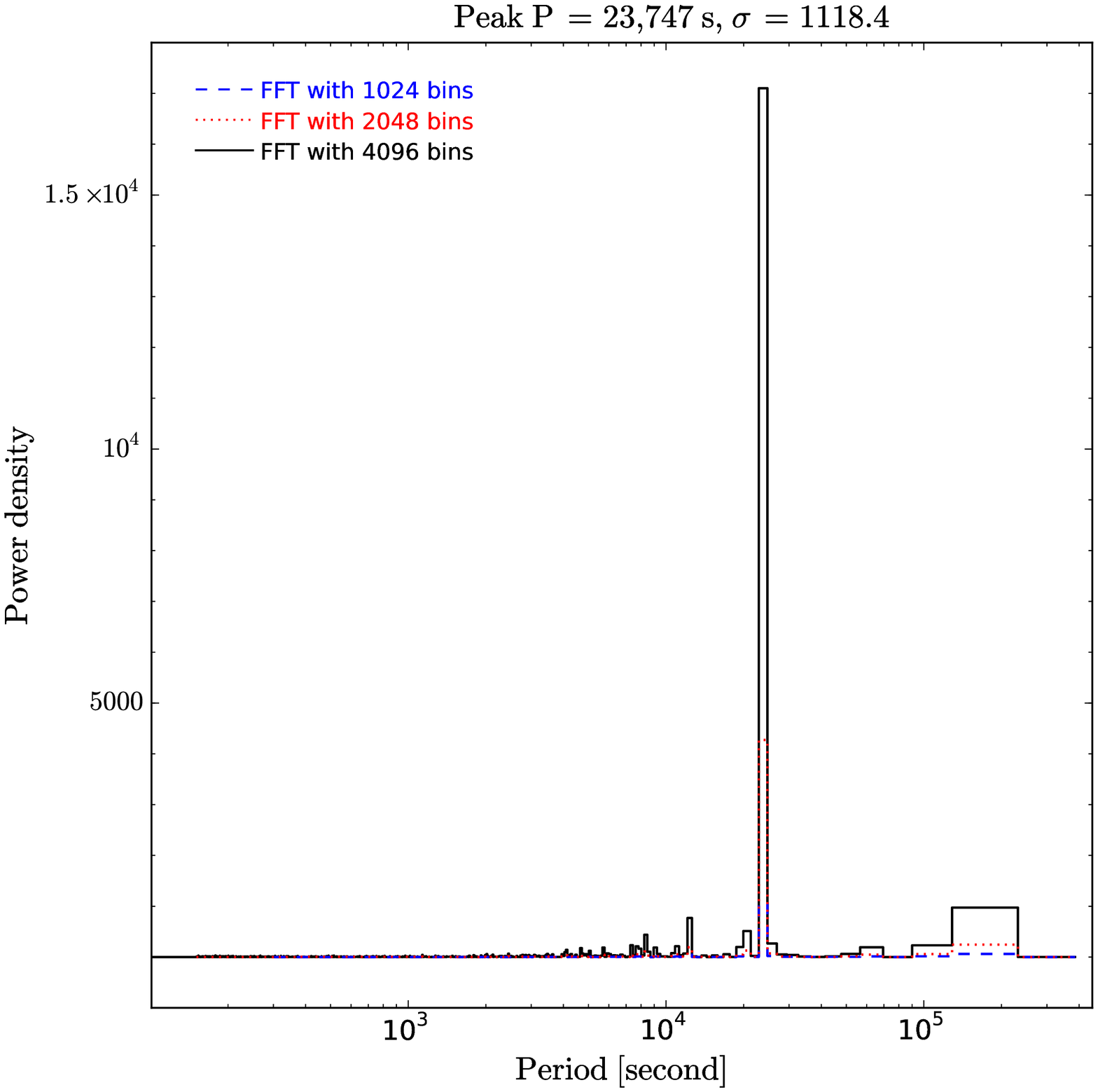}}
\put (-110,-10) {$(3)$}
\hspace{2 mm}
\subfigure{\includegraphics[width=0.45\textwidth]{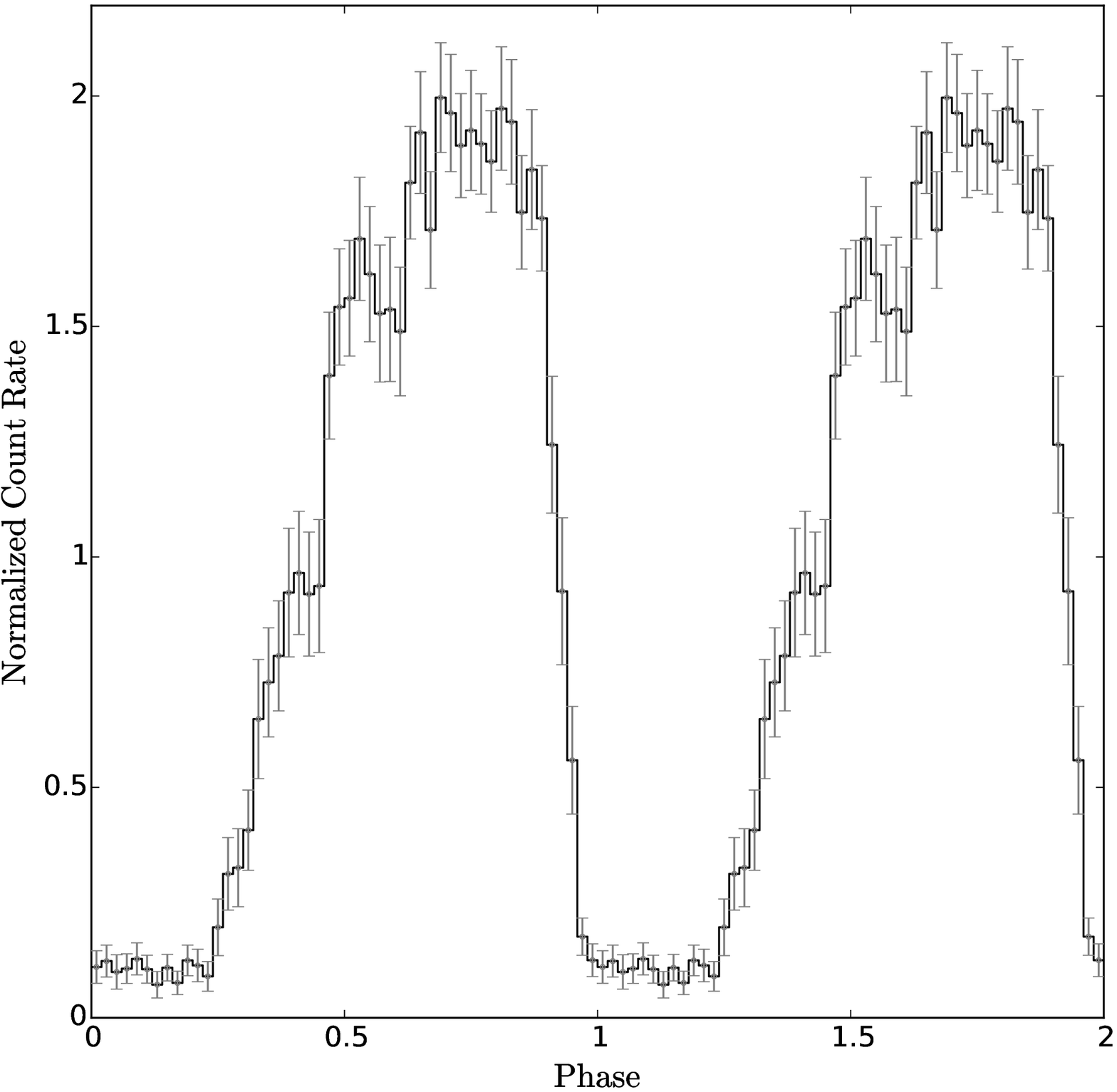}}
\put (-110,-10) {$(4)$}
\caption{
(1) Binned light curve for an X-ray source CXOJ141312.218-652013.81 in the observation Obsid 12823.
The blue and red binned light curves are for all detected sources and the total
background of S3 chip overplotted for comparison.
The three upper panels show the fractions of photon counts in soft (0.3--1 keV),
medium (1--2 keV), and hard (2--8 keV) bands.
(2) K-S tests for the X-ray source, all detected sources, and the total background of S3 chip.
(3) Power spectra of CXOJ141312.218-652013.81 from FFT.
(4) Folded profile of the X-ray source in the 0.3--8 keV band. }
\label{variability.fig}
\end{figure*}

As stated in Section \ref{shortvar.sec}, the $Chandra$ observations are powerful for testing
short timescale variability and the pulsation signal \citep{Esposito2013a, Esposito2013b} of the sources.
In this paper, some products are produced for this purpose, such as
the binned light curves, the null hypothesis probability $P_{K-S}$ from K-S test, and the period significance from FFT.
Take one X-ray source CXOJ141312.218-652013.81 for example (\autoref{variability.fig}),
the binned light curve (acis12823.s3.src50) exhibits several X-ray eclipses during the 150 ks observation.
The light curves for all detected sources and the total background of S3 chip are overplotted for comparison.
The former includes a contribution from this source and distinctly vary with these strong eclipses,
while the latter is nearly invariable in the observation,
indicating that these eclipses are a behavior unique to this source itself instead being caused by background.
The null hypothesis probability $P_{K-S}$ for this source is much smaller than 0.01, indicating strong variability.
A possible signal is found in the power spectrum of the observation Obsid 12823, and the high peak shows
an approximate period of $\sim$ 23.75$\pm$1.84 ks. The phase dispersion minimization method is then
used to derive an accurate period of $\sim$ 26.22$\pm$0.74 ks, which is in agreement with previous studies \citep{Esposito2015}.
The folded light curve is also shown in \autoref{variability.fig}.

\subsection{Inter-observation Variability}
\label{inter.sec}

\begin{figure*}[!htb]
%\figurenum{13}
\center
\includegraphics[width=0.7\textwidth]{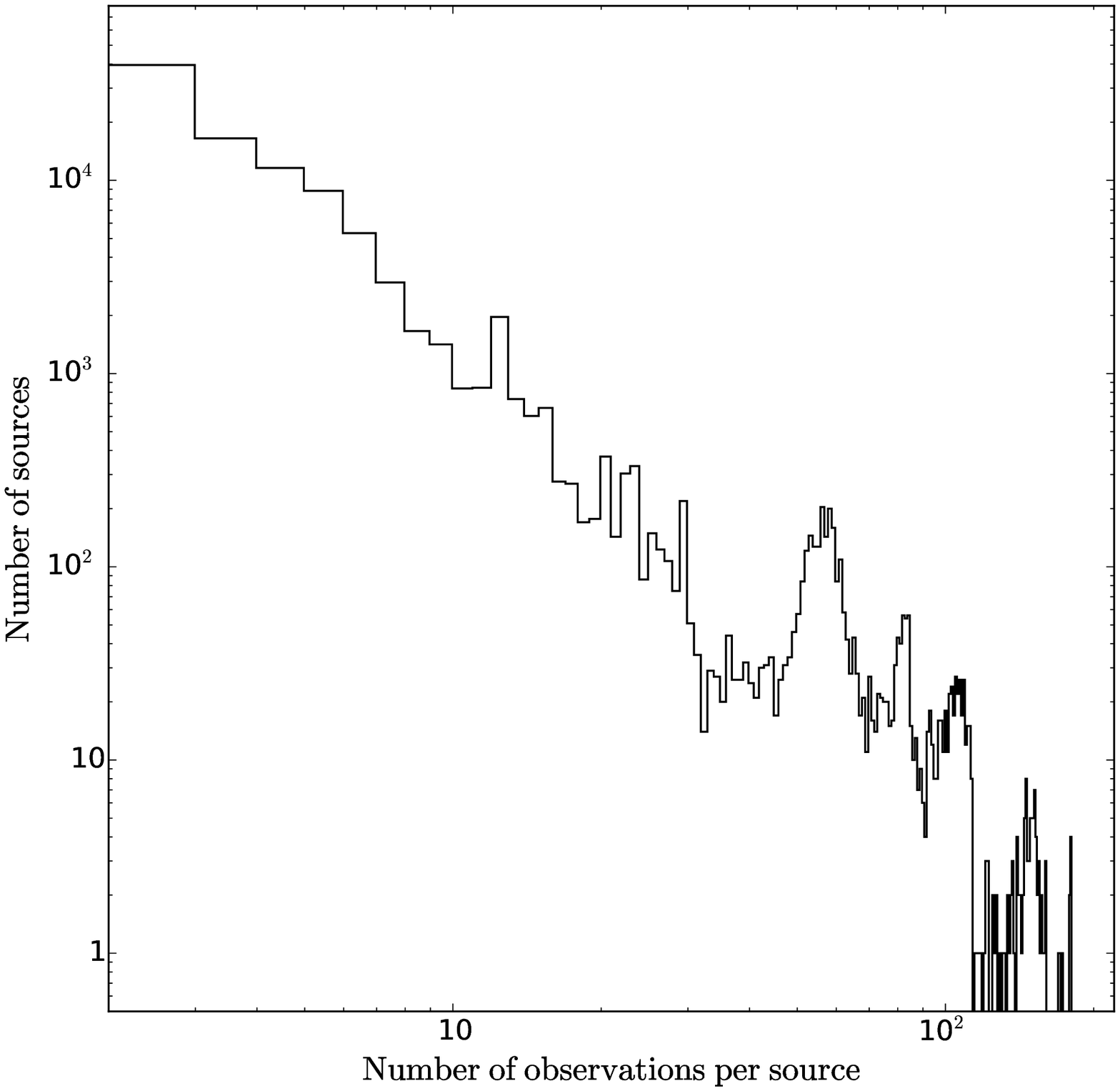}
\caption[]{Distribution of the number of sources observed more than once.}
\label{sourcesta.fig}
\end{figure*}

The inter-observation variability is based on comparison of source fluxes from
multiple observations in which the source is detected.
In this work, about 99,647 sources were observed more than once, with 11,843 sources observed 10
times or more (\autoref{sourcesta.fig}).
The flux ratio $F_{\rm max}/F_{\rm min}$ can be simply used as an extreme indicator of
inter-observation variability, with
$F_{\rm max}$ being the maximum 0.3--8 keV flux from all detections,
and $F_{\rm min}$ being the minimum flux from all detections or upper limits.
About 88,923 sources exhibit different $F_{\rm min}$ and $F_{\rm max}$, including
61,623/33,531/19,324/1496 sources with $F_{\rm max}/F_{\rm min}$ greater than 2/5/10/100, respectively.

\begin{figure*}[!htb]
%\figurenum{14}
\center
\includegraphics[width=0.7\textwidth]{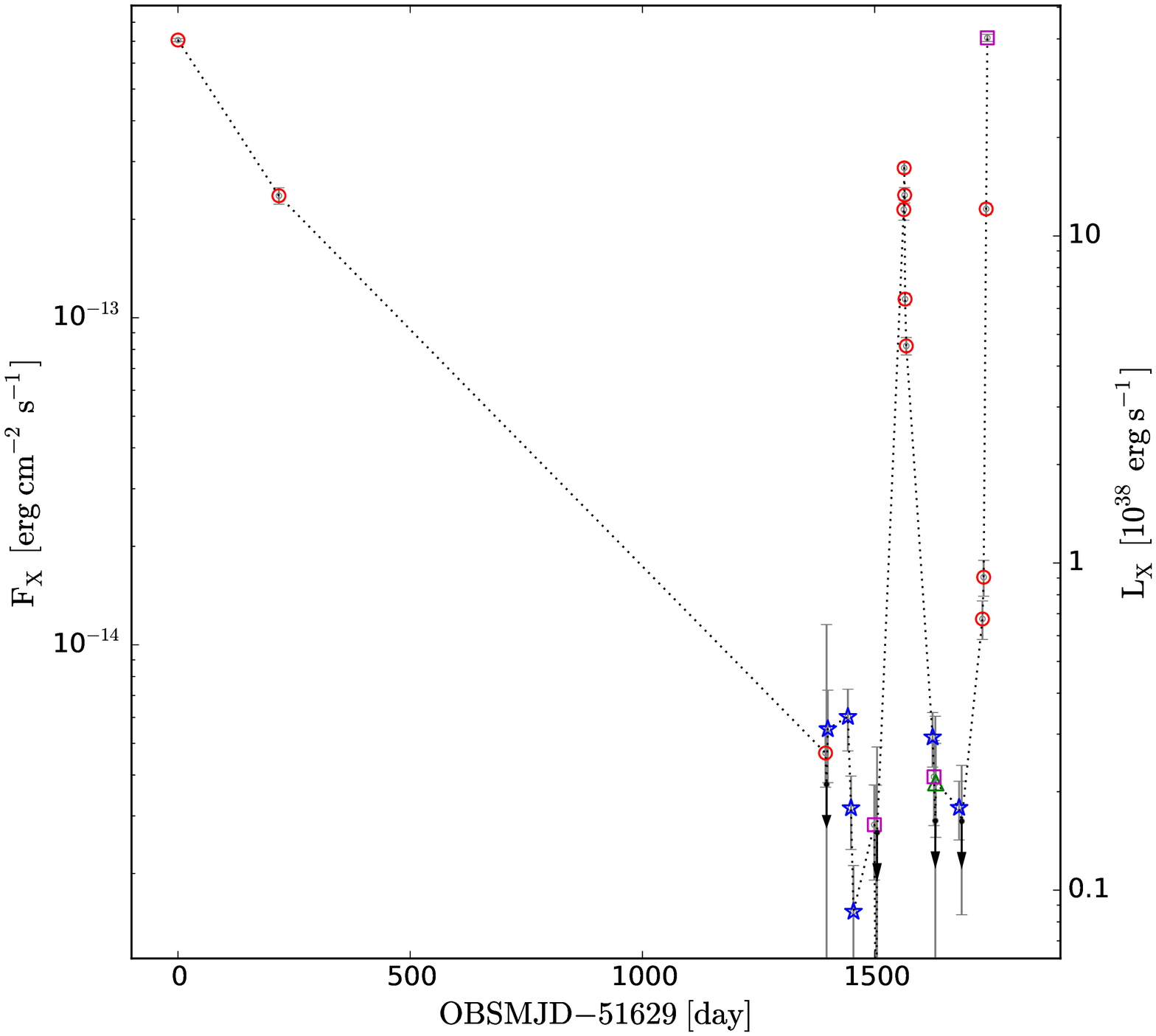}
\caption[]{The light curve for an X-ray source CXOJ140332.371+542102.88.
The fluxes and luminosities in 0.3--8 keV band are computed from the count rates for a
$\Gamma=1.7$ powerlaw model. The source has been observed 26 times
and detected 21 times, with five upper limits (arrow).
The red circles, magenta squares, blue stars, and green triangles indicate
SSS, QSS, hard, and dim phases, respectively. }
\label{longv.fig}
\end{figure*}

As reported by \citet{Evans2010}, the inter-observation variability probability
is quite different from the intra-observation variability probability.
Mostly, the latter criterion can be used to determine a source to be constancy or variable
within the time range of an observation, however,
one can never declare that a source does not vary between different observations.
About 5.9\% (17,092/288,478) of the detected sources above 10 counts
show intra-observation variability, as revealed by the K-S criterion (Section \ref{shortvar.sec}).
This yields 13,040 independent sources, with 7259/4874/3389/439 sources also showing
inter-observation variability with $F_{\rm max}/F_{\rm min} \ge 2/5/10/100$.
An example of long-term and significant variability is shown in \autoref{longv.fig}.
Combining with the long-term light curve, one can easily discover an outburst,
study the temporal X-ray properties,
and explore the emission/accretion mechanisms and
the natures of an X-ray source \citep[e.g.,][]{Kong2005, Mukai2005}.

\subsection{Supersoft X-ray Sources}
\label{soft.sec}

\begin{figure*}[!htb]
%\figurenum{15}
\center
\includegraphics[width=0.7\textwidth]{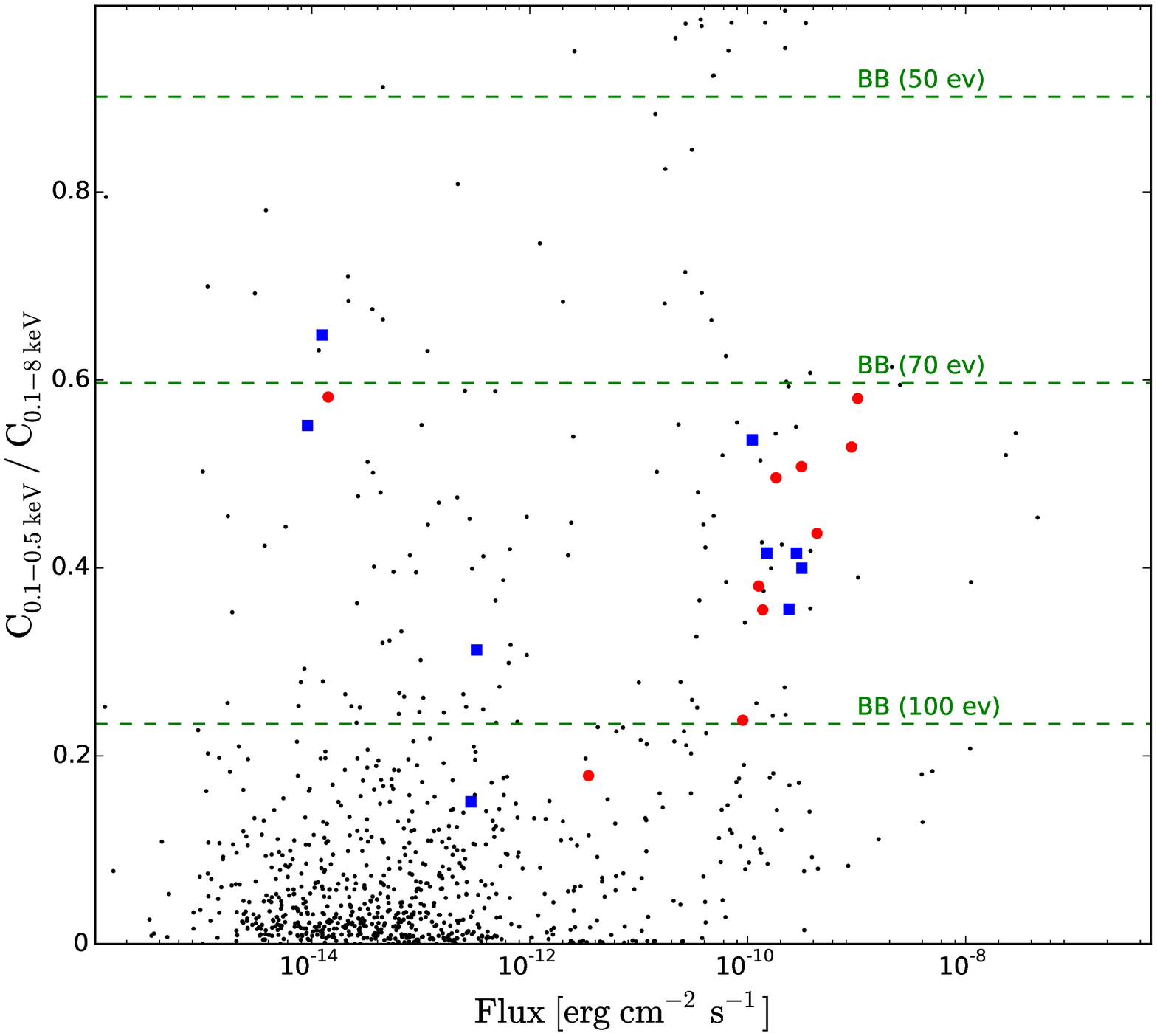}
\caption[]{The ratio of counts $C_{\rm 0.1-0.5~keV}$ to counts $C_{\rm 0.1-8~keV}$ vs.
X-ray flux, which is computed by fitting a blackbody model to the X-ray spectra,
with extinction corrected.
The green lines indicate blackbody models with different temperatures.
M81-ULS1 and M101 X-1 are plotted as examples with red circles and blue squares, respectively.}
\label{b1_5b1_80VSflux.fig}
\end{figure*}

The classification of the very soft X-ray sources have been described in Section \ref{count.sec},
which lead to 1638 individual SSSs.
These sources have extremely soft spectra
with equivalent blackbody temperatures below $\sim$ 100 eV, and are considered as
white dwarfs (WD) burning accreted materials on their surfaces \citep{Kahabka1997}.
\autoref{b1_5b1_80VSflux.fig} shows the distribution of counts ratio
$C_{\rm 0.1-0.5~keV}/C_{\rm 0.1-8~keV}$ with flux for $\sim$ 900 detections of these SSSs.
The flux is computed by fitting a blackbody model to the X-ray spectra,
with extinction corrected.
Three blackbody models with different temperatures (50, 70, and 100 eV) are overplotted
to show the distribution of these SSSs.
Actually, one object can change the hardness greatly in different states, e.g.,
M81-ULS1 and M101 X-1.
About 50 detected sources are located within a very soft region, with
the fraction of counts $C_{\rm 0.1-0.5~keV}$ greater than 50\% of the counts $C_{\rm 0.1-8~keV}$,
which may induce interesting sciences.
Detailed studies on the X-ray spectra and spectroscopic follow-up are needed
to categorize these soft objects and pinpoint their properties.

One special subclass of soft X-ray sources are the ULSs.
These pointlike, extranuclear X-ray sources have peak bolometric
luminosities up to $10^{39}$--$10^{41}$ ergs s$^{-1}$ \citep{Liu2008}.
The ULSs are thought to be high-mass WDs burning matter accreted to the surface
or intermediate-mass black holes (IMBHs) with sub-Eddington accretion
\citep{Kong2004}.
%The ULS studies show that they are likely linked to the physics of two kinds of objects.
%One is a white dwarf (WD) burning accreted materials on its surface \citep{Kahabka1997}, although this model
%cannot well explain the extremely super-Eddington luminosity \citep{Soria2009}.
%The other is an intermediate-mass black hole (IMBH) with sub-Eddington accretion
%\citep{Kong2004}. However, the flux versus temperature relation of ULSs may be a
%strongest argument against this scenario, as the ULSs show a nearly constant temperature
%while the flux varies strongly, which is inconsistent with observations \citep{Tao2012}.
Both scenarios are exciting objects for astronomical studies:
Type Ia supernova progenitors and the seeds of super-mass black holes.
Recently, \citet{Liu2015} reported a discovery of relativistic baryonic jets from M81-ULS1,
and the unusual combination of relativistic jets and persistently supersoft X-ray spectra
raise a new challenge to the conventional theory of jet formation.
Here we define ULSs as those with at least one supersoft phase in which
$L_{X}$ (0.3--8 keV) exceeds $2\times10^{38}$ erg s$^{-1}$ \citep{Liu2011},
e.g., the Eddington luminosity for a WD.
This leads to 166 ULSs within the $D_{25}$ isophotes of 80 galaxies,
and 47 ULSs between the $D_{25}$ and 2$D_{25}$ isophotes of 38 galaxies,
which are listed in Table \ref{independent.tab}.

%and one ULS between the 2$D_{25}$ and 3$D_{25}$ isophotes of one galaxy.
%There are a total 191 ULSs,
%with 19 ULSs projected within two galaxies and two ULXs projected within three galaxies.
%These ULSs are extracted from Table \ref{independent.tab} and listed separately in
%Table \ref{uls.tab} for readers' convenience.
%Cross match using SIMBAD shows 20 objects were classified as galaxy/quasar/AGN, which
%are marked in Table \ref{uls.tab}.

\section{SUMMARY AND CONCLUSIONS}
\label{summary.sec}

The $Chandra$ archival data is a valuable resource for various studies on different topics of X-ray astronomy.
In this paper, we have reduced and analyzed 10,029 observations performed by ACIS,
including 8255 ACIS-I observations and 9388 ACIS-S observations.
For each observation, the on-axis chips are used, including both S2 and S3 chips when the aimpoint is on S3 chip,
and all four I chips when the aimpoint is on an I chip.
The exposure times for the selected observations range from 50 s to 190 ks,
with a total of 221,851 ks.

The purpose of this work is to create a catalog of all point like sources detected
in these observations. To this purpose,
uniform data reduction and analysis procedures were applied to all the fields,
including detecting and visually checking point sources, estimating the source
counts and colors, checking the source variability, and computing the source
spectrum and flux.
The total number of point like objects detected is 363,530, and
the final catalog comprises 217,828 distinct sources,
which is twice the size of the CSC (Version 1.1).
Cross-correlation of these sources with galaxy isophotes yields
17,828 sources within the $D_{25}$ isophotes of 1110 galaxies,
7504 sources between the $D_{25}$ and 2$D_{25}$ isophotes of 910 galaxies,
and 194 sources between the 2$D_{25}$ and 3$D_{25}$ isophotes of 162 galaxies.
Contamination analysis using the log$N$--log$S$ relation shows that 51.3\%
of sources within 2$D_{25}$ isophotes are truly associated with galaxies,
and the fraction increases to 58.9\%, 67.3\%, and 69.1\% for
sources with luminosity above $10^{37}$, $10^{38}$, and $10^{39}$ erg s$^{-1}$.

The catalogs can be used to address a wide range of scientific questions by a broad-based group of scientists.
Among the possible scientific uses, we discussed the possibility to study
intra-observation variability, inter-observation variability, and SSSs.
About 9,650 detected sources above 10 counts are regarded as variable within one observation with the K-S criterion.
Fourier power spectra were computed for 24,247 light curves with more than 200 photons,
which would be very useful to search for coherent or quasi-coherent signals.
There are 99,647 sources observed more than once and 11,843 sources observed 10 times or more,
and about 88,923 sources possess different $F_{\rm min}$ and $F_{\rm max}$, including
61,623, 33,531, 19,324, and 1496 sources
with $F_{\rm max}/F_{\rm min}$ larger than 2, 5, 10, and 100.
%This archival survey leads to 1,995 ULXs with
%$L_X(0.3--8{\rm keV})\ge2\times10^{39}$ erg s$^{-1}$
%within $D_{25}$ isophotes, 1,200 ULXs between $D_{25}$ and 2$D_{25}$ isophotes,
%85 ULXs between 2$D_{25}$ and 3$D_{25}$ isophotes,
%and a total of 3,040 ULXs within 996 host galaxies.
%About 21.5\% of all 363,530 detected sources are quite dim, among the rest about
%87.1\% are classified as hard sources, 8.4\% as QSSs, and 0.8\% as SSSs, which
%lead to 1638 individual SSSs.
%The ULS definition, exhibiting at least one supersoft phase in which
%$L_{X}$ (0.3--8keV) exceeds $2\times10^{38}$ erg s$^{-1}$,
%leads to 166 ULSs within the $D_{25}$ isophotes of 80 galaxies,
%47 ULSs between the $D_{25}$ and 2$D_{25}$ isophotes of 38 galaxies,
%and one ULS between the 2$D_{25}$ and 3$D_{25}$ isophotes of one galaxy.
There are 1638 individual objects classified as SSSs, corresponding to $\sim$ 2350 detections.
About 50 detections show extreme soft X-ray feature, for which
the fraction of counts in 0.1--0.5 keV is greater than 50\% of the counts in 0.1--8 keV.
Detailed studies on the X-ray spectra and spectroscopic follow-up are strongly recommended
to categorize these soft objects and pinpoint their properties.
Furthermore, this survey can enable statistical studies on many aspects,
such as X-ray activities in different types of stars,
X-ray luminosity functions in different types of galaxies,
and multi-wavelength identification and classification on different X-ray populations.

\acknowledgements
\begin{acknowledgements}
This research has made use of data obtained from the $Chandra$ Data Archive,
and software provided by the $Chandra$ X-ray Center (CXC) in the application packages CIAO.
This research has made use of software obtained from the High Energy Astrophysics Science Archive Research Center (HEASARC),
a service of the Astrophysics Science Division at NASA/GSFC and of
the Smithsonian Astrophysical Observatory's High Energy Astrophysics Division.
This research has made use of the SIMBAD database and the VizieR catalogue access tool,
operated at CDS, Strasbourg, France.
Some of the data used in this research are obtained from the Digitized Sky Surveys,
produced at the Space Telescope Science Institute.
The authors acknowledge support from the National Science Foundation of China under grants
NSFC-11273028 and NSFC-11333004, and support from the National Astronomical Observatories,
Chinese Academy of Sciences under the Young Researcher Grant.

\end{acknowledgements}

\clearpage

\begin{table}
\footnotesize
\begin{center}
\caption[]{AVERAGED RELATIVE OFFSET OF THE ECFs DERIVED WITH DIFFERENT RMFs.}\vspace{2mm}
\label{confactor.tab}
\begin{tabular}{ccc}
\tableline
   Chip &  ${\rm (C1-C2)/C2}^a$  & Number of used detections\\
  (1)  &     (2)         &   (3)\\
\hline
i0       &  0.096$\pm$0.195  &   19  \\
i1       & -0.018$\pm$0.100  &   12  \\
i2       &  0.067$\pm$0.125  &   13  \\
i3       &  0.152$\pm$0.070  &   38  \\
s2       &  0.003$\pm$0.081  &   14  \\
s3       &  0.122$\pm$0.038  &   18  \\
\tableline
\end{tabular}
\end{center}
\tablenotetext{a}{${\rm C1}$ represents the ECF derived from RMF at the chip center,
while ${\rm C2}$ represents the ECF derived from RMF at the location of the source.}
\end{table}

\begin{table}
\footnotesize
\begin{center}
\caption[]{FALSE SOURCE RATE ESTIMATES OF THE CATALOG.}\vspace{2mm}
\label{fsr.tab}
\begin{tabular}{cccccccccc}
\tableline
OBSID & ACIS Configuration  & Exposure (ks) & Runs & \multicolumn{3}{c}{Sources} & \multicolumn{3}{c}{False Source Rate}\\
\cmidrule(l){5-7}\cmidrule(l){8-10}
      &                     &                 &      & $\sigma < 3$ & 3 $\leq \sigma <$ 5 & $\sigma \geq$ 5 & $\sigma < 3$ & 3 $\leq \sigma <$ 5 & $\sigma \geq$ 5 \\
(1)       & (2)     & (3)          & (4)     & (5)     & (6)          & (7)      & (8)          & (9)       & (10)\\
\hline
acis379    & ACIS-I (0,1,2,3) & 9     & 50 & 25 &  0 & 0 &  0.5  &  0.0  &  0.0\\
acis1934   & ACIS-I (0,1,2,3) & 29    & 50 & 35 &  7 & 0 &  0.7  &  0.14  &  0.0\\
acis4497   & ACIS-I (0,1,2,3) & 68    & 50 & 34 & 8 & 0 &  0.68  &  0.16  &  0.0\\
acis927    & ACIS-I (0,1,2,3) & 125   & 50 & 45 & 18 & 0 &  0.9  &  0.36  &  0.0\\
acis5337   & ACIS-S (6,7)     & 10    & 50 & 1  &  0 & 0 &  0.02  &  0.0  &  0.0\\
acis4404   & ACIS-S (6,7)     & 30    & 50 & 18 &  0 & 0 &  0.36  &  0.0  &  0.0\\
acis7078   & ACIS-S (6,7)     & 51    & 50 & 40 &  3 & 0 &  0.8  &  0.06  &  0.0\\
acis4613   & ACIS-S (6,7)     & 118   & 50 & 39 &  9 & 0 &  0.78  &  0.18  &  0.0\\
\tableline
\end{tabular}
\end{center}
\end{table}

%\begin{turnpage}
\begin{deluxetable}{ccccccccccrrrccccrcc}
\rotate
\setlength{\tabcolsep}{4pt}
\tablewidth{9.8in}
\tablecolumns{20}
\tabletypesize{\tiny}
\tablecaption{DETECTED X-RAY POINT SOURCES IN THE $Chandra$ ACIS SURVEY.\label{all.tab}}
\tablehead{
   \colhead{Group} & \colhead{CXOGSG J} & \colhead{individual} & \colhead{Expos.} & \colhead{OBSMJD} & \colhead{${\rm D_{\rm edge}}$} & \colhead{OAA} & \colhead{VigF} & \colhead{St. PosErr} & \colhead{PosErr} & \colhead{$\sigma$} & \colhead{Counts} & \colhead{bkgd} & \colhead{$C_{MS}$} & \colhead{$C_{HM}$} & \colhead{$F_X$} & \colhead{$P_{K-S}$} & \colhead{SQH} & \colhead{$\sigma/P_{FFT}$} & \colhead{Notes}\\
   \colhead{No.} & \colhead{} & \colhead{detection} & \colhead{(sec)} & \colhead{} & \colhead{(pixel)} & \colhead{($^{\prime\prime}$)} & \colhead{} & \colhead{($^{\prime\prime},^{\prime\prime}$)} & \colhead{($^{\prime\prime}$)} & \colhead{} & \colhead{} & \colhead{} & \colhead{} & \colhead{(cgs)} & \colhead{} & \colhead{} & \colhead{} & \colhead{} & \colhead{}\\
\colhead{(1)} &  \colhead{(2)}  &  \colhead{(3)} &  \colhead{(4)} & \colhead{(5)} &  \colhead{(6)}  & \colhead{(7)} &  \colhead{(8)}  &  \colhead{(9)} & \colhead{(10)} &  \colhead{(11)} & \colhead{(12)} & \colhead{(13)}  & \colhead{(14)}  &  \colhead{(15)} & \colhead{(16)} & \colhead{(17)}  & \colhead{(18)} & \colhead{(19)}  & \colhead{(20)}
}

\startdata
1 & CXOGSG J003831.2+401711 & 2046.s3.2 & 14762.8 & 51853.696 & 377 & 89 & 0.994 & 0.14,0.07 & 1.0 & 5.7 & 12.4(3.6) & 0.2 & -0.70(0.41) & -0.17(0.50) & 7.35e-15 & 0.32 & QSS & ... & ... \\
 &  & 2047.s3.5 & 14585.9 & 51974.487 & 393 & 193 & 0.974 & 0.17,0.14 & 1.0 & 8.0 & 18.1(4.4) & 0.2 & -0.61(0.44) & -0.21(0.50) & 1.22e-14 & 0.43 & QSS & ... & ... \\
 &  & 2048.s3.11 & 13772.2 & 52093.166 & 359 & 166 & 0.924 & 0.32,0.16 & 1.19 & 3.5 & 7.5(2.8) & 0.1 & 0.17(0.88) & -0.60(0.50) & 5.67e-15 & 0.089 & dim & ... & ... \\
1 & CXOGSG J003834.3+401026 & 2046.s2.1 & 14765.9 & 51853.696 & 107 & 326 & 0.587 & 0.36,0.31 & 1.01 & 15.2 & 43.3(6.8) & 0.9 & 0.40(0.28) & -0.34(0.28) & 5.92e-14 & 0.74 & hard & ... & ... \\
1 & CXOGSG J003837.0+401401 & 2046.s3.5 & 14762.8 & 51853.696 & 123 & 129 & 0.973 & 0.16,0.16 & 1.0 & 4.8 & 10.4(3.3) & 0.2 & -0.09(0.74) & -0.20(0.82) & 6.3e-15 & 0.15 & hard & ... & ... \\
 &  & 2047.s3.1 & 14585.9 & 51974.487 & 123 & 186 & 0.945 & 0.18,0.11 & 1.39 & 3.2 & 6.6(2.6) & 0.1 & -0.41(0.50) & 0.64(0.50) & 4.58e-15 & 0.29 & dim & ... & ... \\
 &  & 2048.s3.u & 13772.2 & 52093.166 & ... & 358 & 0.929 & ... & ... & -0.7 & 4.2(5.7) & ... & ... & ... & 3.17e-15 & ... & ... & ... & ... \\
1 & CXOGSG J003837.7+401357 & 2046.s3.u & 14762.8 & 51853.696 & ... & 138 & 0.970 & ... & ... & -0.4 & 1.0(2.4) & ... & ... & ... & 6.09e-16 & ... & ... & ... & ... \\
 &  & 2047.s3.4 & 14585.9 & 51974.487 & 123 & 185 & 0.917 & 0.16,0.27 & 1.65 & 2.3 & 4.6(2.2) & 0.1 & 0.00(1.00) & 1.00(0.50) & 3.31e-15 & 0.58 & dim & ... & ... \\
 &  & 2048.s3.u & 13772.2 & 52093.166 & ... & 363 & 0.927 & ... & ... & -0.3 & 1.7(5.6) & ... & ... & ... & 1.28e-15 & ... & ... & ... & ... \\
1 & CXOGSG J003838.6+402611 & 2048.s2.8 & 13772.2 & 52093.166 & 131 & 389 & 0.471 & 0.74,0.22 & 3.36 & 2.7 & 6.7(2.8) & 0.3 & -0.09(0.95) & -0.28(0.73) & 1.3e-14 & 0.7 & dim & ... & ... \\
1 & CXOGSG J003838.9+401856 & 2046.s3.4 & 14762.8 & 51853.696 & 366 & 219 & 0.969 & 0.20,0.14 & 1.0 & 11.0 & 30.6(5.7) & 0.9 & -0.06(0.35) & -0.35(0.39) & 1.86e-14 & 0.073 & QSS & ... & ... \\
 &  & 2047.s3.3 & 14585.9 & 51974.487 & 353 & 184 & 0.969 & 0.11,0.07 & 1.0 & 18.5 & 42.1(6.6) & 0.1 & -0.05(0.35) & -0.08(0.39) & 2.84e-14 & 0.94 & hard & ... & ... \\
 &  & 2048.s3.3 & 13772.2 & 52093.166 & 363 & 80 & 0.996 & 0.09,0.05 & 1.0 & 18.1 & 38.5(6.2) & 0.1 & 0.17(0.33) & -0.33(0.34) & 2.71e-14 & 0.74 & hard & ... & ... \\
1 & CXOGSG J003840.6+401956 & 2046.s3.u & 14762.8 & 51853.696 & ... & 282 & 0.915 & ... & ... & -0.9 & 4.8(5.2) & ... & ... & ... & 3.10e-15 & ... & ... & ... & ... \\
 &  & 2047.s3.9 & 14585.9 & 51974.487 & 228 & 231 & 0.946 & 0.26,0.10 & 1.0 & 9.6 & 24.4(5.1) & 0.3 & -0.76(0.46) & 0.24(0.50) & 1.68e-14 & 0.088 & SSS & ... & ... \\
 &  & 2048.s3.4 & 13772.2 & 52093.166 & 261 & 76 & 0.956 & 0.12,0.07 & 1.0 & 5.4 & 10.7(3.3) & 0.0 & -1.00(0.50) & -0.08(1.00) & 7.88e-15 & 0.29 & SSS & ... & ... \\
1 & CXOGSG J003844.1+402407 & 2048.s2.1 & 13772.2 & 52093.166 & 96 & 284 & 0.626 & 0.24,0.19 & 1.08 & 11.0 & 25.9(5.2) & 0.4 & 0.17(0.49) & 0.16(0.37) & 3.81e-14 & 0.39 & hard & ... & ... \\
1 & CXOGSG J003844.3+402409 & 2048.s2.2 & 13772.2 & 52093.166 & 89 & 288 & 0.623 & 0.26,0.20 & 1.17 & 9.9 & 23.0(4.9) & 0.3 & 0.37(0.40) & -0.19(0.39) & 3.39e-14 & 0.64 & hard & ... & ... \\
1 & CXOGSG J003846.5+402259 & 2048.s2.7 & 13772.2 & 52093.166 & 19 & 240 & 0.594 & 0.60,0.38 & 1.94 & 2.5 & 5.4(2.5) & 0.1 & 0.69(0.50) & -0.33(0.84) & 8.41e-15 & 0.36 & dim & ... & ... \\
1 & CXOGSG J003847.0+402422 & 2048.s2.4 & 13772.2 & 52093.166 & 22 & 313 & 0.586 & 0.69,0.47 & 2.54 & 2.6 & 6.0(2.6) & 0.2 & 0.30(0.75) & -0.06(0.96) & 9.49e-15 & 0.034 & dim & ... & ... \\
1 & CXOGSG J003851.0+401806 & 2046.s3.u & 14762.8 & 51853.696 & ... & 279 & 0.945 & ... & ... & -0.7 & 3.4(5.1) & ... & ... & ... & 2.13e-15 & ... & ... & ... & ... \\
 &  & 2047.s3.6 & 14585.9 & 51974.487 & 245 & 114 & 0.946 & 0.23,0.14 & 1.0 & 4.9 & 10.4(3.3) & 0.2 & 0.10(0.87) & -0.11(0.88) & 7.21e-15 & 0.62 & dim & ... & ... \\
 &  & 2048.s3.8 & 13772.2 & 52093.166 & 190 & 224 & 0.954 & 0.41,0.20 & 1.19 & 5.1 & 12.6(3.7) & 0.4 & -0.15(0.79) & 0.14(0.80) & 9.23e-15 & 0.47 & hard & ... & ... \\
1 & CXOGSG J003854.2+401403 & 2047.s2.5 & 14582.7 & 51974.487 & 40 & 163 & 0.677 & 0.19,0.19 & 1.45 & 2.5 & 4.9(2.2) & 0.0 & -0.19(0.84) & 0.20(0.83) & 6.15e-15 & 0.18 & dim & ... & ... \\
1 & CXOGSG J003854.7+402009 & 2046.s3.9 & 14762.8 & 51853.696 & 44 & 390 & 0.919 & 0.48,0.58 & 2.12 & 3.8 & 17.1(5.4) & 8.9 & 0.41(0.48) & -0.12(0.52) & 1.1e-14 & 0.25 & hard & ... & ... \\
 &  & 2047.s3.10 & 14585.9 & 51974.487 & 42 & 246 & 0.851 & 0.27,0.13 & 1.0 & 9.3 & 22.7(4.9) & 0.3 & 0.37(0.44) & -0.13(0.41) & 1.74e-14 & 0.91 & hard & ... & ... \\
 &  & 2048.s3.9 & 13772.2 & 52093.166 & 25 & 241 & 0.879 & 0.34,0.16 & 1.0 & 10.5 & 27.2(5.4) & 0.5 & 0.43(0.36) & -0.14(0.36) & 2.17e-14 & 0.084 & hard & ... & ... \\
1 & CXOGSG J003855.4+401645 & 2046.s3.7 & 14762.8 & 51853.696 & 106 & 296 & 0.900 & 0.21,0.26 & 1.01 & 10.7 & 33.3(6.1) & 1.8 & -0.41(0.33) & -0.02(0.62) & 2.19e-14 & 0.58 & hard & ... & ... \\
 &  & 2047.s3.2 & 14585.9 & 51974.487 & 76 & 98 & 0.984 & 0.08,0.06 & 1.0 & 18.3 & 40.3(6.4) & 0.2 & -0.03(0.29) & -0.45(0.21) & 2.68e-14 & 0.93 & QSS & ... & ... \\
 &  & 2048.s3.6 & 13772.2 & 52093.166 & 148 & 313 & 0.926 & 0.51,0.23 & 1.43 & 6.1 & 19.2(4.8) & 1.7 & -0.51(0.40) & -0.22(0.55) & 1.45e-14 & 0.59 & SSS & ... & ... \\
1 & CXOGSG J003856.6+401831 & 2046.s3.u & 14762.8 & 51853.696 & ... & 348 & 0.916 & ... & ... & -0.2 & 1.0(4.8) & ... & ... & ... & 6.44e-16 & ... & ... & ... & ... \\
 &  & 2047.s3.8 & 14585.9 & 51974.487 & 156 & 170 & 0.948 & 0.32,0.13 & 1.22 & 3.5 & 7.4(2.8) & 0.3 & 0.58(0.67) & -0.59(0.66) & 5.12e-15 & 0.12 & dim & ... & ... \\
 &  & 2048.s3.u & 13772.2 & 52093.166 & ... & 275 & 0.890 & ... & ... & -0.8 & 3.8(4.7) & ... & ... & ... & 2.99e-15 & ... & ... & ... & ... \\
1 & CXOGSG J003900.3+401604 & 2047.s2.1 & 14582.7 & 51974.487 & 42 & 154 & 0.676 & 0.20,0.09 & 1.0 & 5.6 & 10.9(3.3) & 0.1 & 0.19(0.82) & 0.09(0.73) & 1.37e-14 & 0.088 & hard & ... & ... \\
 &  & 2048.s3.u & 13772.2 & 52093.166 & ... & 383 & 0.876 & ... & ... & -2.2 & 15.3(7.0) & ... & ... & ... & 1.22e-14 & ... & ... & ... & ... \\
1 & CXOGSG J003909.7+401705 & 2047.s2.6 & 14582.7 & 51974.487 & 171 & 265 & 0.643 & 0.29,0.49 & 1.58 & 4.6 & 10.2(3.3) & 0.3 & 0.69(0.44) & -0.36(0.71) & 1.36e-14 & 0.3 & dim & ... & ... \\
1 & CXOGSG J003910.8+401215 & 2047.s2.8 & 14582.7 & 51974.487 & 325 & 373 & 0.611 & 0.74,0.47 & 1.73 & 6.5 & 22.1(5.2) & 3.3 & 0.34(0.34) & 0.40(0.36) & 3.09e-14 & 0.64 & hard & ... & ... \\
1 & CXOGSG J003911.2+401742 & 2047.s2.3 & 14582.7 & 51974.487 & 113 & 290 & 0.620 & 0.29,0.24 & 2.59 & 2.3 & 4.7(2.2) & 0.1 & 0.80(0.50) & -0.60(0.70) & 6.43e-15 & 0.48 & dim & ... & ... \\
1 & CXOGSG J003913.4+401353 & 2047.s2.4 & 14582.7 & 51974.487 & 438 & 340 & 0.516 & 0.87,0.64 & 1.73 & 5.9 & 16.5(4.4) & 1.9 & 0.37(0.55) & -0.27(0.57) & 2.73e-14 & 0.14 & hard & ... & ... \\

\enddata
\tablecomments{This table is available in its entirety in machine-readable and Virtual Observatory (VO) forms in the online journal. A portion is shown here for guidance regarding
its form and content. See text for column descriptions.}
\end{deluxetable}
%\end{turnpage}

\clearpage
%\begin{turnpage}
\begin{deluxetable}{cccrrrrrrrrrrrrrrr}
\rotate
\setlength{\tabcolsep}{1pt}
\tablewidth{9.8in}
\tablecolumns{18}
\tabletypesize{\tiny}
\tablecaption{COUNTS IN DIFFERENT BANDS FOR DETECTED X-RAY POINT SOURCES IN THE $Chandra$ ACIS SURVEY.\label{band.tab}}
\tablehead{
   \colhead{Group} & \colhead{CXOGSG J} & \colhead{individual} & \colhead{0.1--1.1} & \colhead{0.3--1.1} & \colhead{0.3--1} & \colhead{0.1--0.5} & \colhead{0.3--0.5} & \colhead{0.1--0.8} & \colhead{0.3--0.8} & \colhead{1--2} & \colhead{1.1--2} & \colhead{0.5--2.4} & \colhead{0.5--2} & \colhead{0.8--2} & \colhead{2--8} & \colhead{0.1--8} & \colhead{0.3--8} \\
   \colhead{No.} & \colhead{} & \colhead{detection} & \colhead{(keV)} & \colhead{(keV)} & \colhead{(keV)} & \colhead{(keV)} & \colhead{(keV)} & \colhead{(keV)} & \colhead{(keV)} & \colhead{(keV)} & \colhead{(keV)} & \colhead{(keV)} & \colhead{(keV)} & \colhead{(keV)} & \colhead{(keV)} & \colhead{(keV)} & \colhead{(keV)} \\
\colhead{(1)} &  \colhead{(2)}  &  \colhead{(3)} &  \colhead{(4)} & \colhead{(5)} &  \colhead{(6)}  & \colhead{(7)} &  \colhead{(8)}  &  \colhead{(9)} & \colhead{(10)} &  \colhead{(11)} & \colhead{(12)} & \colhead{(13)}  & \colhead{(14)}  &  \colhead{(15)} & \colhead{(16)} & \colhead{(17)}  & \colhead{(18)}
}

\startdata
1 & CXOGSG J003831.2+401711 & 2046.s3.2 & 12.0(4.57) & 12.0(4.57) & 11.0(4.43) & 1.0(2.32) & 1.0(2.32) & 6.0(3.6) & 6.0(3.6) & 2.0(2.66) & 1.0(2.32) & 12.0(4.57) & 12.0(4.57) & 7.0(3.78) & -0.17(1.87) & 12.83(4.71) & 12.83(4.71) \\
 &  & 2047.s3.5 & 13.83(4.84) & 14.0(4.84) & 12.0(4.57) & -0.17(1.87) & 0.0(1.87) & 9.83(4.28) & 10.0(4.28) & 3.0(2.94) & 1.0(2.32) & 15.0(4.97) & 15.0(4.97) & 5.0(3.4) & -0.17(1.87) & 14.67(4.97) & 14.83(4.97) \\
 &  & 2048.s3.11 & 3.83(3.18) & 3.83(3.18) & 2.83(2.94) & 1.92(2.66) & 1.92(2.66) & 2.83(2.94) & 2.83(2.94) & 4.0(3.18) & 3.0(2.94) & 4.92(3.4) & 4.92(3.4) & 4.0(3.18) & -0.08(1.87) & 6.75(3.79) & 6.75(3.79) \\
1 & CXOGSG J003834.3+401026 & 2046.s2.1 & 12.75(4.71) & 12.75(4.71) & 7.75(3.96) & 0.92(2.33) & 0.92(2.33) & 6.75(3.79) & 6.75(3.79) & 24.83(6.08) & 19.83(5.56) & 33.67(6.9) & 31.67(6.73) & 25.83(6.17) & 10.42(4.43) & 43.0(7.7) & 43.0(7.7) \\
1 & CXOGSG J003837.0+401401 & 2046.s3.5 & 4.92(3.4) & 4.92(3.4) & 4.92(3.4) & -0.08(1.87) & -0.08(1.87) & 3.92(3.18) & 3.92(3.18) & 3.92(3.18) & 3.92(3.18) & 9.83(4.28) & 8.92(4.12) & 4.92(3.4) & 1.83(2.66) & 10.67(4.43) & 10.67(4.43) \\
 &  & 2047.s3.1 & 3.92(3.18) & 1.92(2.66) & 1.92(2.66) & 2.92(2.94) & 0.92(2.33) & 3.92(3.18) & 1.92(2.66) & -0.08(1.87) & -0.08(1.87) & 1.92(2.66) & 0.92(2.33) & -0.08(1.87) & 3.0(2.94) & 6.83(3.79) & 4.83(3.4) \\
 &  & 2048.s3.u & ... & ... & ... & ... & ... & ... & ... & ... & ... & ... & ... & ... & ... & ... & ... \\
1 & CXOGSG J003837.7+401357 & 2046.s3.u & ... & ... & ... & ... & ... & ... & ... & ... & ... & ... & ... & ... & ... & ... & ... \\
 &  & 2047.s3.4 & 1.0(2.32) & 0.0(1.87) & 0.0(1.87) & 1.0(2.32) & 0.0(1.87) & 1.0(2.32) & 0.0(1.87) & 0.0(1.87) & 0.0(1.87) & 0.0(1.87) & 0.0(1.87) & 0.0(1.87) & 2.0(2.66) & 3.0(2.94) & 2.0(2.66) \\
 &  & 2048.s3.u & ... & ... & ... & ... & ... & ... & ... & ... & ... & ... & ... & ... & ... & ... & ... \\
1 & CXOGSG J003838.6+402611 & 2048.s2.8 & 4.5(3.41) & 4.5(3.41) & 3.5(3.19) & 0.0(1.87) & 0.0(1.87) & 1.83(2.66) & 1.83(2.66) & 2.83(2.94) & 1.83(2.66) & 6.33(3.79) & 6.33(3.79) & 4.5(3.41) & 0.83(2.33) & 7.17(3.97) & 7.17(3.97) \\
1 & CXOGSG J003838.9+401856 & 2046.s3.4 & 16.67(5.22) & 16.92(5.21) & 14.92(4.97) & 1.75(2.66) & 2.0(2.66) & 9.75(4.28) & 10.0(4.28) & 13.0(4.71) & 11.0(4.43) & 26.83(6.27) & 25.92(6.17) & 17.92(5.33) & 2.42(2.95) & 30.08(6.64) & 30.33(6.64) \\
 &  & 2047.s3.3 & 16.83(5.21) & 16.92(5.21) & 14.92(4.97) & 1.92(2.66) & 2.0(2.66) & 10.83(4.43) & 10.92(4.43) & 13.0(4.71) & 11.0(4.43) & 27.92(6.36) & 25.92(6.17) & 17.0(5.21) & 10.0(4.28) & 37.83(7.23) & 37.92(7.23) \\
 &  & 2048.s3.3 & 12.92(4.71) & 12.92(4.71) & 11.92(4.57) & 2.0(2.66) & 2.0(2.66) & 8.0(3.96) & 8.0(3.96) & 18.0(5.33) & 17.0(5.21) & 28.92(6.45) & 27.92(6.36) & 21.92(5.77) & 6.0(3.6) & 35.92(7.06) & 35.92(7.06) \\
1 & CXOGSG J003840.6+401956 & 2046.s3.u & ... & ... & ... & ... & ... & ... & ... & ... & ... & ... & ... & ... & ... & ... & ... \\
 &  & 2047.s3.9 & 23.83(5.98) & 5.83(3.6) & 5.83(3.6) & 24.0(5.97) & 6.0(3.6) & 23.83(5.98) & 5.83(3.6) & 0.0(1.87) & 0.0(1.87) & -0.17(1.87) & -0.17(1.87) & 0.0(1.87) & 1.83(2.66) & 25.67(6.17) & 7.67(3.96) \\
 &  & 2048.s3.4 & 11.0(4.43) & 1.0(2.32) & 1.0(2.32) & 11.0(4.43) & 1.0(2.32) & 11.0(4.43) & 1.0(2.32) & 0.0(1.87) & 0.0(1.87) & 0.0(1.87) & 0.0(1.87) & 0.0(1.87) & -0.08(1.87) & 10.92(4.43) & 0.92(2.33) \\
1 & CXOGSG J003844.1+402407 & 2048.s2.1 & 5.0(3.4) & 5.0(3.4) & 5.0(3.4) & 1.0(2.32) & 1.0(2.32) & 3.0(2.94) & 3.0(2.94) & 10.0(4.28) & 10.0(4.28) & 18.9(5.45) & 14.0(4.84) & 12.0(4.57) & 14.71(4.97) & 29.71(6.55) & 29.71(6.55) \\
1 & CXOGSG J003844.3+402409 & 2048.s2.2 & 6.0(3.6) & 6.0(3.6) & 4.0(3.18) & 1.0(2.32) & 1.0(2.32) & 3.0(2.94) & 3.0(2.94) & 13.81(4.84) & 11.81(4.57) & 19.71(5.56) & 16.81(5.22) & 14.81(4.97) & 8.71(4.13) & 26.52(6.27) & 26.52(6.27) \\
1 & CXOGSG J003846.5+402259 & 2048.s2.7 & -0.08(1.87) & -0.08(1.87) & -0.08(1.87) & -0.08(1.87) & -0.08(1.87) & -0.08(1.87) & -0.08(1.87) & 3.92(3.18) & 3.92(3.18) & 4.92(3.4) & 3.92(3.18) & 3.92(3.18) & 2.0(2.66) & 5.83(3.6) & 5.83(3.6) \\
1 & CXOGSG J003847.0+402422 & 2048.s2.4 & 2.0(2.66) & 2.0(2.66) & 1.0(2.32) & 0.0(1.87) & 0.0(1.87) & 0.0(1.87) & 0.0(1.87) & 3.0(2.94) & 2.0(2.66) & 3.92(3.18) & 4.0(3.18) & 4.0(3.18) & 2.58(2.95) & 6.58(3.79) & 6.58(3.79) \\
1 & CXOGSG J003851.0+401806 & 2046.s3.u & ... & ... & ... & ... & ... & ... & ... & ... & ... & ... & ... & ... & ... & ... & ... \\
 &  & 2047.s3.6 & 4.83(3.4) & 5.0(3.4) & 3.0(2.94) & -0.17(1.87) & 0.0(1.87) & 1.83(2.66) & 2.0(2.66) & 4.0(3.18) & 2.0(2.66) & 8.0(3.96) & 7.0(3.78) & 5.0(3.4) & 2.92(2.94) & 9.75(4.28) & 9.92(4.28) \\
 &  & 2048.s3.8 & 4.92(3.4) & 4.92(3.4) & 4.92(3.4) & 0.0(1.87) & 0.0(1.87) & 3.0(2.94) & 3.0(2.94) & 3.0(2.94) & 3.0(2.94) & 7.92(3.96) & 7.92(3.96) & 4.92(3.4) & 4.83(3.4) & 12.75(4.71) & 12.75(4.71) \\
1 & CXOGSG J003854.2+401403 & 2047.s2.5 & 1.92(2.66) & 1.92(2.66) & 1.92(2.66) & 1.0(2.32) & 1.0(2.32) & 0.92(2.33) & 0.92(2.33) & 1.0(2.32) & 1.0(2.32) & 1.92(2.66) & 1.92(2.66) & 2.0(2.66) & 2.0(2.66) & 4.92(3.4) & 4.92(3.4) \\
1 & CXOGSG J003854.7+402009 & 2046.s3.9 & 1.75(3.45) & 2.83(3.44) & 2.08(3.22) & -1.83(1.95) & -0.75(1.9) & -2.75(1.99) & -1.67(1.94) & 10.25(4.59) & 9.5(4.45) & 13.5(5.25) & 13.08(5.12) & 14.0(5.11) & 7.75(4.77) & 19.0(6.53) & 20.08(6.52) \\
 &  & 2047.s3.10 & 3.92(3.18) & 4.0(3.18) & 3.0(2.94) & 0.92(2.33) & 1.0(2.32) & 0.92(2.33) & 1.0(2.32) & 11.92(4.57) & 10.92(4.43) & 15.92(5.09) & 13.92(4.84) & 13.92(4.84) & 8.92(4.12) & 23.75(5.98) & 23.83(5.98) \\
 &  & 2048.s3.9 & 4.75(3.4) & 4.75(3.4) & 2.75(2.94) & -0.17(1.87) & -0.17(1.87) & -0.17(1.87) & -0.17(1.87) & 15.0(4.97) & 13.0(4.71) & 19.92(5.56) & 17.92(5.33) & 17.92(5.33) & 10.92(4.43) & 28.67(6.46) & 28.67(6.46) \\
1 & CXOGSG J003855.4+401645 & 2046.s3.7 & 21.92(5.88) & 19.42(5.56) & 19.5(5.56) & 4.33(3.41) & 1.83(2.66) & 19.08(5.56) & 16.58(5.22) & 6.5(3.79) & 6.58(3.79) & 25.0(6.18) & 24.17(6.08) & 9.42(4.29) & 5.75(3.8) & 34.25(7.16) & 31.75(6.91) \\
 &  & 2047.s3.2 & 22.83(5.87) & 21.92(5.77) & 18.92(5.44) & 6.83(3.79) & 5.92(3.6) & 15.83(5.09) & 14.92(4.97) & 17.92(5.33) & 14.92(4.97) & 30.92(6.64) & 30.92(6.64) & 21.92(5.77) & 0.92(2.33) & 38.67(7.31) & 37.75(7.23) \\
 &  & 2048.s3.6 & 17.17(5.34) & 16.42(5.22) & 14.42(4.97) & 2.5(2.95) & 1.75(2.66) & 12.33(4.72) & 11.58(4.58) & 4.58(3.41) & 2.58(2.95) & 17.25(5.34) & 17.25(5.34) & 7.42(3.97) & 0.25(2.35) & 20.0(5.79) & 19.25(5.68) \\
1 & CXOGSG J003856.6+401831 & 2046.s3.u & ... & ... & ... & ... & ... & ... & ... & ... & ... & ... & ... & ... & ... & ... & ... \\
 &  & 2047.s3.8 & 1.0(2.32) & 1.0(2.32) & 1.0(2.32) & 0.0(1.87) & 0.0(1.87) & 1.0(2.32) & 1.0(2.32) & 5.0(3.4) & 5.0(3.4) & 6.0(3.6) & 6.0(3.6) & 5.0(3.4) & 0.92(2.33) & 6.92(3.79) & 6.92(3.79) \\
 &  & 2048.s3.u & ... & ... & ... & ... & ... & ... & ... & ... & ... & ... & ... & ... & ... & ... & ... \\
1 & CXOGSG J003900.3+401604 & 2047.s2.1 & 2.92(2.94) & 2.92(2.94) & 1.92(2.66) & -0.08(1.87) & -0.08(1.87) & 0.92(2.33) & 0.92(2.33) & 4.0(3.18) & 3.0(2.94) & 7.0(3.78) & 6.0(3.6) & 5.0(3.4) & 5.0(3.4) & 10.92(4.43) & 10.92(4.43) \\
 &  & 2048.s3.u & ... & ... & ... & ... & ... & ... & ... & ... & ... & ... & ... & ... & ... & ... & ... \\
1 & CXOGSG J003909.7+401705 & 2047.s2.6 & 0.92(2.33) & 0.92(2.33) & -0.08(1.87) & 0.0(1.87) & 0.0(1.87) & 0.0(1.87) & 0.0(1.87) & 6.0(3.6) & 5.0(3.4) & 5.75(3.6) & 5.92(3.6) & 5.92(3.6) & 2.83(2.94) & 8.75(4.13) & 8.75(4.13) \\
1 & CXOGSG J003910.8+401215 & 2047.s2.8 & -0.58(1.89) & -0.58(1.89) & -0.58(1.89) & 0.0(1.87) & 0.0(1.87) & -0.42(1.89) & -0.42(1.89) & 7.58(3.96) & 7.58(3.96) & 9.92(4.44) & 7.0(3.97) & 7.42(3.97) & 17.33(5.46) & 24.33(6.29) & 24.33(6.29) \\
1 & CXOGSG J003911.2+401742 & 2047.s2.3 & 1.0(2.32) & 1.0(2.32) & 0.0(1.87) & 0.0(1.87) & 0.0(1.87) & 0.0(1.87) & 0.0(1.87) & 4.0(3.18) & 3.0(2.94) & 5.0(3.4) & 4.0(3.18) & 4.0(3.18) & 1.0(2.32) & 5.0(3.4) & 5.0(3.4) \\
1 & CXOGSG J003913.4+401353 & 2047.s2.4 & 3.83(3.18) & 3.83(3.18) & 2.83(2.94) & -0.08(1.87) & -0.08(1.87) & 1.92(2.66) & 1.92(2.66) & 8.92(4.12) & 7.92(3.96) & 12.75(4.71) & 11.83(4.57) & 9.83(4.28) & 4.5(3.62) & 16.25(5.35) & 16.25(5.35) \\
\enddata
\tablecomments{This table is available in its entirety in machine-readable and Virtual Observatory (VO) forms in the online journal. A portion is shown here for guidance regarding
its form and content. See text for column descriptions.}
\end{deluxetable}
%\end{turnpage}

\clearpage
%\begin{turnpage}
\begin{deluxetable}{ccclrrrrrrcrrccc}
\rotate
\setlength{\tabcolsep}{4pt}
\tablewidth{8.5in}
\tablecolumns{16}
\tabletypesize{\tiny}
\tablecaption{INDEPENDENT X-RAY POINT SOURCES IN THE $Chandra$ ACIS SURVEY.\label{independent.tab}}
\tablehead{
   \colhead{Group} & \colhead{CXOGSG J} & \colhead{Pos.} & \colhead{source} & \colhead{$\alpha$} & \colhead{$r_{25}$} & \colhead{Dis.} & \colhead{V/D} & \colhead{$L_{Xmax}$} & \colhead{$F_{Xavg}$} & \colhead{$F/F$} & \colhead{$\sigma_{max}$} & \colhead{$C_{max}$} & \colhead{SQH} & \colhead{Var/K-S} & \colhead{Notes} \\
   \colhead{No.} & \colhead{} & \colhead{Err.} & \colhead{Name} & \colhead{($\prime$)} & \colhead{} & \colhead{(Mpc)} & \colhead{} & \colhead{(erg/s)} & \colhead{(cgs)} & \colhead{} & \colhead{} & \colhead{} & \colhead{} & \colhead{} & \colhead{} \\
\colhead{(1)} &  \colhead{(2)}  &  \colhead{(3)} &  \colhead{(4)} & \colhead{(5)} &  \colhead{(6)}  & \colhead{(7)} &  \colhead{(8)}  &  \colhead{(9)} & \colhead{(10)} &  \colhead{(11)} & \colhead{(12)} & \colhead{(13)}  & \colhead{(14)}  &  \colhead{(15)} & \colhead{(16)}
}

\startdata
1 & CXOGSG J003831.2+401711 & 1 & NGC224-X610 & 75.962 & 0.82 & 0.7516 & 3/3 & 8.27e+35 & 8.41e-15 & $2.2^{+2.1}_{-0.96}$ & 7.959 & 18.120 & q2 d1 & ... & ... \\
1 & CXOGSG J003834.3+401026 & 1.01 & NGC224-X314 & 80.991 & 0.85 & 0.7516 & 1/1 & 4.01e+36 & 5.92e-14 & 1 & 15.200 & 43.326 & h1 & ... & ... \\
1 & CXOGSG J003837.0+401401 & 1 & NGC224-X719 & 77.792 & 0.82 & 0.7516 & 3/2 & 4.27e+35 & 5.44e-15 & $2^{+0.63}_{-0.63}$ & 4.800 & 10.392 & d1 h1 & ... & ... \\
1 & CXOGSG J003837.7+401357 & 1.65 & NGC224-X800 & 77.774 & 0.82 & 0.7516 & 3/1 & 2.24e+35 & 3.31e-15 & $5.4^{+2.6}_{-2.6}$ & 2.264 & 4.644 & d1 & ... & ... \\
1 & CXOGSG J003838.6+402611 & 3.36 & NGC224-X597 & 68.220 & 0.78 & 0.7516 & 1/1 & 8.81e+35 & 1.30e-14 & 1 & 2.715 & 6.650 & d1 & ... & ... \\
1 & CXOGSG J003838.9+401856 & 1 & NGC224-X426 & 73.674 & 0.79 & 0.7516 & 3/3 & 1.93e+36 & 2.47e-14 & $1.5^{+0.65}_{-0.44}$ & 18.499 & 42.120 & q1 h2 & ... & ... \\
1 & CXOGSG J003840.6+401956 & 1 & NGC224-X542 & 72.699 & 0.78 & 0.7516 & 3/2 & 1.14e+36 & 1.23e-14 & $5.4^{+1.1}_{-1.1}$ & 9.572 & 24.363 & s2 & ... & ... \\
1 & CXOGSG J003844.1+402407 & 1.08 & NGC224-X372 & 69.070 & 0.76 & 0.7516 & 1/1 & 2.58e+36 & 3.81e-14 & 1 & 10.973 & 25.900 & h1 & ... & ... \\
1 & CXOGSG J003844.3+402409 & 1.17 & NGC224-X399 & 69.012 & 0.76 & 0.7516 & 1/1 & 2.3e+36 & 3.39e-14 & 1 & 9.885 & 22.992 & h1 & ... & ... \\
1 & CXOGSG J003846.5+402259 & 1.94 & NGC224-X675 & 69.626 & 0.76 & 0.7516 & 1/1 & 5.7e+35 & 8.41e-15 & 1 & 2.529 & 5.432 & d1 & ... & ... \\
1 & CXOGSG J003847.0+402422 & 2.54 & NGC224-X659 & 68.518 & 0.75 & 0.7516 & 1/1 & 6.43e+35 & 9.49e-15 & 1 & 2.618 & 6.040 & d1 & ... & ... \\
1 & CXOGSG J003851.0+401806 & 1 & NGC224-X661 & 72.922 & 0.77 & 0.7516 & 3/2 & 6.26e+35 & 8.22e-15 & $4.3^{+1.3}_{-1.3}$ & 5.056 & 12.552 & d1 h1 & ... & ... \\
1 & CXOGSG J003854.2+401403 & 1.45 & NGC224-X721 & 75.843 & 0.80 & 0.7516 & 1/1 & 4.17e+35 & 6.15e-15 & 1 & 2.516 & 4.873 & d1 & ... & ... \\
1 & CXOGSG J003854.7+402009 & 1 & NGC224-X480 & 70.858 & 0.75 & 0.7516 & 3/3 & 1.47e+36 & 1.67e-14 & $2^{+1.5}_{-0.77}$ & 10.491 & 27.210 & h3 & ... & ... \\
1 & CXOGSG J003855.4+401645 & 1 & NGC224-X442 & 73.516 & 0.77 & 0.7516 & 3/3 & 1.82e+36 & 2.11e-14 & $1.8^{+1}_{-0.6}$ & 18.284 & 40.300 & q1 s1 h1 & ... & ... \\
1 & CXOGSG J003856.6+401831 & 1.22 & NGC224-X742 & 71.954 & 0.76 & 0.7516 & 3/1 & 3.47e+35 & 5.12e-15 & $8^{+3}_{-3}$ & 3.450 & 7.423 & d1 & ... & ... \\
1 & CXOGSG J003900.3+401604 & 1 & NGC224-X582 & 73.523 & 0.77 & 0.7516 & 2/1 & 9.29e+35 & 1.37e-14 & $1.1^{+0.34}_{-0.34}$ & 5.587 & 10.859 & h1 & ... & ... \\
1 & CXOGSG J003909.7+401705 & 1.58 & NGC224-X585 & 71.679 & 0.75 & 0.7516 & 1/1 & 9.22e+35 & 1.36e-14 & 1 & 4.579 & 10.228 & d1 & ... & ... \\
1 & CXOGSG J003910.8+401215 & 1.73 & NGC224-X415 & 75.603 & 0.80 & 0.7516 & 1/1 & 2.09e+36 & 3.09e-14 & 1 & 6.545 & 22.100 & h1 & ... & ... \\
1 & CXOGSG J003911.2+401742 & 2.59 & NGC224-X717 & 71.003 & 0.74 & 0.7516 & 1/1 & 4.36e+35 & 6.43e-15 & 1 & 2.282 & 4.662 & d1 & ... & ... \\
1 & CXOGSG J003913.4+401353 & 1.73 & NGC224-X438 & 73.956 & 0.78 & 0.7516 & 1/1 & 1.85e+36 & 2.73e-14 & 1 & 5.879 & 16.491 & h1 & ... & ... \\
1 & CXOGSG J003916.4+400826 & 4.09 & NGC224-X272 & 78.326 & 0.84 & 0.7516 & 1/1 & 5.98e+36 & 8.82e-14 & 1 & 7.930 & 26.979 & h1 & ... & ... \\
1 & CXOGSG J003954.9+414424* & 1.83 & NGC205-X7 & 6.070 & 0.90 & 0.815 & 1/1 & 2.75e+36 & 3.45e-14 & 1 & 7.288 & 25.503 & h1 & ... & ... \\
1 & CXOGSG J003954.9+414424* & 1.83 & NGC224-X394 & 42.496 & 1.37 & 0.7516 & 1/1 & 2.34e+36 & 3.45e-14 & 1 & 7.288 & 25.503 & h1 & ... & ... \\
1 & CXOGSG J003957.4+414436* & 2.39 & NGC205-X15 & 5.797 & 0.83 & 0.815 & 1/1 & 1.17e+36 & 1.47e-14 & 1 & 4.155 & 13.108 & h1 & ... & ... \\
1 & CXOGSG J003957.4+414436* & 2.39 & NGC224-X570 & 42.284 & 1.36 & 0.7516 & 1/1 & 9.96e+35 & 1.47e-14 & 1 & 4.155 & 13.108 & h1 & ... & ... \\
1 & CXOGSG J004001.4+414509* & 1.18 & NGC205-X3 & 5.593 & 0.72 & 0.815 & 1/1 & 4.42e+36 & 5.55e-14 & 1 & 12.889 & 49.626 & h1 & ... & ... \\
1 & CXOGSG J004001.4+414509* & 1.18 & NGC224-X321 & 42.111 & 1.35 & 0.7516 & 1/1 & 3.76e+36 & 5.55e-14 & 1 & 12.889 & 49.626 & h1 & ... & ... \\
1 & CXOGSG J004007.3+414100* & 1.42 & NGC205-X19 & 2.838 & 0.52 & 0.815 & 1/1 & 5.68e+35 & 7.12e-15 & 1 & 3.286 & 6.682 & d1 & ... & ... \\
1 & CXOGSG J004007.3+414100* & 1.42 & NGC224-X702 & 38.528 & 1.24 & 0.7516 & 1/1 & 4.83e+35 & 7.12e-15 & 1 & 3.286 & 6.682 & d1 & ... & ... \\
1 & CXOGSG J004007.7+413750* & 1.71 & NGC205-X17 & 4.336 & 0.65 & 0.815 & 1/1 & 9.88e+35 & 1.24e-14 & 1 & 3.466 & 8.022 & d1 & ... & ... \\
1 & CXOGSG J004007.7+413750* & 1.71 & NGC224-X609 & 36.507 & 1.18 & 0.7516 & 1/1 & 8.41e+35 & 1.24e-14 & 1 & 3.466 & 8.022 & d1 & ... & ... \\
1 & CXOGSG J004009.9+414040* & 1 & NGC205-X5 & 2.402 & 0.44 & 0.815 & 1/1 & 3.81e+36 & 4.78e-14 & 1 & 20.136 & 45.201 & h1 & ... & ... \\
1 & CXOGSG J004009.9+414040* & 1 & NGC224-X342 & 37.942 & 1.23 & 0.7516 & 1/1 & 3.24e+36 & 4.78e-14 & 1 & 20.136 & 45.201 & h1 & ... & ... \\
1 & CXOGSG J004013.7+414236* & 1 & NGC205-X8 & 2.169 & 0.29 & 0.815 & 1/1 & 2.32e+36 & 2.91e-14 & 1 & 13.136 & 27.546 & h1 & ... & ... \\
1 & CXOGSG J004013.7+414236* & 1 & NGC224-X422 & 38.690 & 1.24 & 0.7516 & 1/1 & 1.97e+36 & 2.91e-14 & 1 & 13.136 & 27.546 & h1 & ... & ... \\
1 & CXOGSG J004020.0+414449* & 1.61 & NGC205-X18 & 3.670 & 0.34 & 0.815 & 1/1 & 8.13e+35 & 1.02e-14 & 1 & 3.990 & 9.106 & h1 & ... & ... \\
1 & CXOGSG J004020.0+414449* & 1.61 & NGC224-X646 & 39.420 & 1.25 & 0.7516 & 1/1 & 6.91e+35 & 1.02e-14 & 1 & 3.990 & 9.106 & h1 & ... & ... \\
1 & CXOGSG J004024.1+413620* & 1.84 & NGC205-X14 & 4.843 & 0.45 & 0.815 & 1/1 & 1.4e+36 & 1.76e-14 & 1 & 3.546 & 7.504 & d1 & ... & ... \\
1 & CXOGSG J004024.1+413620* & 1.84 & NGC224-X532 & 33.167 & 1.07 & 0.7516 & 1/1 & 1.19e+36 & 1.76e-14 & 1 & 3.546 & 7.504 & d1 & ... & ... \\
\enddata
\tablecomments{This table is available in its entirety in machine-readable and Virtual Observatory (VO) forms in the online journal. A portion is shown here for guidance regarding
its form and content. See text for column descriptions.}
\end{deluxetable}
%\end{turnpage}

\end{document}